\documentclass[traditabstract]{aa}

\usepackage{graphicx}

\begin{document}

\title{Cool carbon stars in the halo and in dwarf galaxies: H$\alpha$, colours, and variability
\thanks{Based  on observations made with the NTT and 3.6 m telescope at the European Southern Observatory 
(La Silla, Chile; programs 084.D-0302 \& 070.D-0203), with the TAROT telescopes at La Silla \& at
 Observatoire de la C\^{o}te d'Azur (France), and on the exploitation
of  the Catalina Sky Survey and the LINEAR variability databases.
}}

\author{N.~Mauron\inst{1}, K.S.~Gigoyan\inst{2}, P.~Berlioz-Arthaud\inst{3}, A.~Klotz\inst{4}}

\offprints{N.~Mauron}

\institute{Laboratoire Univers et Particules de Montpellier, UMR 5299 CNRS \& Universit\'e  
Montpellier II,  Place Bataillon, 34095 Montpellier, France;  
 \email{nicolas.mauron@univ-montp2.fr}
\and
V.A. Ambartsumian Byurakan Astrophysical Observatory \& Isaac Newton Institute of Chile, 
Armenian Branch, 0213 Aragatsotn Marz, Armenia;
\email{kgigoyan@bao.sci.am}
\and
Observatoire de Lyon, CRAL, UMR 5574 CNRS \& Universit\'e de Lyon I, 9 av. Charles Andr\'e, 
69230 Saint-Genis Laval, France;
\email{paul.berlioz-arthaud@univ-lyon1.fr}
\and
Universit\'e Paul Sabatier  \& Institut de Recherche en Astrophysique et Plan\'etologie, 
UMR 5277 CNRS \& UPS, 14 Av. Edouard Belin, 31400 Toulouse, France;
\email{Alain.Klotz@irap.omp.eu}
}

\date{}

\abstract {

The population of  cool carbon (C) stars located far from the galactic plane is probably made of debris of small 
galaxies such as the \object{Sagittarius dwarf spheroidal} galaxy (Sgr), which are   disrupted by the gravitational
field of the Galaxy. 
We aim to know this population better through spectroscopy, 2MASS photometric colours, and variability data. 
When possible, we compared the halo results to
C star populations in the \object{Fornax dwarf spheroidal} galaxy, Sgr, and the solar neighbourhood. 
We first present a few new discoveries of C stars in the halo 
and in  Fornax. The number of spectra of halo C stars is now 125. Forty percent show
 H$\alpha$ in emission. 
The narrow location in the $JHK$ diagram of the halo C stars is found to 
differ from that of similar C stars in the above galaxies. The light curves of the Catalina and LINEAR variability
databases were exploited  to derive the pulsation periods  of 66 halo C stars. A few supplementary periods
were obtained with the TAROT telescopes. We confirm that the period
distribution of the halo strongly resembles that of Fornax, and we found that it is very different from the C 
stars in the solar neighbourhood. There is a larger proportion of  short period Mira/SRa variables in 
the halo than in Sgr, but the survey for C stars in this dwarf galaxy is not complete, and the study of 
their variability needs to be continued to investigate the link between Sgr and the cool halo C stars.}

\titlerunning{Cool carbon stars in the halo and in dwarf galaxies}
\authorrunning{N.~Mauron et al.}
\maketitle

% ============================================================================

\section{Introduction} 
%%%%%%%%%%%%%%%%%%%%%%  table des coordonnees ===================================
\begin{table*}[!ht]
	\caption[]{Observed carbon stars in the halo and in the dwarf galaxies Carina and Fornax. 
  Coordinates $\alpha$ and $\delta$ (J2000) are given in the object names (2MASS Jhhmmss.ss $\pm$\,ddmmss.s). 
 Galactic coordinates $l$ and $b$ are given in degrees. $B$ and $R$ in mag. are generally taken from the USNO-A2 catalogue,
with 1-$\sigma$ uncertainties of 0.25\,mag. The 
 $J,H,K$ data are near-infrared magnitudes from the 2MASS catalogue, with uncertainties of 0.02 to 0.04 mag.}
	\label{table1}
	\begin{flushleft}
	\begin{tabular}{lcrrrrrrrrrl}
	\noalign{\smallskip}
	\hline
	\hline
	\noalign{\smallskip}
No.&  2MASS name  & $l$~~~~ & $b$~~ & $B$ & $R$~~ & $B$-$R$ & $J$~~ & $H$~~ & $K$~~ & $J$-$K$&Note\\
	\noalign{\smallskip}
	\hline
        \noalign{\smallskip}

                    \multicolumn{12}{c}{New halo C stars found in 2MASS}\\
 102& \object{2MASS J053053.22$-$182524.5} & 221.382 & $-$25.773  & 20.2 &12.9 & 7.3 & 10.064 &  8.628 &  7.636 &  2.428& \\
 103& \object{2MASS J062806.04$-$531105.2} & 261.889 & $-$24.788  & 18.6 &14.6 & 4.0 & 12.744 & 11.744 & 11.268 &  1.476& \\
 104& \object{2MASS J071218.53$-$633809.3} & 274.472 & $-$21.877  & 18.2 &16.5 & 1.7 & 13.583 & 12.156 & 11.067 &  2.516& \\
 105& \object{2MASS J072703.58$-$645912.3} & 276.392 & $-$20.791  & 20.4 &15.4 & 5.0 & 12.973 & 11.723 & 10.859 &  2.114& \\
 106& \object{2MASS J200144.00$-$302446.5} &  10.954 & $-$27.553  & 16.9 &14.0 & 2.9 & 11.659 & 10.832 & 10.358 &  1.301& \\
 107& \object{2MASS J200424.84$-$300651.1} &  11.466 & $-$28.021  & 16.8 &14.8 & 2.0 & 12.400 & 11.473 & 10.815 &  1.585& \\
 108& \object{2MASS J201818.84$-$665057.9} & 328.838 & $-$33.281  & 16.4 &13.9 & 2.5 & 10.818 &  9.677 &  8.977 &  1.841& \\
 109& \object{2MASS J203347.68$-$463620.5} & 353.398 & $-$36.543  & 16.0 &12.5 & 3.5 & 10.505 &  9.538 &  8.920 &  1.585& \\

\noalign{\smallskip}

\multicolumn{12}{c}{Observed halo C stars from the Byurakan survey}\\

- & \object{2MASS J134226.79$-$071522.9} & 324.513 & $+$53.464   & 15.9 &13.1 & 2.8 & 10.520 &  9.617 &  9.238 &  1.282& (1)\\
- & \object{2MASS J151840.24$+$145903.1} & 20.976  & $+$53.729   & 14.9 &11.1 & 3.8 &  8.793 &  7.845 &  7.342 &  1.451& (2)\\
- & \object{2MASS J161817.11$-$045641.9} &  8.321  & $+$30.670   & 14.9 &12.6 & 2.3 &  9.862 &  8.943 &  8.629 &  1.233& (3)\\
- & \object{2MASS J162136.27$-$085318.8} &  5.246  & $+$27.666   & 15.8 &12.1 & 3.7 &  9.334 &  8.332 &  7.826 &  1.508& (4)\\

\noalign{\smallskip}

 \multicolumn{12}{c}{Observed C stars in Carina}\\
 -& \object{2MASS J064113.53$-$505425.0} & 260.033 & $-$22.265  & 18.7 &16.7 & 2.0 & 13.925 & 13.073 & 12.658 &  1.267& (5) \\
 -& \object{2MASS J064141.45$-$505808.0} & 260.119 & $-$22.212  & 18.2 &16.2 & 2.0 & 13.742 & 12.805 & 12.340 &  1.402& (6) \\
 -& \object{2MASS J064144.48$-$510020.9} & 260.160 & $-$22.215  & 19.0 &16.3 & 2.7 & 14.246 & 13.229 & 12.693 &  1.553& (7) \\

\noalign{\smallskip}
 \multicolumn{12}{c}{Observed C  stars in Fornax}\\

 F00& \object{2MASS J023740.58$-$342008.1} & 237.037 & $-$66.141  & 19.7 &17.3 & 2.4 & 15.557 & 14.493 & 13.937 &  1.620& \\
 F06& \object{2MASS J023853.09$-$344919.9} & 238.097 & $-$65.806  & 19.9 &17.4 & 2.5 & 15.474 & 14.280 & 13.452 &  2.022& \\
 F08& \object{2MASS J023857.01$-$344634.0} & 237.981 & $-$65.802  & 22.2 &17.5 & 5.7 & 14.789 & 13.664 & 13.076 &  1.713& \\
 F29& \object{2MASS J023951.79$-$341717.4} & 236.729 & $-$65.705  & 18.0 &16.2 & 1.8 & 15.265 & 14.224 & 14.115 &  1.150& \\
 F31& \object{2MASS J023953.89$-$344402.8} & 237.791 & $-$65.618  & 21.1 &18.9 & 2.2 & 16.470 & 15.177 & 14.277 &  2.193& \\
 F36& \object{2MASS J024009.47$-$340625.7} & 236.271 & $-$65.674  & 21.2 &17.1 & 4.1 & 15.790 & 14.556 & 13.668 &  2.122& \\
 F52& \object{2MASS J024041.09$-$342354.9} & 236.923 & $-$65.519  & 18.3 &16.5 & 1.8 & 15.496 & 14.660 & 14.418 &  1.078& \\
 F58& \object{2MASS J024103.56$-$344805.4} & 237.845 & $-$65.372  & 23.3 &17.6 & 5.7 & 14.441 & 13.365 & 12.694 &  1.747& \\

           \noalign{\smallskip}
   \hline
\end{tabular}
\end{flushleft}

{\small Notes: (1) FBS\,1339$-$070;  (2) FBS\,1516$+$151; (3) FBS\,1615$-$048; (4) FBS\,1618$-$087; 
(5) ALW Carina 2 (Azzopardi et al.\,1986); (6) C3 of Mould et al.\,(1982), also ALW Carina 6; 
(7) C4 of Mould et al.\,(1982), also ALW Carina 7  }

\end{table*}

Carbon stars found at high galactic latitude comprise several types of objects, 
such as asymptotic giant star (AGB) N-type carbon (C) stars, CH-type giants,  
carbon dwarfs, or very metal-poor  carbon-rich objects. 
The AGB C stars 
in the Galactic disc have been known for a long time, are present  in large quantity
(several thousands) and are well documented (for reviews, see, e.g. 
Wallerstein \& Knapp \cite{walknapp98}, Lloyd Evans \cite{lloydevans10}). 
In contrast, their counterparts found 
in the halo are rare, with $\sim$\,150 objects, and their origin is not
entirely clear.  The goal of this paper is to consider these halo AGB C stars as a 
population and to study some of its properties.

 The first discoveries  of  faint, red C-rich objects residing 
 out of the galactic plane were achieved with objective prism surveys
 (Mac Alpine \& Lewis\ \cite{macalpine78}; Sanduleak \& Pesch\ \cite{sanduleak88}),
and Bothun et al.\ (\cite{bothun91}) emphasized the importance of these objects 
for studying the halo properties.
Searching for debris of tidally captured  and dislocated 
systems,  Totten and Irwin (\cite{ti98}; see also Totten et al.\,\cite{tiw00}) 
carried out a systematic survey for 
faint high galactic latitude C stars, with a selection of objects based on 
their very red colour measured on Schmidt plates. They achieved slit 
spectroscopy of candidates  and  found $\sim$\,40  objects comprised of 
CH-type and N-type stars.  Together with results of previous observations, 
the distances and radial velocities of these objects were a decisive step 
that led Ibata et~al.~(\cite{ibata01}) to discover the Sagittarius (Sgr) Stream 
and discuss the halo oblateness. 

These developments underscored the need to increase  the number of 
halo C stars as much as possible, whether they are AGB or CH-type objects. 
Our previous works (Mauron et al.\ \cite{mauron04}, \cite{mauron05}, \cite{mauron07}; 
Mauron~\cite{mauron08}, hereafter Papers I to IV) showed that these C stars can 
be selected on the basis of their near-infrared 2MASS photometry
followed by slit spectroscopy. More than 100 new  C stars were discovered in this way, 
with some of them located as far as $\sim$\,80-130 kpc from us and  used
 as probes of the distant halo (Deason et al.\,\cite{deason12}). At the same time, 
the search based  on the Byurakan prism-objective plates still continues  (Gigoyan et 
al.~\cite{gigoyan01}, \cite{gigoyan12}), providing us with a similar number of 
interesting cases, although less distant in general. The Sloan survey has also 
produced several distant N-type stars (Green \cite{green13}).

These surveys for halo C stars are not completed yet, but it is interesting to study 
the properties of this population, and to compare it with other C populations 
in the Local Group. The AGB C stars are generally used as metallicity indicators
and/or  tracers of intermediate-age star formation episodes (see e.g. 
Mouhcine \& Lan\c{c}on  \cite{mouhcine03}; Groenewegen \cite{groenewegen07}). 
Located in the halo that is old, our C stars are trespassers 
(Battinelli et Demers \cite{battinelli12}). From their sample of halo carbon stars of 
CH or AGB type, Ibata et al.\,(\cite{ibata01}) showed that
a large portion of the AGB objects  trace the Sgr stream, but one can ask whether
they have the same history as those in Sgr, and from where the others originate. 
For almost all galaxies close to the Milky Way, specific surveys of AGB stars 
(including  C stars) have been achieved, but much remains to be done. For example,
the surveys of C stars in Sgr and in Fornax are not complete, despite large efforts
devoted to these systems recently (Whitelock et al.\,\cite{whitelock09}; 
 Battinelli \& Demers \cite{battinelli13}).

In this paper, we focus on three properties of the halo C stars: H$\alpha$ emission, the positions
in the 2MASS $JHK$ two-colour diagram, and  pulsation periods. Our approach is also to compare 
when possible the halo population to Fornax, Sgr, the Magellanic Clouds, and the solar neighbourhood.
Concerning variability, we build on the
pioneering works of Battinelli and Demers (\cite{battinelli12}, \cite{battinelli13}) 
and take advantage of two huge, recently released  databases providing stellar light curves, that is
the Catalina Sky  Survey (hereafter simply Catalina; Drake et al.~\cite{drake09}, \cite{drake13}) 
and the LINEAR survey (Sesar et al.\,\cite{sesar11}, \cite{sesar13}). We also acquired 
some photometric monitoring  with  the TAROT telescopes. These data allow us to compare 
the period distributions of halo C stars and other 
populations with a better statistics thans previously available.

In Sect.\,2, we first present a few new discoveries of C stars in the halo and in
the Fornax dwarf galaxy from ESO observations. The optical monitoring of halo C stars 
is also described. Some spectroscopic results are given in Sect.~3.
 In Sect.~4, we study the location of halo C stars in the 2MASS $JHK$\footnote{The 's' of 
the $K_{\rm s}$ band of 2MASS is omitted in this paper} two-colour diagram.
The light curves of the Catalina and  LINEAR databases and those obtained with the TAROT 
telescopes are exploited in Sect.~5 to derive the variability 
classification\footnote{We classified an object as a Mira if it is a regular (periodic)
 variable with a peak-to-peak amplitude of at least 2.5 mag in the V band,
or 1.5, 0.9, and 0.4 mag in the R band, I band and K band, respectively. 
SRa-type variables are periodic with lower amplitudes.}
 and the period  of 74 halo C stars. In the discussion (Sect.\,6),
 we confirm that the period distribution for Fornax and the halo are very similar. 
Comparison with Sgr and the solar neighbourhood is also provided.  The implications 
on the origin of halo C stars are discussed, before we conclude in Sect.~7.

%%%%%%%%%%%%%%%%%%%%%%%%%%%%%%%%%%%%%%%%%%%%%%%%%%%%%%%%%%%%%%%%%%%%%%%%%%%%%%
 
 \section{Observations}

\subsection{Spectroscopic observations}

  Spectroscopy of halo candidate\footnote{Candidates  obey $J-K \geq 1.2$, $|b| \geq 20$\degr\, 
 and $K \geq 7$. See Sect.~4 for more details on our selection of candidate C stars.} 
 C stars was achieved at ESO (La Silla) 
 on 17--18 October 2009 at the NTT telescope 
 equipped with the EFOSC2 instrument. Grism\,\#\,5 of 300\,gr\,\,mm$^{-1}$ was used, 
 providing spectra in the range 5200--9300\,\AA.  
 A slit of 1\farcs0 was chosen, leading to a resolution of $\sim$\,16~\AA.
 The detector was a Loral 2060\,$\times$\,2060 CCD chip with 15-$\mu$m pixels.
 The frames were binned 2$\times$2, and the resulting dispersion is
 4.1~\AA\, per binned pixel. We were able to secure the spectra of 25 candidates 
 with exposure times of generally a few minutes, and eventually, eight
 were found to be C-rich.  We also observed three carbon stars in 
 the \object{Carina dwarf}  galaxy 
 because they were erroneously believed to be in the halo, and for comparison 
\object{APM~2225$-$1401}, a C star  from the list of Totten and Irwin (\cite{ti98}).

 The reductions included 
 bias subtraction, flat-fielding, extraction of object and sky one-dimension spectra, 
 cleaning of cosmic-ray hits, and wavelength calibration. No
 spectrophotometric standard star was observed, but an approximate calibration  
 was achieved as follows: one of our candidates, 
 \object{2MASS~J234442.57$+$090902.0},  turned out to be a brown dwarf. Its spectral features
 were almost identical to those of the L1 template 2MASS~J143928.36+192914.9,
 for which  spectrophotometry is available (Kirkpatrick et al.\,\cite{kirkpatrick99}).  
 We therefore smoothly rectified  all our spectra in the red 
 region ($\geq$ 6700-9300\,\AA) so that the L-star  spectrum fitted  the L1 template 
 relative intensity. 
  In the region 5700-6700\,\AA, 
 we again smoothly modified the general shape of our spectra so  that the intensity 
 ratio between 7000\,\AA\, and 5700\,\,\AA\, for APM\,2225$-$1401 became identical 
 to that in Fig.\,5 of Totten and Irwin (1998). Finally, to derive an absolute
 calibration, we considered five  (presumably 
 non-variable) candidates found to be M dwarfs, and we used their USNOC-A2 
 $R$ magnitudes  to derive an average scaling factor and obtain $f_{\lambda}$(7000\AA).
 The  spectra  (in the appendix and Fig.\,\ref{figure01}) are  plotted in relative intensity, but the 
 factors to convert them to units of  erg~s$^{-1}$\,cm$^{-2}$\,\AA$^{-1}$  
 are given in the appendix.

We found spectra that covered the H$\alpha$ region for four halo stars in the
 Byurakan Astrophysical Observatory archive. They 
were obtained with the BAO 2.6\,m telescope and the ByuFOSC2 spectrograph. A slit of 2$''$ was used,
together with a 600\,gr\,mm$^{-1}$ grating, and the detector was a Tektronix 1024 CCD 
chip with 24-$\mu$m pixels.  The dispersion was 2.7~\AA\, per pixel, and the resolution is $\sim$\,8~\AA. 
These spectra were taken on 28 March 1999, 12 June 2002, 11 May 2000,
and 11 June 2000.

 Concerning  Fornax, spectra of C stars were found in the ESO Archive
 (program 70.D-0203, P.I. Marc Azzopardi). They were obtained on  5 November  2002  
 with the ESO 3.6\,m telescope 
 and the EFOSC instrument. The slit was 1\farcs5  wide and grism\,\#6 was used. The 
 detector was a  Loral chip with 2048$\times$2048 15-$\mu$m pixels that was binned 2$\times$2, so that
 the dispersion  was 4.19\,\AA\,per binned pixel, and the resolution is 23\,\AA.
 The  spectral coverage is from 4000\,\AA\,\,to 7950\,\AA. Reductions of the raw data
 were carried out as mentioned above. Flux calibration was achieved with  LTT\,2415. 
 All the spectra are shown in the appendix.

  Table\,\ref{table1} gives information first on the eight C stars discovered 
in the halo by searching in the 2MASS catalogue. Then are listed the four halo C stars found 
in Byurakan. The C stars  in Carina and Fornax follow. For halo stars that were 
selected through their position in the $JHK$ two-colour diagram, the first  column 
of Table\,\ref{table1} gives the running number following those of Mauron\,(\cite{mauron08}). 
For Fornax, the first column is an internal  designation. The $B$ and $R$ magnitudes of F29 and F36
are not available in USNO-A2.0 and were derived from Supercosmos (Hambly et al.\,\cite{hambly01}). 
Similarly, those of F31 and F58 were derived from the USNO-B1.0 catalogue. 

The $K$  column of Table\,\ref{table1}  deserves some comment. The $K$ magnitudes 
are around 13.5 for stars in Fornax, and this dwarf galaxy is located at about 140 kpc 
from us (van den Bergh \cite{vandenbergh00}). Our stars are at distances between 10 and 50 kpc
if identical to those in Fornax regarding luminosity. Objects \#103, \#104, and 
\#105 are at the periphery of the Large Magellanic Cloud, and a measurement of their 
radial velocity is necessary to check whether they belong to the halo or to  this galaxy. 

%%%%%%%%%%%%%%%%%%%%%  figure montrant un spectre; les autres en Appendix
    \begin{figure}[!ht]
    \begin{center}
    \resizebox{8cm}{!}{\rotatebox{-90}{\includegraphics{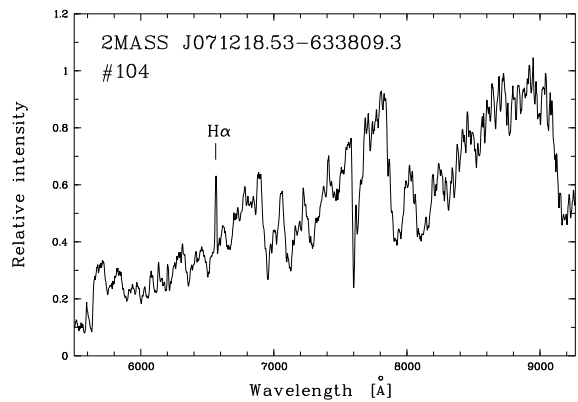}}}
    
    \caption[]{Representative spectrum of halo C stars. Most of the 
features are due to C$_2$ and CN bands. H$\alpha$ is in emission. The strong absorption 
band at 7600\,\AA\, is due to telluric O$_2$.}
    \label{figure01}
    \end{center}
    \end{figure}
%%%%%%%%%%%%%%%%%%%%%%%%%%%%%%%%%%%%%%%%%%%

\subsection{Photometric monitoring}

Sixteen  C stars were monitored with the ground-based 25\,cm diameter TAROT telescopes 
(Klotz et al.\,\cite{klotz08}, \cite{klotz09}).  This monitoring took place irregularly 
at ESO La Silla and Observatoire de la C\^{o}te d'Azur (France) beginning in 2010. 
The studied objects were chosen from the list of Mauron\,(\cite{mauron08}) complemented with
a few sources known to have a very red $J$\,$-$\,$K$. The $V$ and $I$ passbands were 
used, and often $R$ as well. The goal cadence was two or three measurements per week
 and per filter, but this was affected by  technical problems.
The reduction method of the  TAROT observations is described in Damerdji 
et al.\,(\cite{damerdji07}). Periods were estimated by considering the dates of two 
maxima or two minima. A few objects were monitored both in Chile and in France,
allowing a verification of this estimation.  Seven objects were eventually found to have 
Catalina/LINEAR data, and this enabled us to find that most TAROT periods are known 
within 4\% of relative uncertainty, which is quite enough for deriving histograms. The 
results are listed in Table\,\ref{table2}, and all light curves  are given in the Appendix. 
 An example  is shown in Fig.\,\ref{figure02}.

%%%%%%%%%%%%%%%%%%%%%%%%%%%% petite table avec les periodes
         \begin{table*}[!th]
         \caption[]{TAROT observations of cool carbon stars and obtained periods ($P$ in days).
When available, periods derived from Catalina/LINEAR databases are also given ($P_{\rm CL}$ in days).}
        \label{table2}
        \begin{center}
        \begin{tabular}{llrccc}
        \hline
        \hline
        \noalign{\smallskip}
 Name & ~~~2MASS coord. & $K$~~ & $J$-$K$ &  $P$ &  $P_{\rm CL}$\\
        \noalign{\smallskip}
        \hline
        \noalign{\smallskip}
        \noalign{\smallskip}
        
  m85       & J021926.95$+$355058.9 & 8.52 & 3.20 &    474  &    \\
  m86       & J022432.00$+$372933.1 & 8.75 & 2.74 &    327  & 337 \\
  m87       & J023904.88$+$345507.6 & 8.38 & 3.88 &    426  & 424 \\
  CGCS 6306 & J084522.27$+$032711.2 & 6.25 & 3.41 &    325  &     \\
  m34       & J085418.70$-$120054.1 & 8.02 & 2.73 &    373  & 389 \\
  APM~1256+1656 & J125833.51$+$164012.2 & 7.82 & 3.48 &    381  &     \\
  m91       & J133557.07$+$062354.9 &10.57 & 1.69 &    242  & 243 \\
  CGCS 3716 & J163631.69$-$032337.5 & 6.10 & 3.91 &    498  &     \\
  m15       & J172825.74$+$700829.9 & 9.02 & 2.52 &    317  & 301 \\
  m96       & J181329.44$+$453117.5 & 6.71 & 3.81 &    348  &     \\
  IZ Dra = m97 & J184950.90$+$621725.4 & 7.49 & 1.31 &    316  &     \\
  m16       & J194219.01$-$351937.6 &10.04 & 2.62 &    222  & 230 \\
  m77       & J195840.17$+$774526.2 & 8.69 & 3.36 &    382  &     \\
  m100      & J202000.44$-$053550.6 & 8.61 & 3.43 &    397  & 424 \\
  m82       & J215526.98$+$234214.4 & 5.50 & 3.39 &    318  &     \\
APM~2223+2548 & J222619.32$+$260338.5 & 4.52 & 2.73 &    295: &     \\
  
        \noalign{\smallskip}
        \noalign{\smallskip}
        \hline
        \end{tabular}
        \end{center}
{\small Note: $K$ from 2MASS is indicated and is lower than 7.0 in five cases.
These stars are too bright to be considered as halo C star 
with our criteria and are included here for completeness. These objects are ignored in the remainder
of the paper. All objects have $|b|$\,$>$20\degr.
For m16, Battinelli \& Demers (\cite{battinelli12}) found $P$=229\,days.} 
        \end{table*}
%%%%%%%%%%%%%%%%%%%%%%%%%%%%%%%%%%%%%%%%%%%%%%%%%%%%

%%%%%%%%%%%%%%%%%%%%%  figure montrant une courbe de lumiere TAROT; les autres en Appendix
    \begin{figure}[!ht]
    \begin{center}
    \resizebox{8cm}{!}{\rotatebox{-90}{\includegraphics{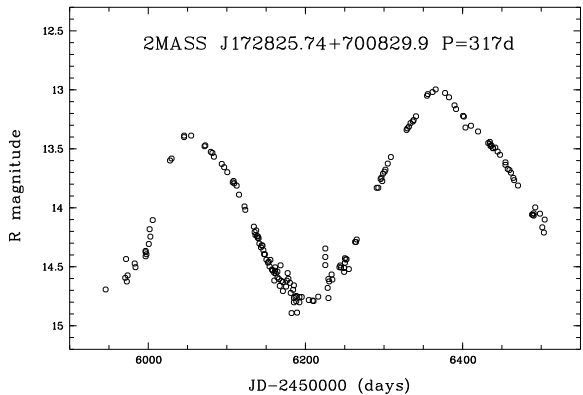}}}
    
    \caption[]{An example of light curve obtained with the TAROT telescopes.}
    \label{figure02}
    \end{center}
    \end{figure}
%%%%%%%%%%%%%%%%%%%%%%%%%%%%%%%%%%%%%%%%%%%

\section{Spectroscopic results}

\subsection{Spectra of Fornax C stars}

The dearchived spectra of Fornax C stars need to be seen in perspective with previous works. 
Studying the  red stars of this galaxy,   Demers \& Kunkel (\cite{demers79}) suggested 
that some of them could be C stars, which was verified by Aaronson \& Mould (\cite{aaronson80}). 
Subsequently, many AGB C stars were discovered thanks to objective-prism spectroscopy  
(Frogel et al.\,\cite{frogel82}; Westerlund et al.\,\cite{westerlund87}). The 2MASS survey
provided homogeneous near-infrared photometry, which was exploited by 
 Demers et al.\,(\cite{demers02}, hereafter DDB02), 
 who identify candidate C stars with a $J$\,$-$\,$K$ between 1.4 and 2.0.  More recently,
 Whitelock et al.\,(2009) presented a comprehensive study of the variable or non-variable 
AGB stars based on multi-epoch near-infrared photometry. Fornax  C stars were also 
specifically sought by  Groenewegen et al.\,(\cite{groenewegen09}, hereafter GLM09) 
through infrared spectroscopy. 

The eight spectra presented here  reveal five new C stars. We collect in Table\,\ref{table3}
various information about them. DDB02 included three of our objects: our spectra 
confirm  their suggested classification as C stars.  The five other stars are 
not in the DDB02 list because of their too faint or too strong $J$\,$-$\,$K$ colours.
GLM09 also included three of our eight objects: their target selection criteria 
excluded F29 and F52, which are too blue. F31 is also absent from their sample, 
because the $J$-band uncertainty is larger than 0.12.
It is unclear why F00 and F36 are not in GLM09, because their colours and flux 
uncertainties obey their criteria. We note, however, that the right ascension of 
F00 is lower than that of all GLM09 objects.
F58 was found in Paper\,\,I to be a C star without H$\alpha$ in
emission (on 30 August 2002), but this line is  in emission in the spectrum 
presented here. This star is a low-amplitude Mira with  a period of 280~days according 
to  Whitelock et al.\,(\cite{whitelock09}).

%%%%%%%%%%%%%%%%%%%%%%  table des proprietes ... et Notes %%%%%%%%%%%%
\begin{table}[!ht]
        \caption[]{Properties of Fornax C stars. Cross identifications with the lists 
of Demers et al.\,(\cite{demers02}, noted DDB02) and Groenewegen 
et al.\,(\cite{groenewegen09}, noted GLM09) 
are given, followed by the indication that the star is variable, with or without period, 
from near-infrared observations by Whitelock et al.\,(\cite{whitelock09}, noted WMF09).}
 	\label{table3}
        \begin{center}
         \begin{tabular}{lllll}
         \noalign{\smallskip}
         \hline
         \hline
         \noalign{\smallskip}
      &  DDB02 &   GLM09  &  WMF09 & Status\\
         \noalign{\smallskip}
         \hline
         \noalign{\smallskip}
 	
F00   & \#1     &  -      &  non variable             &  New C \\
F06   & -       &  -      &  periodic red var.    &  New C \\
F08   & \#2     &  \#32   &  var. with no per.    &  Known C \\
F29   &  -      &  -      &  non variable             &  New C \\
F31   &  -      &  -      &  var. with no per.    &  New C \\
F36   &  -      & \#15    &  -                    &  Known C \\
F52   &  -      &   -     & non variable              &  New C \\
F58   & \#25    & \#27    & periodic red var.     &  Known C \\
 
    \noalign{\smallskip}
    \hline
 \end{tabular}
 \end{center}
 \end{table}
%%%%%%%%%%%%%%%%%%%%%%%%%%%%%%%%%%%%%%%%%%%%%%%%%%%%%%%%%%%%%%%%%%%%%%%%%%%%%%%%

\subsection{H$\alpha$ emission in the spectra of halo C stars}

All eight spectra of 2MASS-selected C stars  are presented in the appendix. 
In Fig.\,\ref{figure01}, we show one representative spectrum. 
Of the eight 2MASS halo objects, five present
H$\alpha$ in emission, although faint in two cases (objects \#105 and \#109). The four FBS spectra
are also shown in the appendix, with one displaying H$\alpha$ in emission. 
One can obtain a 
more meaningful statistics by considering  all the halo C stars discovered in Paper I to IV, and the
present paper as well. We can also add the 26 N-type stars found by Totten and Irwin (their Fig.\,5),
out of which six present H$\alpha$ in emission. We derive a total of 125 spectra of halo C stars 
(with a single spectrum per object), and 50 
have this line in emission, which represents a proportion of  40$\%$. 

 This relatively high 
proportion is connected with the  presence of shock waves in the stellar atmospheres and
the fraction of regularly pulsating  stars in our  sample. As a comparison,  
Lloyd Evans (2013, private comm.) performed a spectroscopic survey of carbon stars 
in the solar neighbourhood at moderate resolution (2\,\AA). 
This survey yields the following results when one considers the variability 
classifications  from the {\it General Catalogue of Variable Stars} (herafter GCVS; 
 Samus et al.\,\cite{samus13}): for Miras, 55 out of 79 
spectra show H$\alpha$ in emission, that is 70 $\%$.
For SRa type stars, the proportion is almost similar, 66$\%$ (31/47). For SRb and Lb type stars, 
it is $\sim$\,20$\%$ (8/47 and 7/31, respectively). 
Finally, among spectra of stars classified simply "SR" or "SR?", only 5$\%$ (2/40) show  
H$\alpha$ in emission. 

%%%%%%%%%%%%%%%%%%%%%%%%%%%%%%%%%%%%%%%%%%%%%%%%%%%%%%%%%%%%%%%%%%%%%%%%%%%%%%%%%%%%%%%%%%%%%%%%%%%%%

\section{Colour properties of halo C stars and comparison with C stars of some galaxies}

\subsection{Colour properties of halo C stars}

In this section, we address the colour properties of cool halo C stars.  
Given that in our previous survey (Paper I to IV)  the halo C stars were selected 
as lying within a very narrow  band of the $JHK$ two-colour diagram,  we investigated 
whether we could have systematically  missed other cool C stars that would be 
located outside this colour region.

\begin{figure}[!ht]
    \begin{center}

    \resizebox{9cm}{!}{\rotatebox{-90}{\includegraphics{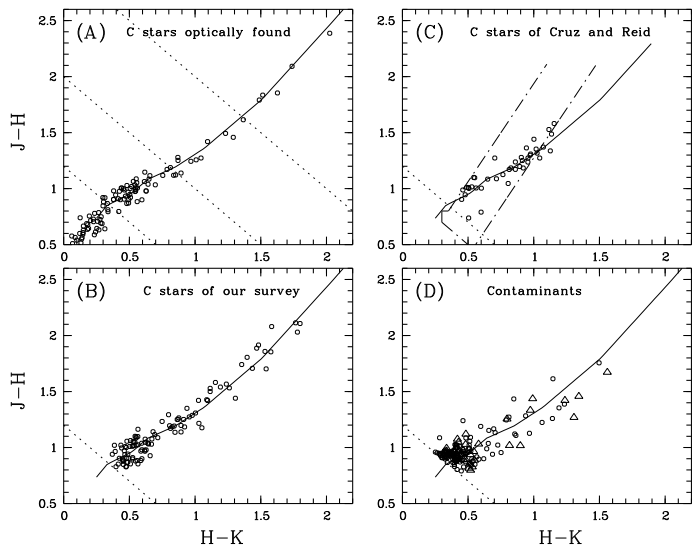}}}

    \caption[]{ {\it Panel A:} Halo C stars optically selected either through 
    objective-prism plates or from optical photometry. The dotted lines indicate
    $J$\,$-$\,$K$\,$=$1.2, 2, and 3. The  $J$\,$-$\,$K$\,$=$1.2 line is repeated
    in panels B, C, and D.
   {\it Panel B:} Our 107 near-infrared (2MASS) selected C stars. {\it Panel C:} Carbon stars 
    discovered by Cruz and Reid during their survey for M, and L-type dwarfs; 
    the colour domain searched by them is indicated by dash-dotted lines. 
    {\it Panel D:} Contaminants in our survey for halo C stars; 
    most are M-type dwarfs, occasionally giants; a few non-M-type objects (L dwarfs, QSOs) are 
    indicated by triangles. The solid line drawn
    in all panels is the best fit for all currently known halo C stars.}
    \label{figure03}
    \end{center}
    \end{figure}

 In Fig.\,\ref{figure03} (Panel A), we show  the position in the $JHK$ colour-colour diagram of
the C stars listed in Totten \& Irwin (\cite{ti98}) and in 
 Gigoyan et al.\,(\cite{gigoyan01}). The vast majority of these
 108 C stars were discovered through their properties at optical wavelengths, either
thanks to their spectra on objective-prism plates or through their $B$\,$-$\,$R$ colour.
All objects obey $|b| > 20$\degr. The 2MASS $JHK$ colours are not corrected for
the galactic extinction because it is negligible in most cases. 
It can be seen that these  optically selected C stars form a narrow sequence in this diagram. 
Our survey for new C stars is based on this property, and
our discoveries in Papers I to IV are plotted in Panel (B). The solid line 
  drawn in all panels
is the mean locus of all known halo C stars with $K > 7$, $J-K$\,$>$\,1.2 
and $|b| > 20$\degr\,(this line is tabulated in the appendix). The  distance 
of representative points from this line is 
called $\epsilon$, with $\epsilon$ negative when the point is located below the line.
It is found that $|\epsilon|$\,$<$\,0.13 mag for most objects. 
So far, of 294 candidate objects for which slit spectra were obtained, we found 107 
 to be C objects. 

 In panel (C), we plot  the C stars   listed in Cruz 
 et\,\,al.\,(\cite{cruz03}, \cite{cruz07})  
and in Reid et al.\,(\cite{reid08}).  These authors performed a survey for cool dwarfs of 
 type M and L close to the Sun, and found these C stars as interlopers. 
Forty of these objects obey $|b|>$\,20\degr\, and $J-K$\,$>$\,1.2. The selection criteria 
used by these authors to find cool dwarfs  involve optical and 
infrared bands. In the $JHK$ plane, the region where their
targets are located is delineated with a dashed-dotted line. It can be noted that 
most of their C-type interlopers are clustered along the mean C line. 
 Of their 40 objects, there are only two stars lying 
at some distance from this line (in the lower part of panel C). 
Their $\epsilon$ is $-0.19$ with $J-K =$  1.24 and 1.39 (lowest points in panel C). 
This suggests that with $|\epsilon| < 0.13$, we miss a very low fraction of halo C stars.

In panel (D) of Fig.\,\ref{figure03}, we plot 
the  non-C stars (contaminants) that were examined, comprising objects found to be 
of type M  and objects of other type (young stellar objects, and more rarely
active galactic nuclei or L-type dwarfs). Their distances to the mean C line is 
larger than in panel (B). It can be seen also that these contaminants are more
concentrated near the limit $J$\,$-$\,$K$\,$=$\,$1.2$ than the C stars. This is because
the majority of contaminants are M dwarfs and their main location in the $JHK$ two-colour 
diagram is below the  $J$\,$-$\,$K$\,$= 1.2$ line (see for example Fig.\,2 of Cruz 
 et al.\,\cite{cruz03}). Despite the high proportion of candidates with   
$J$\,$-$\,$K$\,$=$\,$1.2$ to $1.3$, we found  relatively few C stars in this colour interval.

%--------------------------  figure 2   montrant les diagrammes JHK
%-------------------------------------------------------

\subsection{Comparison of $JHK$ colours between the halo and some galaxies}

\begin{figure}
   \begin{center}
%   \vspace{5mm}
    \includegraphics*[width=6cm,angle=-90]{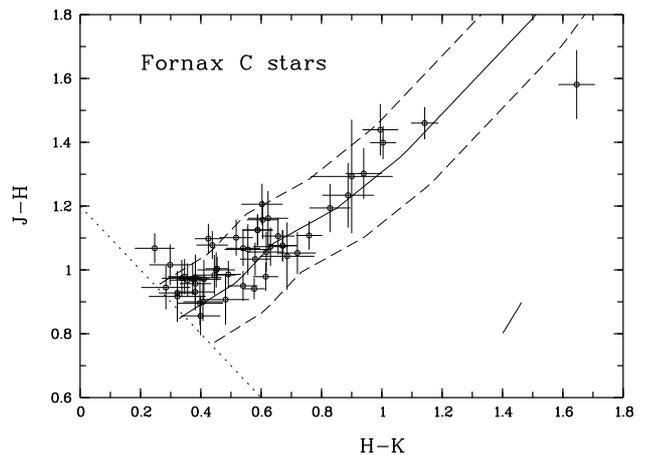}
    \caption[]{Two-colour $JHK$ diagram of the known Fornax C stars. 
The solid line indicates the mean locus of halo C stars ($\epsilon$\,$=$\,0), 
and the dashed lines indicate $\epsilon$\,$=$\,$\pm$0.13 mag. The small inclined
bar in the lower right corner is a reddening vector for $A_V$=1.}
   \label{figure04}
   \end{center}
\end{figure}

We have seen previously that  we used $|\epsilon|$\,$<$\,$0.13$\,mag as a compromise to find  halo C stars as 
efficiently as possible while limiting the number of candidates to be spectroscopically confirmed.
Would this criterion select the C stars in Fornax? We show in Fig.\,\ref{figure04} that this
 is mostly the case: of 52 Fornax 
C stars, 46 are selected, and some of the unselected ones were possibly discarded
 because of poor-quality 2MASS data. However,
Fig.\,\ref{figure04} also suggests that there is an offset between the Fornax and the halo distributions. 
We found that this offset cannot be reduced to zero by introducing the
foreground interstellar extinction of Fornax for two reasons: first, the line of reddening makes a small angle
with  the $JHK$ C star locus; secondly, the interstellar reddening to Fornax is low,  
 $E(B-V)$\,$\sim$\,$0.03$ (van den Bergh \cite{vandenbergh00}). 

\begin{figure}
   \begin{center}
%   \vspace{5mm}
    \includegraphics*[width=6cm,angle=-90]{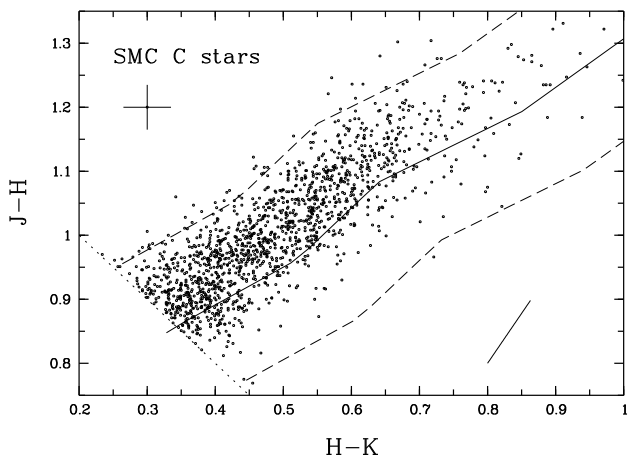}
    \caption[]{Two-colour $JHK$ diagram of optically identified  C stars
in the Small Magellanic Cloud  (small dots). Typical error bars are shown in the upper left corner. 
The solid and dashed lines indicate $\epsilon$\,$=$\,0 and $\epsilon$\,$=$\,$\pm$0.13 mag, 
 respectively, as in Fig.\,4.
The reddening bar in the lower right corner is for $A_V$=1.}
   \label{figure05}
   \end{center}
\end{figure}

To have a comparison with a galaxy with an order of magnitude more C stars but still a low 
foreground extinction, we  considered the Small Magellanic Cloud (hereafter \object{SMC}). 
The C star catalogues of Morgan \& Hatzidimitriou\,(\cite{morgan95}) and of Rebeirot et al.\,(\cite{rebeirot93}) were
cross-matched with 2MASS with a matching radius of 2\arcsec, producing a sample of $\sim$\,1400 objects with relatively
accurate photometry ($JHK$ uncertainties $<$ 0.03~mag.). We find in Fig.\,\ref{figure05}
 that a vast majority (97\%) of C stars in the \object{SMC} verify 
$|\epsilon| < 0.13$\,mag, but again there is a systematic offset of the mean $\epsilon$. This would be reduced to
zero by adopting $E(B-V)$\,$=$\,$0.3$-$0.4$ mag, but this amount of extinction is far above 
the currently admitted value $E(B-V) \sim 0.06 $\,mag  (van den Bergh \cite{vandenbergh00}). 

\begin{figure}
%   \vspace{5mm}
    \includegraphics*[width=8cm,angle=-90]{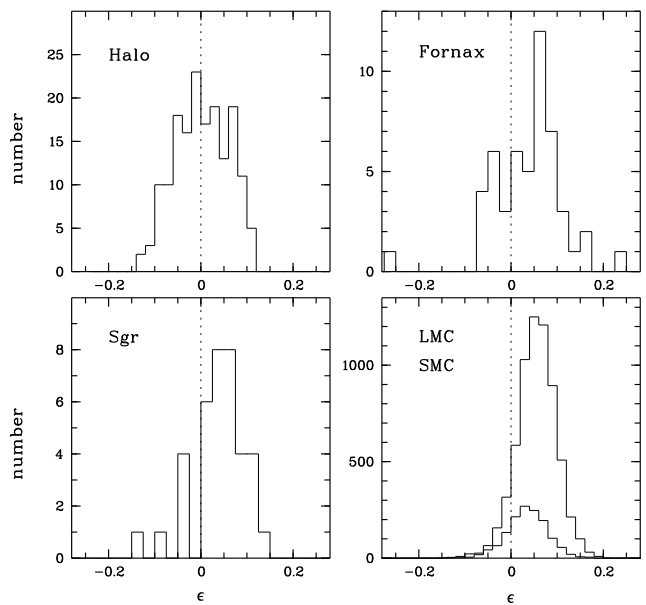}
    \caption[]{Histograms of $\epsilon$ for C stars in the halo, Fornax, Sgr, and the Magellanic
Clouds. $\epsilon$ is the distance to the mean locus of halo C stars in the $JHK$ two-colour diagram.
The distribution of $\epsilon$ is symmetric for the halo by construction. 
Galactic extinction has been taken into account for the four galaxies, but requires very small 
corrections.}
    \label{figure06}
    \end{figure}

 In Fig.\,\ref{figure06}, we plot the histograms of $\epsilon$ for the halo, Fornax, Sgr, and the Magellanic Clouds (MC). 
Extinction was taken into account: for C stars in Sgr and the halo, the $JHK$ colors were derredenned 
using the Schlegel et al.\,(1998) galactic extinctions to individual objects 
before calculating the distance $\epsilon$ to the halo mean line. For C stars in Fornax, in 
the \object{SMC} and in the \object{LMC}, we
adopted $E(B-V)$\,$=$ 0.03, 0.06, and 0.13~mag, respectively (van den Bergh \cite{vandenbergh00}).
Our compilation of C stars in Sgr comes from the lists of Whitelock et al.\,(\cite{whitelock99}), Lagadec 
et al.\,(\cite{lagadec09}), and McDonald et al.\,(\cite{mcdonald12}).
The C stars of the \object{LMC} were taken from Kontizas et al.\,(\cite{kontizas01}). 
This figure shows that the colours of C stars in Fornax, Sgr, and the MCs are quite similar, but 
those of the halo are slightly different. 

We  also attempted to compare of halo C stars with AGB C stars of the solar neighbourhood 
regarding their position in the  $JHK$ diagram. However, we met with two major difficulties: first, 
 the nearest AGB C stars have 2MASS measurements of very poor quality because of saturation. 
 Secondly, when one imposes good-quality 2MASS data, the selected objects are  fainter and therefore 
 more distant. These distances, typically 2\,kpc, imply that extinction is of the  order of 0.3 mag  
 in the $K$ band. Then, the uncertainty on this extinction implies that the positions in the $JHK$ 
 diagram are also too uncertain, and this does not allow us to conclude.

 To summarize, the halo C stars are bluer in $J-H$ for a given $H-K$ than those in Fornax, 
 Sgr, and the MCs.  This difference can be explained if halo C stars are slightly (0.03 mag) 
 fainter  only in the H band (for a given J-K) than in for instance Fornax. Alternatively, one could 
 propose that the J- or K-band flux  are brighter (by $\sim$ 0.05 mag) in the halo 
 C stars than in extragalactic C stars. It might be interesting to search for absorption 
 bands in the H band  of halo C stars that would not be present (or be fainter) in the C 
 stars of the above mentioned galaxies.   Because 2MASS photometry saturates for nearby, 
 bright galactic C stars, a direct comparison is not possible with the results of 
 Cohen et al.\,(\cite{cohen81}). These authors found that \object{SMC} and \object{LMC} C stars are bluer 
 in $J-H$ at a given $H-K$ with respect to galactic C stars. The same effect is seen for three 
 high-velocity N-type stars in the Galaxy (Lloyd Evans \cite{lloydevans11}). 
 This offset is interpreted as an effect of metallicity involving the CN absorption band in 
 the $J$ filter, because N is less abundant. If this is the case here, the location of the halo 
 stars suggests that some of them might be more metal-poor than those in Fornax. This point 
 certainly deserves further investigation.

%%%% figure colcol Fornax %%%%%%%%%%%%%%%%%%%%%%%%%%%
%%%%%%%%%%%%%%%%%%%%%%%%%%%%%%%%%%%%%%%%%%%%%%%%%%%%

%%%% figure colcol SMC %%%%%%%%%%%%%%%%%%%%%%%%%%%

%                --------------------
%%%%%%%%%%%%%%%%%%%%%%%%%%%%%%%%%%

% -----------------------------------------------------------------------------------------------------
\section{Variability of halo C stars: exploiting the  Catalina and LINEAR databases}

%%%%%%%%%%%%%%%%%%%%%%%%% light curve catalina et LINEAR %%%%%%%%%%%%%%%5
\begin{figure*}
\begin{center}
\includegraphics*[width=15cm,angle=-00]{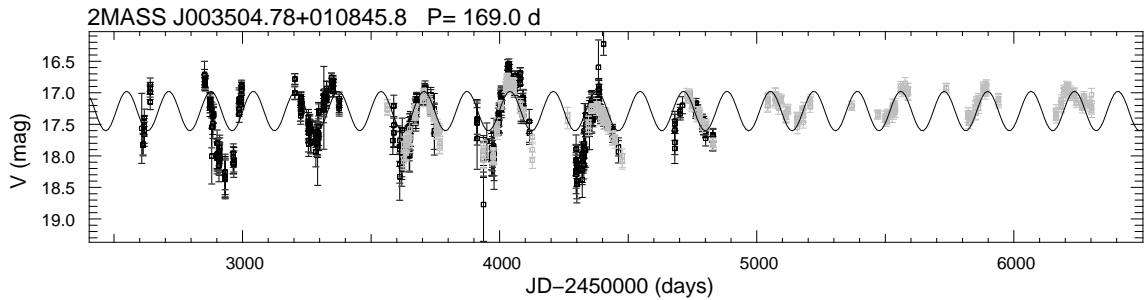}
\caption{ Light curve of m59 with points in black from LINEAR and in grey from Catalina. 
The  solid line is the best-fitting sinusoid. The obtained period is 169 days. 
 This object has $K=13.28$ and is as far away as
 $\sim$~110~kpc  from the Sun (Paper III), illustrating the probed distance of 
 Catalina/LINEAR databases with C stars. All LINEAR/Catalina light curves are shown 
in the appendix.}
\label{figure07}
\end{center}
\end{figure*}
% %%%%%%%%%%%%%%%%%%%%%%%%%%%%%%%%%%%%%%%%%%%%%%%%%%%%%%%%%%%%%%%%%

In addition to specific  infrared colours, cool C stars have the property of 
being      
variable on long time scales, typically $\sim$\,50 to 1000 days. In this section, 
we  use this characteristic to  study known halo C stars. A  variability study has been  
conducted previously by Battinelli \& Demers\,(\cite{battinelli12}). 
By monitoring 23 sources, these authors discovered 13 Miras and 
derived their periods. They also found that the Mira period distribution in the halo 
is quite similar to that of the Fornax dwarf spheroidal galaxy, although the statistics 
 were poor.  Here, we take the 
opportunity to  investigate this similarity further by exploiting the recent release of 
 two databases  useful for variability studies:
the Catalina Sky Survey database\footnote{http://nesssi.cacr.caltech.edu/DataRelease} and the  
 LINEAR  database\footnote{http://astroweb.lanl.gov/lineardb/}. 

Our goal is to extend  the  comparison of the halo and Fornax to more Miras and to SRa-type stars 
(i.e. small-amplitude Miras) with the benefit that  better statistics are obtained. 
A disavantage of considering SRa stars is that one has in general less information (than for Miras) 
of how the periods of these variables are related to the characteristics of a stellar 
population such as  its age. 
  
The Catalina database offers  data for $\sim$\,500~million objects. The  sky coverage 
 is limited by $-75$\degr\,\,$< \delta < 70$\degr\,\, and $|b|$\,$\ga$\,15\degr, for  an area
of $\sim$\,33\,000 square degrees. The database provides for each object 
 the observation epoch, a Catalina V magnitude, and its error. Depending on
the object position, a range of 30 to 400 epochs are given, and the observations cover
a time span of about seven years (2005-2013). This database, initially aimed at the study 
of near-earth objects, is also of
unique value for studying variable stars. For example, Drake et al.\,(\cite{drake13}) presented an
analysis of a large number of RR Lyrae stars in the outer galactic halo  
with typical magnitudes $V$\,$\sim$\,19.
More details on the Catalina instrumentation and photometry can be found at the web site of the database 
  or in the Drake et al. papers.

 The  LINEAR  database offers measurements for $\sim$\,25~million objects over the period  
 1998--2009. The sky coverage is smaller than that of Catalina, but extends over more than  
$\sim$\,10\,000 deg$^2$ in the 
northern hemisphere. Light curves include an average of 250 measurements, and the errors are 
typically 0.2\,mag at Sloan red magnitude $r$\,$\sim$\,18. There is an overlapping interval 
of about three years between the Catalina data and the LINEAR measurements. More details on LINEAR 
can be found in Sesar et al. (\cite{sesar11}, \cite{sesar13}).

To study the  variability of halo C stars, we started with a list of 147 objects
obeying our criteria and searched the Catalina database. Of these 147 objects, 143 were found
to have data. This was complemented  in about half of the cases
by LINEAR measurements. No  shift in magnitudes was found to be needed between Catalina 
and LINEAR magnitudes. 
A periodogram of the data was obtained first. Then, the best four periods were considered
as starting points for four separate non-linear least-squares fits of a sinusoid on the data. 
We used the Levenberg-Marquardt method, and the four fitted quantities are
period, amplitude, maximum date, and maximum magnitude. 
 The fit providing the smallest $\chi^2$ was 
then chosen and inspected by eye. We retained 66 objects with a clear periodicity,
  which represents a proportion of 45$\%$.
Other cases were rejected for several reasons: 1) the number of data points was too small ($n  < 15)$; 2) 
only weak light variation occured given the magnitude errors; 3) the light curve was
 irregular and could not be fitted with a sinusoid.

The list of the retained 66 objects is provided in the appendix, where the light curves
are also shown. An example of the data and fitted sinusoid is given in Fig.\,\ref{figure07}, where one
can appreciate the quality of the data provided by  Catalina and  LINEAR.
The periods  we obtained range from  148 to 515 days with typical uncertainties 
 of $\leq$ 3 days.

There are eight objects in common
with the list of Battinelli \& Demers (\cite{battinelli12}). Our periods generally agree 
 well with theirs 
(see Table\,\ref{table4}) and the differences are not large enough to affect their histograms. However,  
for m18, this difference reaches 45 days. 
For that particular case, the Catalina light curve is of very good quality, and inspection of 
their Fig.~2 suggests that they may have  overestimated  $P$  due to an ill-defined first maximum.

The  amplitudes  are generally 0.3 to 2.1 magnitudes, 
not taking into account flux drops probably caused by dust obscuration events 
(see e.g. Whitelock et al.\,\cite{whitelock06}). If taken at face value, 
these amplitudes are small and no Miras would be identified. 
 However, all Catalina images are taken unfiltered and our C stars are very red.
Consequently, the effective passband  of Catalina light curves 
may be significantly redder than the V band for our objects.  
This shift to the red of the effective 
 wavelength is supported by the fact that Catalina magnitudes agree very well 
 with those of LINEAR, which are $r$-band magnitudes. In addition, the Catalina amplitudes of 
 the eight Miras of Battinelli \& Demers (\cite{battinelli12}) mentioned above
 are between 0.7 and 2.1 magnitudes (see Table\,\ref{table4}). We have indicated in the appendix 
 Table\,A.4 the 19 objects with Catalina amplitudes larger  than  1.5 mag which 
  we assume are Miras, the others in this table are of the SRa type.

%%%%%%%%%%%%%%%%%%%%%%%%%%%%%%%%%%%%%%%%%%%%%%%%%%%%%%%%%%%% %%%%%%%%%%%%%%%%%%%%%%%%%%%%%%%%
\begin{table}[!ht]
\caption[]{Periods and peak-to-peak amplitudes for halo C Miras found by 
Battinelli and Demers (2012). The quantities are noted $P_{\rm BD}$ (in days) 
and $A_K$ (in K-band magnitudes), as determined by these authors. 
The periods found in this work are 
$P$ (in days), and the Catalina amplitudes are $A_C$ (in mag).
The Catalina effective passband is discussed in the text.}
\label{table4}
\begin{center}
\begin{tabular}{cccccc}
\hline 
\hline
\noalign{\smallskip}
\noalign{\smallskip}

name & 2MASS coord. &  $P$  &  $P_{\rm BD}$& $A_C$  & $A_K$\\
%     &  (J2000)     & (days)&  (days)      &      & \\

\noalign{\smallskip}
\hline
\noalign{\smallskip}

m31 &  J001655.77$-$440040.6   &   442    &    465 & 2.1 & 0.7\\
m41 &  J134723.04$-$344723.4   &   307    &    309 & 0.9 & 0.5\\
m52 &  J193734.13$-$353237.7   &   368    &    358 & 0.7 & 0.5\\
m16 &  J194219.01$-$351937.7   &   230    &    229 & 1.5 & 0.9\\
m17 &  J194221.31$-$321104.1   &   233    &    247 & 1.0 & 0.6\\
m18 &  J194850.65$-$305831.7   &   337    &    382 & 1.8 & 1.0\\
m19 &  J195330.18$-$383559.3   &   256    &    256 & 1.6 & 0.7\\
m24 &  J220653.67$-$250628.2   &   331    &    327 & 1.6 & 0.6\\

\noalign{\smallskip}
\hline
\end{tabular}
\end{center}
\end{table}

%%%%%%%%%%%%%%%%%%%%%%%%%%%%%%%%%%%%%%%%%%  DISCUSSION %%%%%%%%%%%%%%%%%%%%%%%%%%%%%%%%%%%%%

\section{Discussion}

%                --------------------
\begin{figure*}
  \begin{center}
%   \vspace{5mm}
    \includegraphics*[width=5cm,angle=-90]{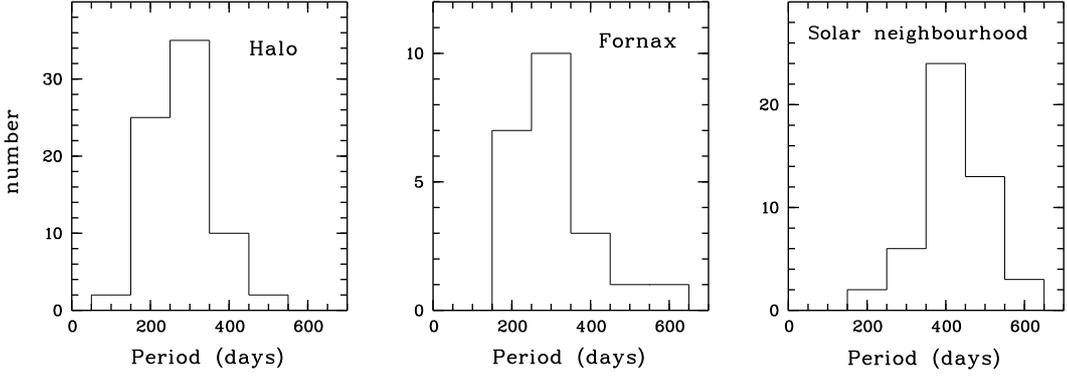}
    \caption[]{Period histograms of Mira or SRa  C stars in the halo (74 objects),
  Fornax (22 objects), and the solar neighbourhood (48 objects). For the latter, 
 the flux-limited sample of  Claussen et al. (\cite{claussen87}) is considered. 
 Note that the period bins are identical to those of Battinelli \& Demers 
(\cite{battinelli12},  \cite{battinelli13}), and that ordinates 
 indicate the number of objects in each bin. See text for more details.}
    \label{figure08}
    \end{center}
    \end{figure*}
%%%%---------------------------------------------------

After collecting halo C star periods from  the Catalina, LINEAR, and TAROT experiments,
and adding those of Battinelli \& Demers\,(2012),  we obtain a total of 74 periodic 
objects (Mira and SRa types) and we can discuss their period distribution, with particular
focus on a comparison with other populations. Here, we  considered the AGB C population of 
Fornax, of  the solar neighbourhood, and of Sagittarius.

\subsection{Comparison with the long-period variables of Fornax}
 
The Catalina  sky coverage includes the Fornax field, but  LINEAR does not. A  
list of C stars in Fornax was first built from the papers by Whitelock et 
 al.\,(\cite{whitelock09}) and Groenewegen et al.\,(\cite{groenewegen09}), 
 in which both variable and non-variable sources were considered. 
We added the five objects discovered by us (see Sect.~2),
yielding a total number of 63 C stars. We emphasize that  no sources lacking 
 slit-spectroscopic confirmation 
of being C rich were included. Data were extracted from the Catalina database and 
periods were searched as described previously. The result is reported 
 in Table\,\ref{table5}, where 
 we have included objects not seen by Catalina, but studied by 
 Whitelock et al.\,(\cite{whitelock09}).  A total of 22 C stars have a period.
When Catalina data are acceptable, there is a reasonable agreement between periods 
 derived by us and periods of Whitelock et al.\,(2009). More precisely, for objects 
 \#38, 47, 58 and 62,  our $P$ are  298, 334, 225, and 240 days,
while Whitelock et al.\,(\cite{whitelock09}) found 303, 320, 235, and 230 days. 
 As shown in Fig.\,\ref{figure08}, the $P$ distributions of Fornax and the halo are almost identical. 
 This fully confirms the findings of Battinelli \& Demers\,(2012).

 Following a suggestion of the referee, we can also compare the number ratio N(Miras)/N(C stars) 
in the halo and in Fornax. There are 19 halo Miras indicated in Table\,A.4, and four detected with TAROT
(m85, APM\,1256$+$1656, m97, m77). Battinelli and Demers (\cite{battinelli12}) 
classified m06, m11, m35, m49, and m55 as Miras. They also classified m17, m41, and m52 as Miras,
but we classify them SRa on the basis of Catalina amplitudes. Finally, m34, and m15 are 
Miras or SRa, depending on whether TAROT or Catalina observations are considered. In summary,
we obtain between 28 and 33 Miras in the halo for a total of 147 halo C stars known. Therefore, the
halo proportion N(Miras)/N(C stars) is $\sim$ 20$\%$.  In Fornax, for 63 C stars known, there are
7 Miras listed in Whitelock et al.\,(\cite{whitelock09}). No additional Miras were found in this work from
Catalina data because all amplitudes are smaller than 1.5 mag. Thus, in Fornax, N(Miras)/N(C stars) is
only 11$\%$. This shows that although the period distributions of Miras\,$+$\,SRa variables are similar, 
there are clear differences between these two populations. This point deserves further study.

%%%%%%%%%%%%%%%%%%%%%%%%%%%%%%%%%%%%%%%%%%%%%%%%%%%%%%%%%%%%5
    \begin{table}[!ht]
  \caption[]{Fornax C stars with periods. $K$ and $J$$-$$K$ are from 2MASS. The period $P$
is in days. The first column is an internal number. W indicates that $P$ is from Whitelock et al.\,(2009),
otherwise $P$ is from this work. }
  \label{table5}
  \begin{center}
  \begin{tabular}{llllll}
  \hline 
  \hline
   \noalign{\smallskip}  
 Id. &~~~2MASS coord.  & ~~~$K$   & $J$$-$$K$  &   ~$P$  & \\
   \noalign{\smallskip}
  \hline
  \noalign{\smallskip}
For04 & J023850.56$-$344031.9 &    12.88 &     3.22  &    350 &  W\\
For05 & J023853.10$-$344919.9 &    13.45 &     2.02  &    242 &  \\
For06 & J023857.01$-$344634.0 &    13.07 &     1.71  &    277 &  \\
For07 & J023857.05$-$344748.9 &    13.56 &     1.25  &    190 &  \\
For09 & J023912.33$-$343245.0 &    12.12 &     2.60  &    470 &   W\\
For18 & J023937.39$-$343626.9 &    12.98 &     1.76  &    255 &   W\\
For22 & J023941.60$-$343556.7 &    14.16 &     3.49  &    400 &   W\\
For23 & J023948.45$-$343507.9 &    12.84 &     1.78  &    284 &  W\\
For33 & J023958.62$-$342528.0 &    13.65 &     1.12  &    634 & \\
For35 & J024001.36$-$342018.9 &    13.34 &     1.34  &    230 & \\
For36 & J024002.52$-$342742.7 &    13.32 &     1.80  &    267 &  W\\
For38 & J024002.74$-$343149.0 &    13.18 &     1.60  &    298 & \\
For43 & J024006.66$-$342322.3 &    12.61 &     1.86  &    335 & \\
For47 & J024010.17$-$343321.9 &    12.54 &     1.51  &    334 & \\
For48 & J024011.79$-$343245.5 &    13.43 &     1.32  &    220 & \\
For49 & J024012.08$-$342625.5 &    12.72 &     2.40  &    375 &  W\\
For51 & J024015.62$-$343403.0 &    13.23 &     1.70  &    280 & \\
For52 & J024017.79$-$342735.8 &    13.18 &     2.24  &    275 & \\
For54 & J024019.95$-$343309.7 &    13.90 &     1.35  &    171 &  W\\
For58 & J024025.05$-$342858.3 &    13.50 &     1.31  &    225 & \\
For62 & J024053.33$-$341213.0 &    13.26 &     1.77  &    240 & \\
For63 & J024103.56$-$344805.4 &    12.69 &     1.74  &    280 &  W\\
  \noalign{\smallskip}
   \hline
  \end{tabular}
   \end{center}
   \end{table}
%%%%%%%%%%%%%%%%%%%%%%%%%%%%%%%%%%%%%%%%%%%%%%%%%%%%%%%%%%%%%%%%%%%%

\subsection{Comparison with  the solar neighbourhood}

Concerning periodic C-rich variables in the solar neighbourhood, we considered the flux-limited sample
of Claussen et al.\,(\cite{claussen87}). This sample is composed of 215 C stars with $K < 3$, with $K$ 
from the {\it Two Micron Sky Survey} that have  $-33$\degr\,$<$\,$\delta$\,$<$\,$+81$\degr.
It is a  statistically complete sample. Information on variability classification and especially periods
can be found in the GCVS, but periods are not defined or not available for all objects.  We found periods for
 33 Miras and  15 SRa-type stars. Fig.\,\ref{figure08} shows the period histogram for these 
solar-neighbourhood objects.
It can be seen that this histogram strongly peaks at $\sim$ 400 days, with very few objects with $P < 350$ days,
in contrast to the halo and Fornax distributions.

\subsection{Comparison with Sagittarius}

Concerning the link between halo and Sgr C stars, it would of course be desirable to compare the period 
distributions of Mira/SRa variables as rigorously as possible, that is, with the two populations 
observed and measured  with the same instrument. Unfortunately,  the Sgr galaxy is too close 
to the galactic plane and is not covered by the Catalina or the  LINEAR experiments.
We tentatively propose, however, to proceed as follows. Four C-rich Miras and one SR in Sgr have 
been reported by  Whitelock et al.\,(1999) with measured periods, and six other C Miras were
found by Lagadec et al.\,(\cite{lagadec09}). Recently, Battinelli \& Demers\,(\cite{battinelli13}) 
identified and monitored  13 Miras and one SR in Sgr. Not all of the latter objects are 
spectroscopically confirmed C stars, but their near-infrared 
colours strongly favour this chemistry. One object of  Battinelli \& Demers\,(\cite{battinelli13}),
called Sgr 13  (2MASS\,J185329.37$-$293824.1), is also in the list 
of  Whitelock et al.\,(\cite{whitelock99}), and the period determinations agree. 
In the end, 22 Miras and two SRs  are known with periods available. 

\begin{figure}
  \begin{center}
%   \vspace{5mm}
    \includegraphics*[width=7.5cm,angle=-90]{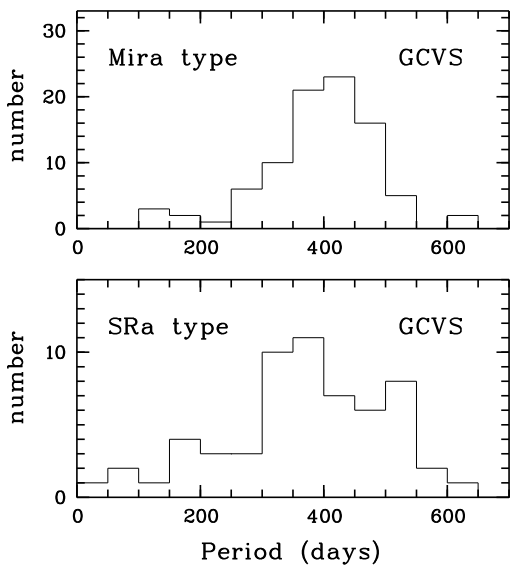}
    \caption[]{Period distributions for carbon variables of Mira-type (89 objects) 
and SRa-type (59) in the General Catalogue of Variable stars (GCVS).}
    \label{figure09}
    \end{center}
    \end{figure}

One point of concern is the low number of SRa stars (low-amplitude Miras) obtained above. Indeed,
we have little information on the population of SRa-type 
C stars in Sgr, to say nothing of their $P$ distribution.  Here, we tentatively  assume 
that this distribution is roughly
similar to that of Miras. This point is mildly  supported by the Miras and SRa of 
the GCVS as shown in  Fig.\,\ref{figure09}.  At least, the median periods of the two distributions are quite similar 
at $\sim$\,400~days. To summarize, we have to keep in mind that the Sgr sample is incomplete.

\begin{figure}
  \begin{center}
%   \vspace{5mm}
    \includegraphics*[width=7.5cm,angle=-90]{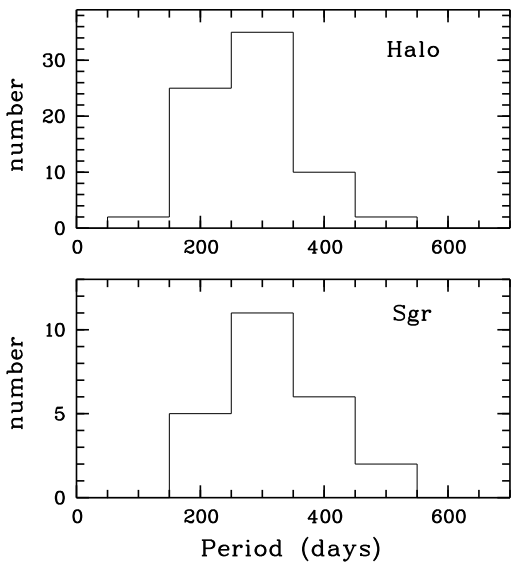}
    \caption[]{Period distribution of Mira/SRa C stars in the halo (74 objects) 
 and in Sgr (24 objects).}
    \label{figure10}
    \end{center}
    \end{figure}

The result is shown in Fig.\,\ref{figure10}, where we compare Miras/SRa in the halo and in Sgr.
The distributions are somewhat different:  
 the Sgr objects with  $P$\,$\ga$\,$350$\,days are relatively more numerous, and there are fewer objects 
with $P$\,$\la$\,$250$\,days.

 %%%%%%%%%%%%%%%%%%%%%%%%%
%%%%%%%%%%%%%%%%%%%%%%%%%%%%%%%%%%

%%%%%%%%%%%%%%%%%%%%%%%%
%%%%%%%%%%%%%%%%%%%%%%%%%%%%%%%%%%

\subsection{Interpretation}

The origin of the halo AGB C stars is not entirely clear. It is often assumed that these stars 
belong to the Sgr Stream, but this may not be the case of them all.
  Ibata et al.\,(\cite{ibata01}) considered
an optically selected sample of 75 faint, high-latitude C stars, with radial velocities and 
spectral classification CH or N-type. They showed that half of this sample (38 objects) 
traces the Sgr Stream. Similarly, Mauron et al.\,(\cite{mauron04}) identified another sample
of 28 halo AGB C stars by using their infrared colours, measured their radial velocities, and
concluded that again half of them belong to this Stream. These findings suggest that
halo C stars may have  different origins. 
To explain the presence of the very red 
dust-enshrouded C star IRAS~08546$+$1732 far from the galactic plane, Cutri et al.\,(\cite{cutri89}) 
mentioned several possibilities, and among them the hypothesis that some progenitors of C stars 
might be ejected from the galactic disc. Our comparison of period distributions for 
the halo and the solar neighbourhood 
strongly suggests that ejection from the disc is relatively minor, because if this were the case,
we should see in the halo a larger portion of objects with $\sim$\,400-day periods. Another
source of C stars in the halo could originate in blue stragglers evolving up to the AGB phase that in turn 
originate in dissolved globular clusters. This is qualitatively supported by the discovery of 
a long-period C Mira ($P=515$\,d) in the cluster Lynga~7 (Matsunaga \cite{matsunaga06}, 
Feast et al.\,\cite{feast13}).

Concerning the similarity of $P$ histograms of the halo and Fornax, it has to be noted that
this similarity was previously noted by Battinelli \& Demers (\cite{battinelli12}). 
Their histograms (in their Fig.~5) can be directly compared with ours in Fig.\,\ref{figure07}, since 
the bins in $P$ are identical. The strongest point
to note is that we have found a short-period population (with $P \leq 250$\,days) that was not 
seen by them due in large part to their small-number statistics and in part because the
optical light curves we used as opposed to near-infrared ones, permitted us to discover 
low-amplitude variables (about half of the  variables with $P \leq 250$\,days have Catalina
 amplitudes lower than 1.0~mag).

We know  the link between C Miras of the galactic disc and their estimated ages  relatively 
well (Feast et al.\,\cite{feast06}), but this link is less well established for SRa type stars. However, 
assuming roughly the same relation, the $P$ histogram of Fornax and the halo indicate, with most
objects having $P \leq 350$\,days, that their ages are older than $\sim$\,3~Gyr. There is
one halo object in our list with $P$ as long as 515\,days. This is 2MASS\,J124337.31+022130.2,
with $K$\,$=$\,$11.45$, $J-K$\,$=$\,$1.54$, and $b$\,$=$\,$+65$\degr. Its amplitude is 
only $\sim$\,0.3\,mag. Inspection of the light curve data suggests that this period might 
be a long secondary period of an irregular variable. We believe that  this object is very 
different from a Mira.

The Sgr histogram contains a larger proportion of $\sim$\,400-day periods than the halo or Fornax.
This is qualitatively consistent with more star formation in Sgr in recent times due probably to a
current disturbance. The best colour-magnitude diagrams  of Sgr indeed show  that multiple, young and 
intermediate-age populations exist in this galaxy with different metallicities
 (e.g. Siegel et al.\,\cite{siegel07}, Giuffrida et al.\,\cite{giuffrida10}). 
 However, the Sgr histogram numbers are small.  Whitelock et al.\,(\cite{whitelock99}) estimate 
the number of AGB C stars in this galaxy to be $\sim$\,100. If about half of these 100 stars 
are Mira/SRa type, as we found for the halo (cf. Sect.~5), the total number of these objects 
could be $\sim$\,50 and the quality of the $P$ distribution could be significantly improved. 
A systematic search and monitoring survey of cool variable populations in Sgr is of course 
highly desirable.

%%%%%%%%%%%%%%%%%%%%%%%%%%%%%%%%%%%%%%%%%%%%%%%   CONCLUSION %%%%%%%%%%%%%%%%%%%%%%%%%%%%%%%%%%%%%%%%%%%

\section{Conclusions}

The main conclusions of this paper are listed below.

\smallskip

{\noindent}(1) Several new AGB C stars were found in the halo and in Fornax, 
and their spectra were presented.

\smallskip

{\noindent}(2) By considering the 125 spectra of halo C stars taken previously or 
in this work, we found that
the halo C stars present H$\alpha$ in emission with a percentage of 40\%.
This fraction is smaller than that found for galactic AGB stars of  Mira or SRa type, but 
clearly larger than found for galactic SRb or irregulars.

\smallskip

{\noindent}(3) Our near-infrared criteria to search for halo C stars, 
in particular $| \epsilon | < 0.13$, do not 
 exclude  those that could be identical to the C stars of  Fornax, of Sagittarius, or
 of the Magellanic Clouds. However, we found that the $JHK$ colours of halo C stars differ 
slightly from those of C stars of these galaxies. 

\smallskip

{\noindent}(4) Thanks to the recently released Catalina and  LINEAR  databases,
we were able to examine the light curves of 143 halo C stars and found 66 new periodic (Mira 
or SRa-type)  variables among them, meaning that $\sim$\,45$\%$ of 
these objects are periodic. Of these 66 objects, we find 19 objects with Catalina
amplitudes larger than 1.5 mag,  which we propose are Mira variables.

\smallskip

{\noindent}(5) We found  13 new red periodic variables in the Fornax dwarf galaxy. 
When these findings on the halo and Fornax are added to 
previous works, the distribution of periods in the halo  and in Fornax are very similar, 
confirming with larger  numbers the previous results of  Battinelli \& Demers (2012).  

\smallskip

{\noindent}(6) The halo period distribution is very different from that of the solar
neighbourhood, implying that little pollution of halo C population arises from the disc 
of the Galaxy.

\smallskip

{\noindent}(7) Finally, there is also a slight indication  that Miras/SRa are older in 
the halo than in Sgr, but additional monitoring and confirmative spectroscopy of Sgr 
AGB stars are needed.

\smallskip

%%%%%%%%%%%%%%%%%%%%%%%%%%%%%%%%%%%%%%%%%%      
\begin{acknowledgements}
It is a pleasure to thank Tom Lloyd Evans for providing us with his list of
carbon stars showing H$\alpha$ in emission (or not), and for  very useful 
remarks. We thank the anonymous referee for questions and comments that 
clarified and improved the paper. We also thank Eric Thi\'ebaut for giving us the 
Levenberg-Marquardt software written in Yorick. N.M. is indebted to
Olivier Richard for his generous help concerning computers.
This publication makes use of data products from the Two Micron All Sky Survey  
2MASS (University of Massachusetts and IPAC/California Institute of Technology, 
funded by  NASA \& NSF), the Catalina Sky Survey (California Institute of 
Technology, NASA), and the Lincoln Near-Earth Asteroid Research  LINEAR 
program (Massachusetts Institute of Technology Lincoln Laboratory, NASA \& 
United States Air Force). This research has made use of Simbad and Vizier
tools offered by the Centre de Donn\'{e}es de Strasbourg (Institut National des 
Sciences de l'Univers, CNRS, France). In particular, we used the General Catalogue of Variable Stars,
developed at Sternberg Astronomical Institute and at the Institute of Astronomy
of Russian Academy of Sciences.
\end{acknowledgements}
%-------------------BIBLIOGRAPHY---------------------------------------------------

%%%%%%%%%%%%%%%%%%%%%%%%%%%%%%%%%%%%%%%%%%%%%%%%%%%%%%%%%%%%%%%%%%%%%%%%%%%%%%%%%%%%%%%%%%%

 \Online
 \appendix
 \section{}
 
 This appendix presents  the spectra of  C stars listed in Table\,\ref{table1} of the paper
 (eight halo C stars, four FBS halo stars, three C stars in Carina, and eight Fornax C stars).
 The CN and C$_2$ bands dominate these spectra, occasionally with H$\alpha$ in emission (indicated).
 The strong absorption feature at 7600 \AA\, is due to telluric O$_2$.
 The scaling factors $f$ to obtain the ordinates in erg~s$^{-1}$~cm$^{-2}$~\AA$^{-1}$ is 
 given in  Table\,A.1. The FBS stars have no flux calibration.
 Note that almost all these stars are subject to variability.

%%%%%%%%%%%%%%%%%%%%%%%%%%%%%%%%%%%%%%%%%%%%%%%%%%%%%%%%%%%%5
    \begin{table}[!ht]
  \caption[]{Scaling factors $f$ for spectra shown in this appendix}
  \begin{center}
  \begin{tabular}{lr}
  \hline 
  \hline
   \noalign{\smallskip}  
  ~~~Name   & $f$~~~~~ \\
            & (erg s$^{-1}$ cm$^{-2}$ A$^{-1}$)\\
   \noalign{\smallskip}
  \hline
  \noalign{\smallskip}
 Halo \#102 &    1.6 10$^{-15}$ \\
 Halo \#103 &    2.1 10$^{-15}$ \\
 Halo \#104 &    4.4 10$^{-15}$ \\
 Halo \#105 &    5.0 10$^{-16}$ \\
 Halo \#106 &    5.5 10$^{-15}$ \\
 Halo \#107 &    5.8 10$^{-15}$ \\
 Halo \#108 &    1.6 10$^{-14}$ \\
 Halo \#109 &    3.3 10$^{-14}$ \\
 \noalign{\smallskip}
 \noalign{\smallskip}
 Carina ALW~2      &   0.9 10$^{-15}$ \\
 Carina [MCA] C3   &   5.5 10$^{-16}$ \\
 Carina [MCA82] C4 &   1.0 10$^{-15}$ \\
 \noalign{\smallskip}
 \noalign{\smallskip}
 Fornax F00        &   2.4 10$^{-16}$ \\
 Fornax F06        &   1.6 10$^{-16}$ \\
 Fornax F08        &   2.4 10$^{-16}$ \\

 Fornax F29        &   3.0 10$^{-16}$ \\
 Fornax F31        &   0.97 10$^{-16}$ \\
 Fornax F36        &   1.55 10$^{-16}$ \\

 Fornax F52        &   2.7 10$^{-16}$ \\
 Fornax F58        &   4.0 10$^{-16}$ \\

 \noalign{\smallskip}
 \noalign{\smallskip}
   \hline
  \end{tabular}
   \end{center}
   \end{table}
%%%%%%%%%%%%%%%%%%%%%%%%%%%%%%%%%%%%%%%%%%%%%%%%%%%%%%%%%%%%%%%%%%%%

Table\,A.2 presents the list of objects in the field of Fornax for which
EFOSC2 spectroscopy was analysed and which are not C stars. 
These stars are mostly M-type foreground dwarfs,
except for the object 2MASS~J024000.78$-$341812.2, which appears fuzzy along the slit
in the spectra and is probably a galaxy.

 In Table\,A.4, the  periodic variables found with Catalina \& LINEAR data are listed.
In the first column, the objects named ``SDSS Green $\#$nnn'' are taken from
Table\,1 of Green (2013), where nnn is the object rank in his table. Similarly,
the object named 2MASS Gizis $\#32$ comes from Table\,1 of Gizis (2002). The names ``mxxx'' come
from the lists in Paper I to IV.

  %%%%%%%%%%%   Table des objets non C dans Fornax  &&&&
    \begin{table*}[!ht]
  \caption[]{Objects in the direction of Fornax that are not carbon stars. 
$K$ and $J-K$ are taken from 2MASS. $R$ and $B-R$ are taken from the USNO-A2 catalogue. }
  \begin{center}
  \begin{tabular}{ccccc}
  \hline 
  \hline
   \noalign{\smallskip}  
    Identification  & $K$ & $J-K$ & $R$ & $B-R$\\
   \noalign{\smallskip}
  \hline
  \noalign{\smallskip}

\object{2MASS J023830.242$-$344503.07}   & 14.319 & 1.157   &  17.2 &   1.6 \\
\object{2MASS J023850.934$-$341619.25}   & 14.012 & 1.218   &  16.4 &   1.8 \\
\object{2MASS J023855.471$-$344916.60}   & 13.103 & 1.135   &  16.4 &   2.3 \\
\object{2MASS J023859.919$-$344526.81}   & 14.268 & 1.002   &  17.1 &   1.7 \\
\object{2MASS J023908.628$-$342740.78}   & 13.444 & 1.067   &  16.4 &   0.3 \\
\object{2MASS J023911.288$-$345302.11}   & 14.112 & 0.990   &  16.7 &   1.5 \\
\object{2MASS J023912.019$-$343741.44}   & 14.256 & 0.929   &  17.1 &   1.9 \\
\object{2MASS J023921.337$-$342329.22}   & 14.121 & 1.038   &  17.2 &   1.6 \\
\object{2MASS J023929.844$-$342723.21}   & 13.760 & 1.124   &  16.9 &   2.3 \\
\object{2MASS J023930.656$-$342407.86}   & 13.956 & 1.160   &  16.9 &   2.0 \\
\object{2MASS J023931.452$-$342258.67}   & 13.987 & 1.180   &  17.2 &   1.6 \\
\object{2MASS J023937.523$-$342500.07}   & 13.675 & 1.111   &  15.6 &   1.3 \\
\object{2MASS J023940.112$-$343402.43}   & 14.094 & 1.146   &  17.2 &   1.9 \\
\object{2MASS J023940.941$-$343256.92}   & 13.747 & 1.092   &  16.1 &   1.5 \\
\object{2MASS J023943.364$-$341348.59}   & 14.096 & 1.052   &  16.7 &   2.1 \\
\object{2MASS J023944.088$-$342610.11}   & 13.857 & 1.088   &  16.9 &   1.4 \\
\object{2MASS J023951.447$-$343912.65}   & 14.025 & 1.188   &  16.9 &   1.6 \\
\object{2MASS J023952.001$-$341712.40}   & 14.101 & 1.204   &  16.6 &   2.2 \\
\object{2MASS J023959.516$-$343243.29}   & 14.017 & 1.133   &  16.9 &   1.7 \\
\object{2MASS J024000.782$-$341812.29}   & 14.465 & 1.947   &  16.9 &   0.3 \\
\object{2MASS J024001.815$-$344403.93}   & 14.276 & 0.970   &  16.9 &   1.5 \\
\object{2MASS J024008.194$-$342605.63}   & 13.576 & 1.216   &  17.0 &   1.7 \\
\object{2MASS J024012.469$-$343905.80}   & 13.727 & 1.094   &  17.1 &   2.1 \\
\object{2MASS J024014.037$-$342331.66}   & 13.868 & 1.224   &  16.9 &   1.9 \\
\object{2MASS J024015.956$-$343529.39}   & 13.895 & 1.220   &  16.9 &   1.9 \\
\object{2MASS J024022.672$-$344037.32}   & 14.049 & 1.130   &  17.1 &   1.8 \\
\object{2MASS J024023.541$-$341520.42}   & 14.201 & 1.157   &  17.0 &   1.7 \\ 
\object{2MASS J024023.898$-$341810.82}   & 13.929 & 1.179   &  17.0 &   1.8 \\
\object{2MASS J024031.954$-$342048.75}   & 13.854 & 1.154   &  17.1 &   1.8 \\
\object{2MASS J024036.689$-$341713.73}   & 14.100 & 1.069   &  17.0 &   1.7 \\
\object{2MASS J024039.071$-$342646.15}   & 14.321 & 0.987   &  16.4 &   1.2 \\
\object{2MASS J024039.570$-$343609.51}   & 14.066 & 1.135   &  16.7 &   2.0 \\
\object{2MASS J024040.040$-$342937.08}   & 13.955 & 1.031   &  16.9 &   1.4 \\
\object{2MASS J024050.051$-$342558.07}   & 13.560 & 1.237   &  16.8 &   1.9 \\
\object{2MASS J024052.884$-$342830.39}   & 14.343 & 1.363   &  16.0 &   0.0 \\
\object{2MASS J024055.364$-$342725.27}   & 14.066 & 1.347   &  17.1 &   2.0 \\
\object{2MASS J024056.877$-$342709.31}   & 13.943 & 1.207   &  17.1 &   1.5 \\
\object{2MASS J024105.935$-$342731.89}   & 13.870 & 1.204   &  17.1 &   1.7 \\
\object{2MASS J024122.424$-$341129.64}   & 14.173 & 1.057   &  17.0 &   1.8 \\
 \noalign{\smallskip}
 \noalign{\smallskip}
   \hline
  \end{tabular}
   \end{center}
   \end{table*}

%%%%%%%%%%%%%%%%%%%%%%%%%%%%%%%%%%%%%%%%%%%%%%%%%%%%%%%%%%%%%%%%%

    \begin{table}[!ht]
  \caption[]{Average 2MASS $JHK$ colours of halo carbon stars. This table defines
the line from which the distance $\epsilon$ is determined.}
  \begin{center}
  \begin{tabular}{cc}
  \hline 
  \hline
   \noalign{\smallskip}  
    $H-K$   & $J-H$\\
   (mag)    &  (mag)\\
   \noalign{\smallskip}
  \hline
  \noalign{\smallskip}
  0.330 & 0.850 \\
  0.520 & 0.962\\
  0.640 & 1.083 \\
  0.850 & 1.193 \\
  1.070 & 1.360\\
  1.510 & 1.798 \\
  2.050 & 2.495 \\
  2.300 & 2.830 \\
  2.860 & 3.340 \\

 \noalign{\smallskip}
 \noalign{\smallskip}
   \hline
  \end{tabular}
   \end{center}
   \end{table}
% -----------------------------------------------------------------------

\clearpage

  \begin{figure*}
\caption{Spectra of halo carbon stars}
\resizebox{8.5cm}{!}{\rotatebox{-90}{\includegraphics{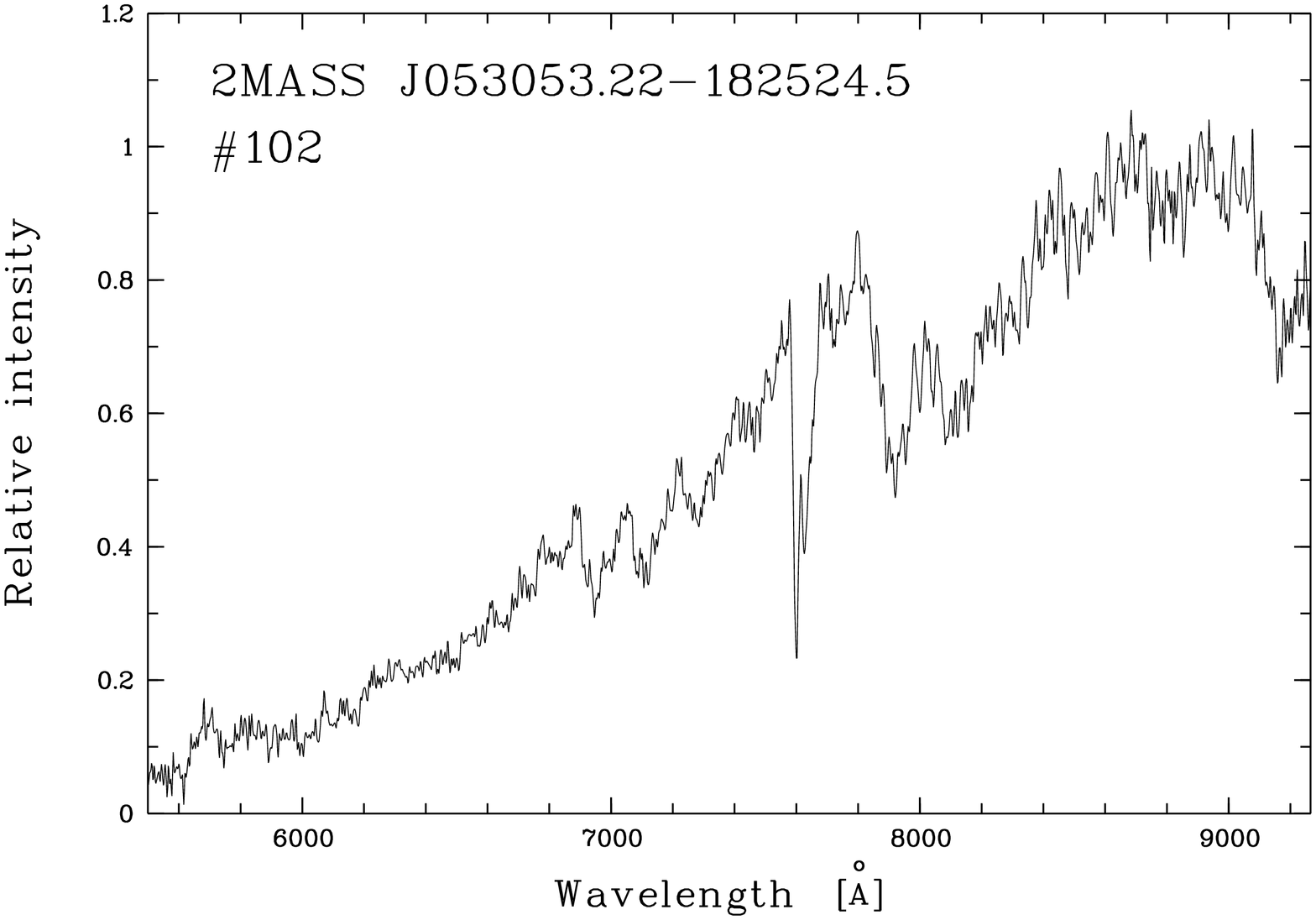}}}
\resizebox{8.5cm}{!}{\rotatebox{-90}{\includegraphics{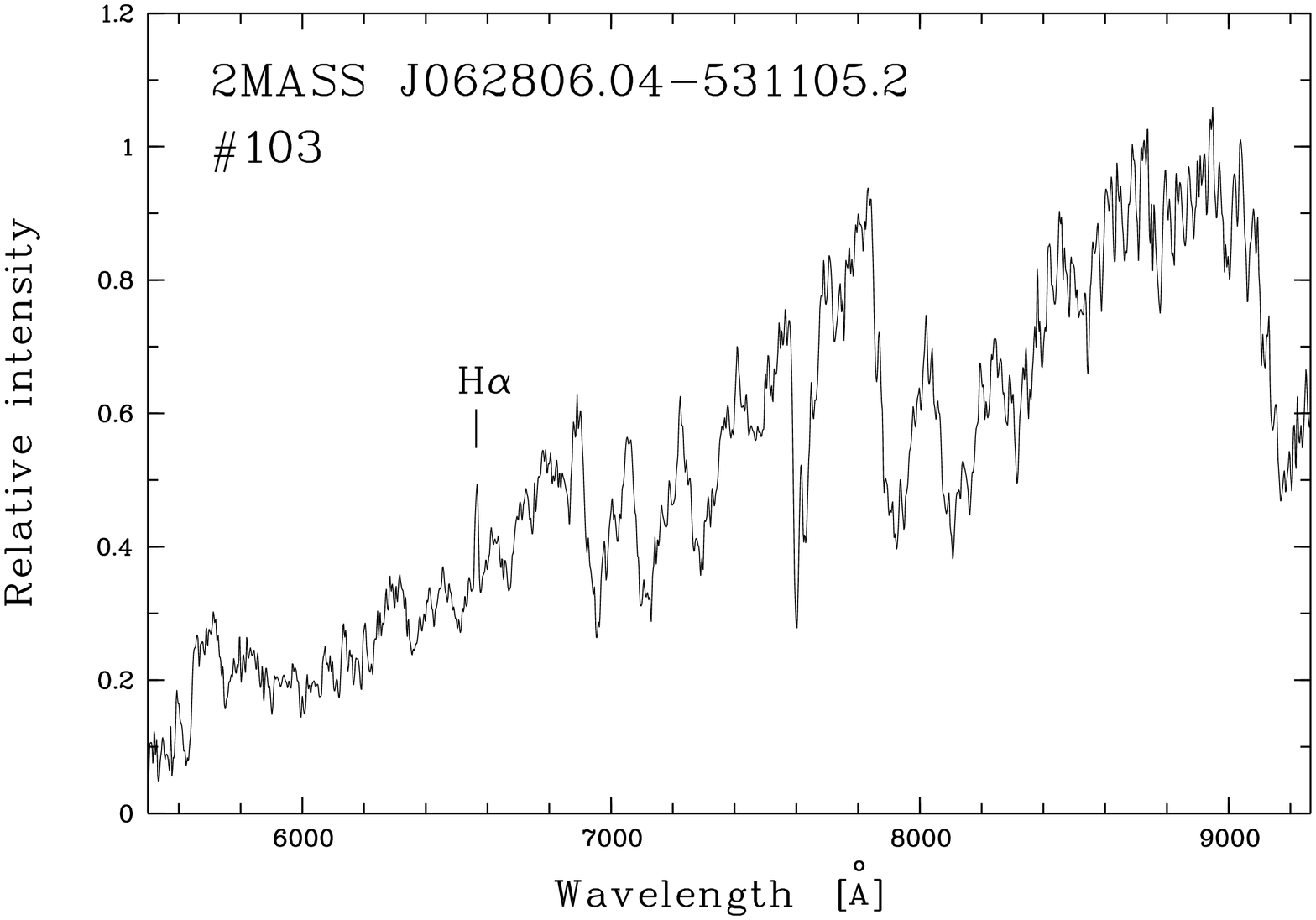}}}
\resizebox{8.5cm}{!}{\rotatebox{-90}{\includegraphics{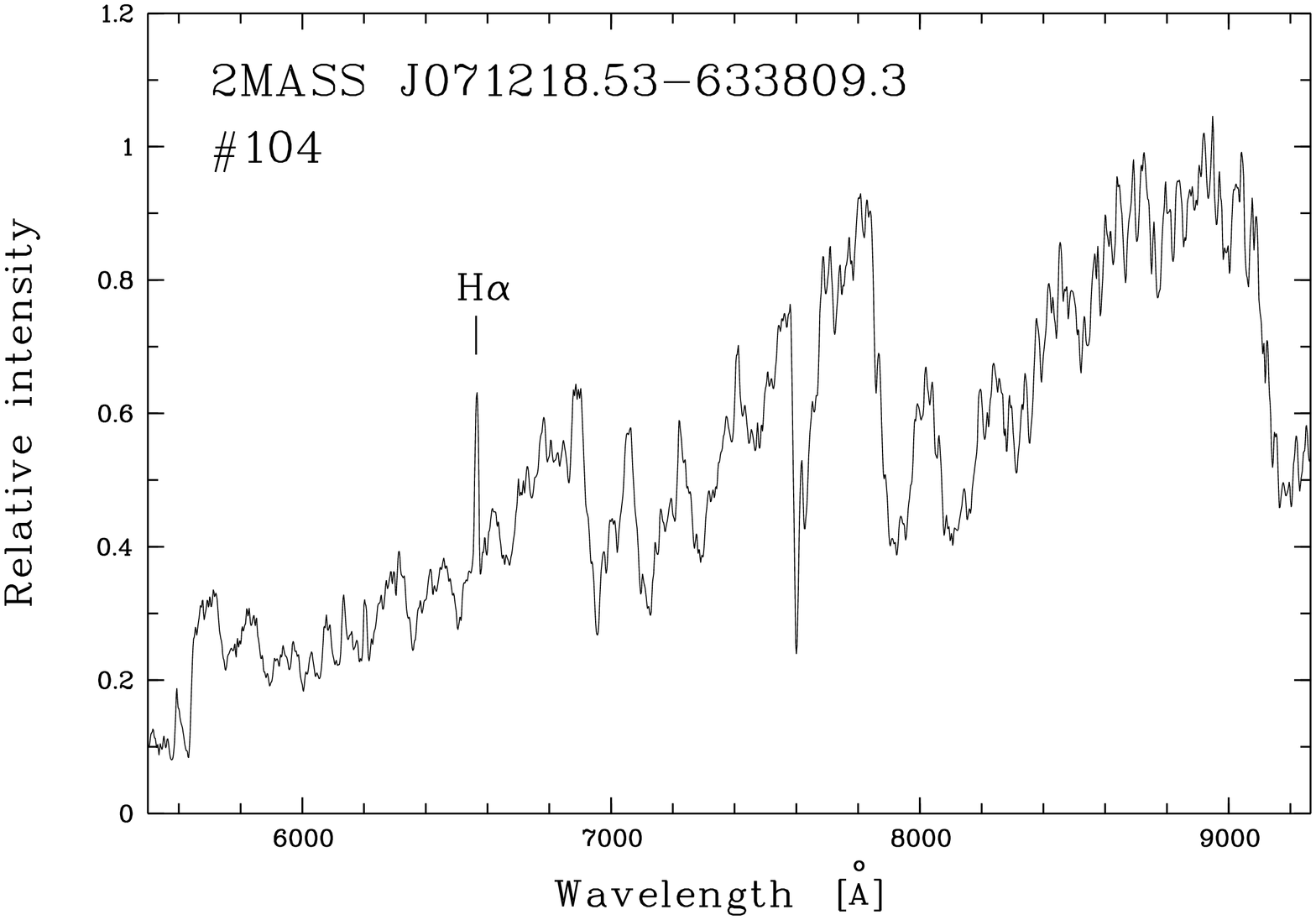}}}
\resizebox{8.5cm}{!}{\rotatebox{-90}{\includegraphics{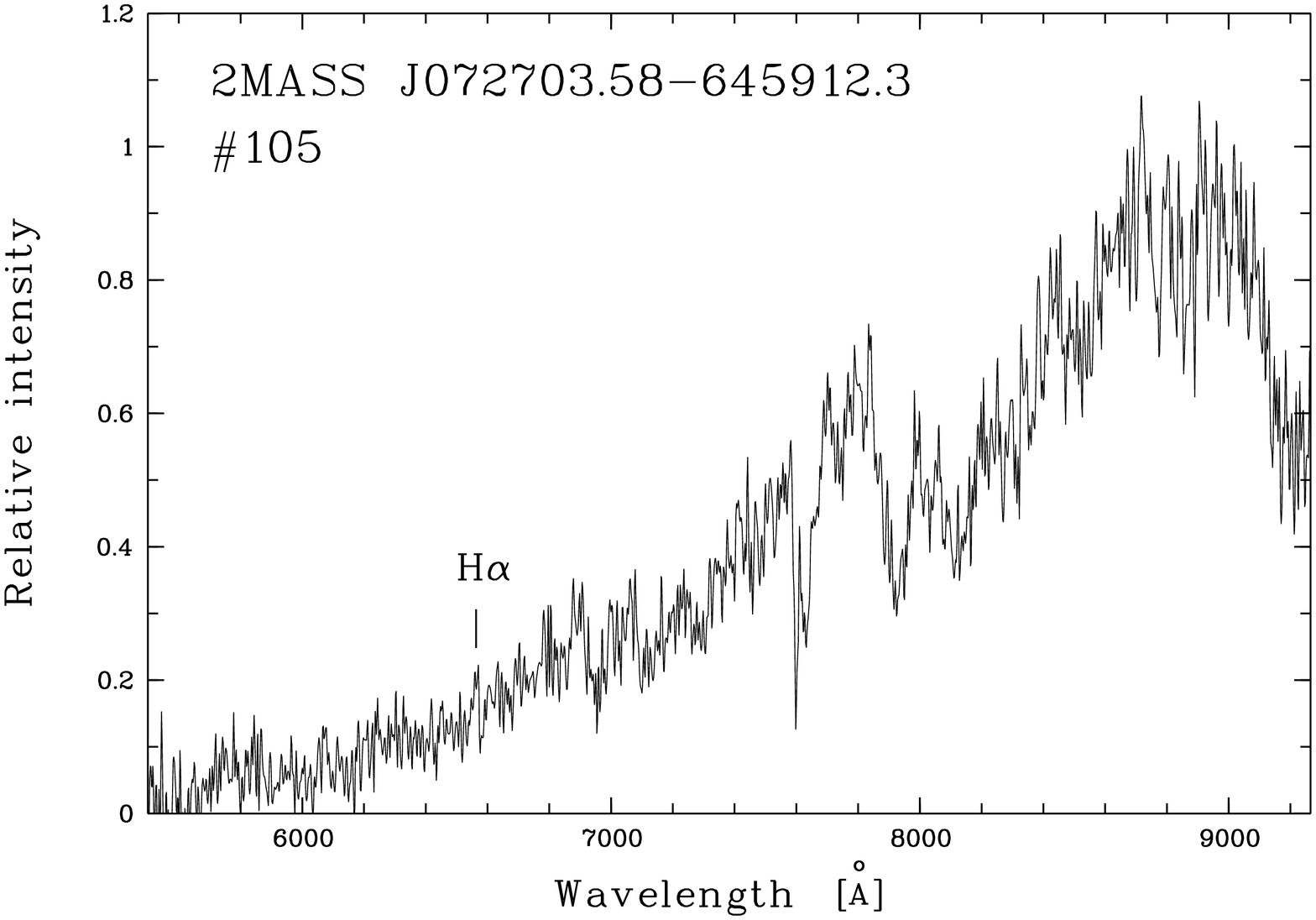}}}
\resizebox{8.5cm}{!}{\rotatebox{-90}{\includegraphics{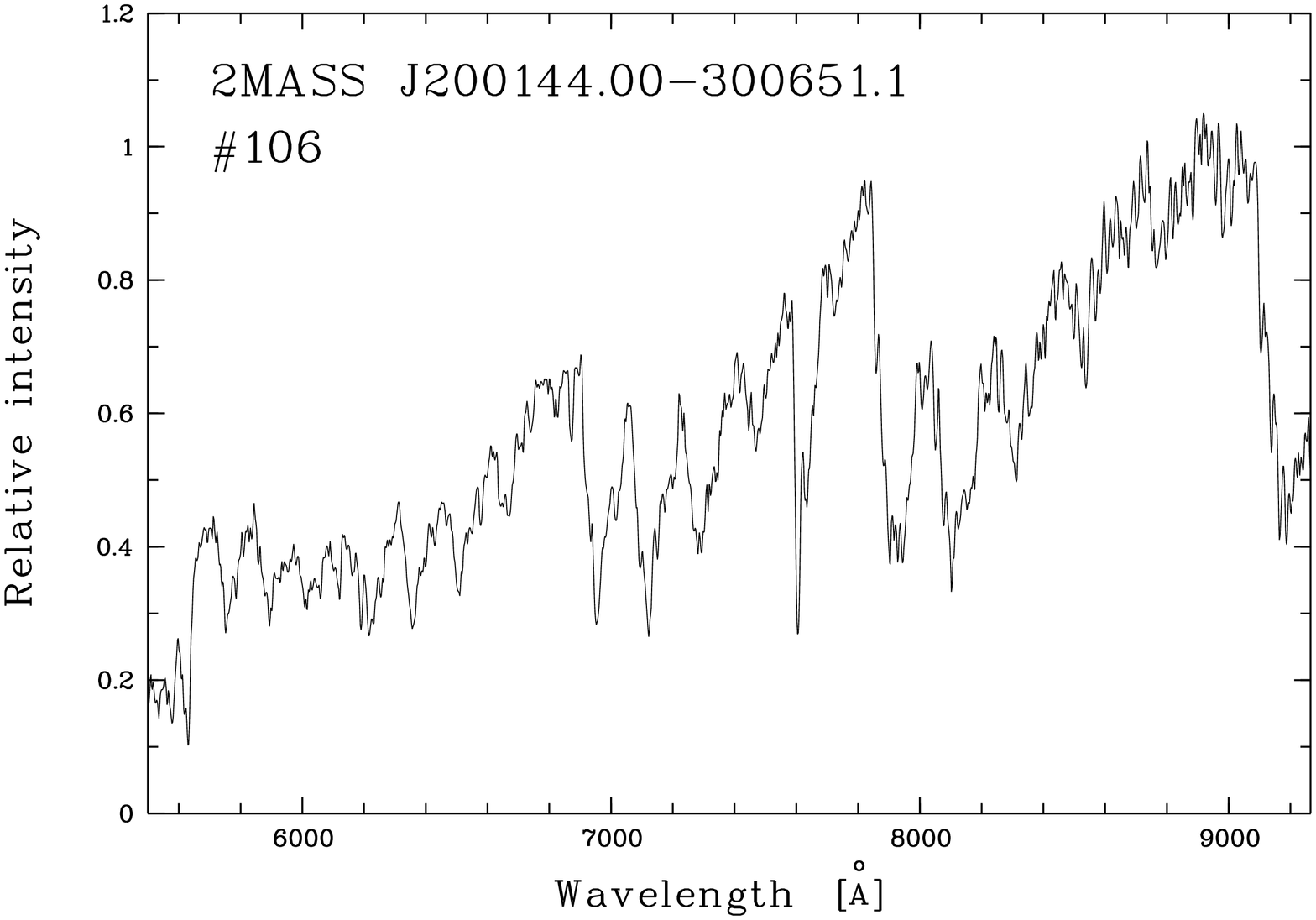}}}
\resizebox{8.5cm}{!}{\rotatebox{-90}{\includegraphics{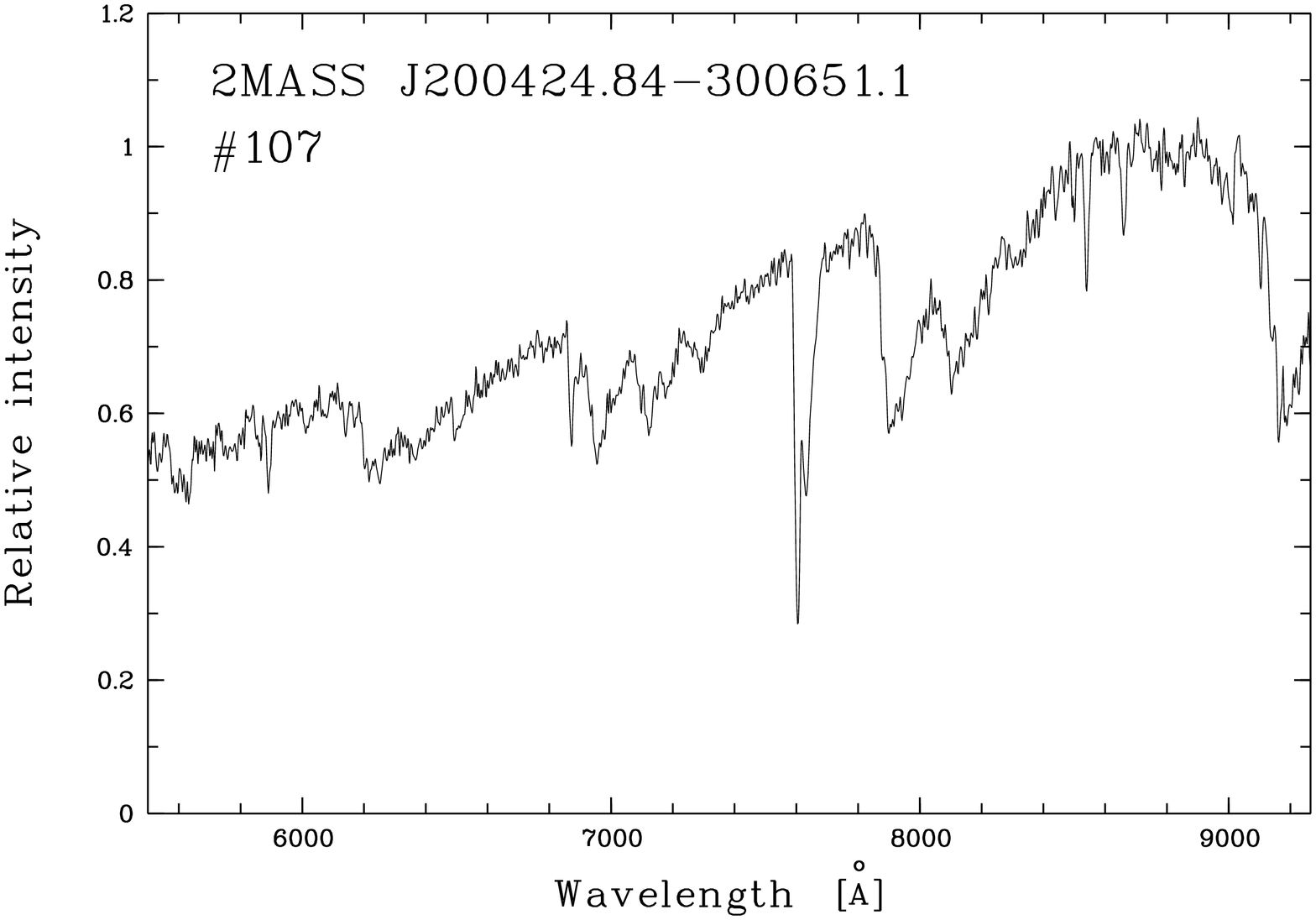}}}
\resizebox{8.5cm}{!}{\rotatebox{-90}{\includegraphics{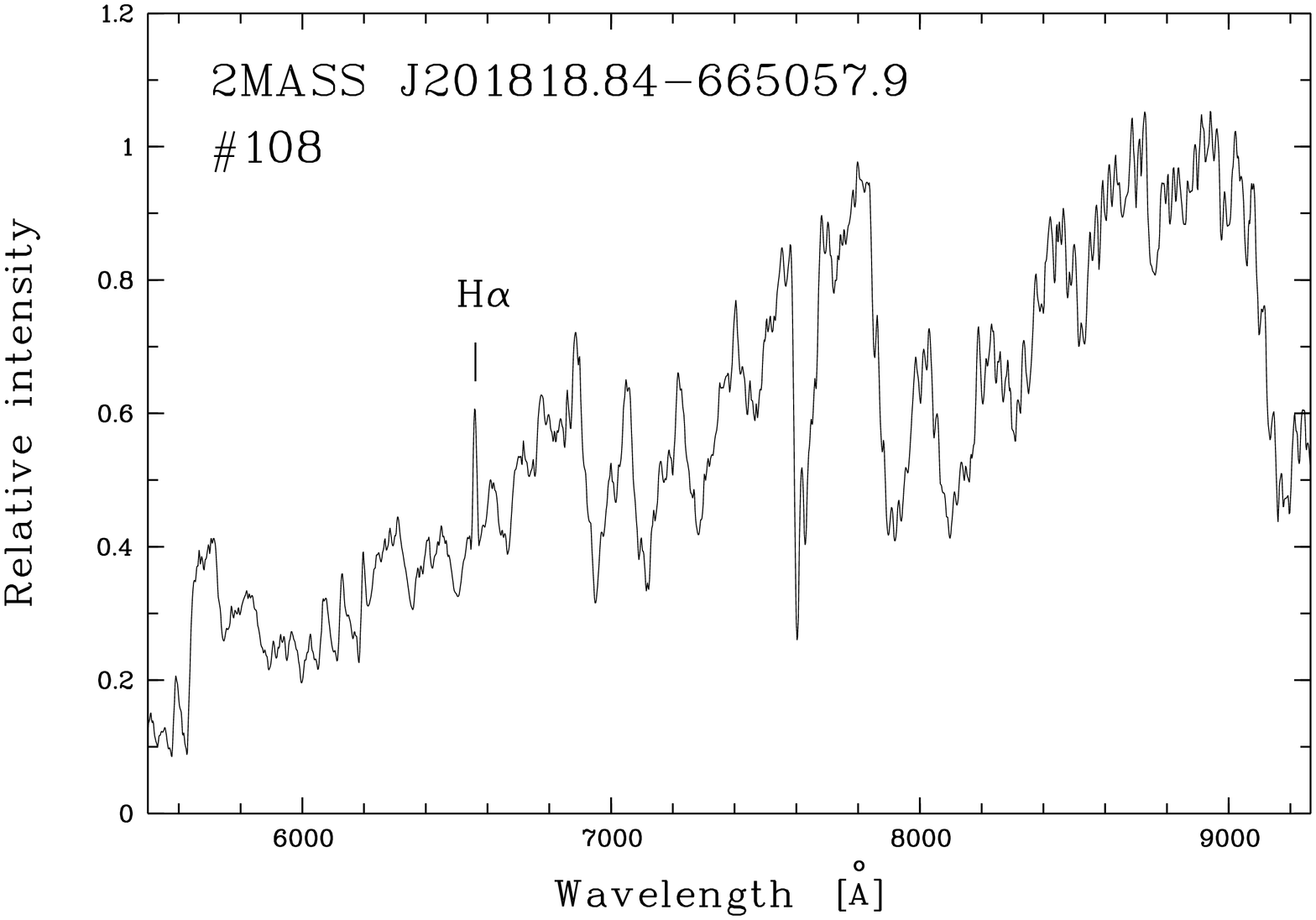}}}
\hspace{0.85cm}
\resizebox{8.5cm}{!}{\rotatebox{-90}{\includegraphics{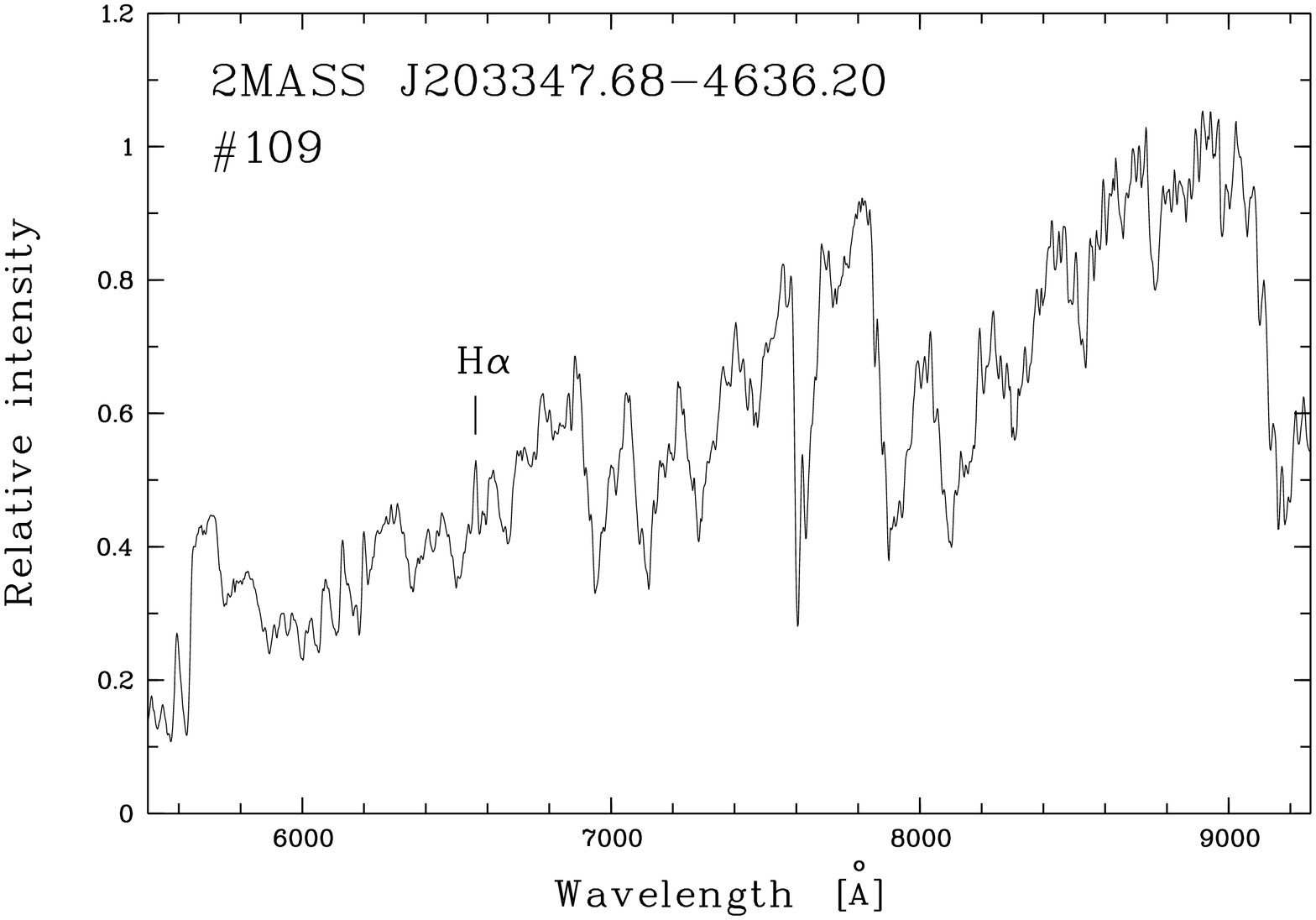}}}
\end{figure*}

\begin{figure*}
\caption{Spectra of FBS halo carbon stars}
\resizebox{8.5cm}{!}{\rotatebox{-90}{\includegraphics{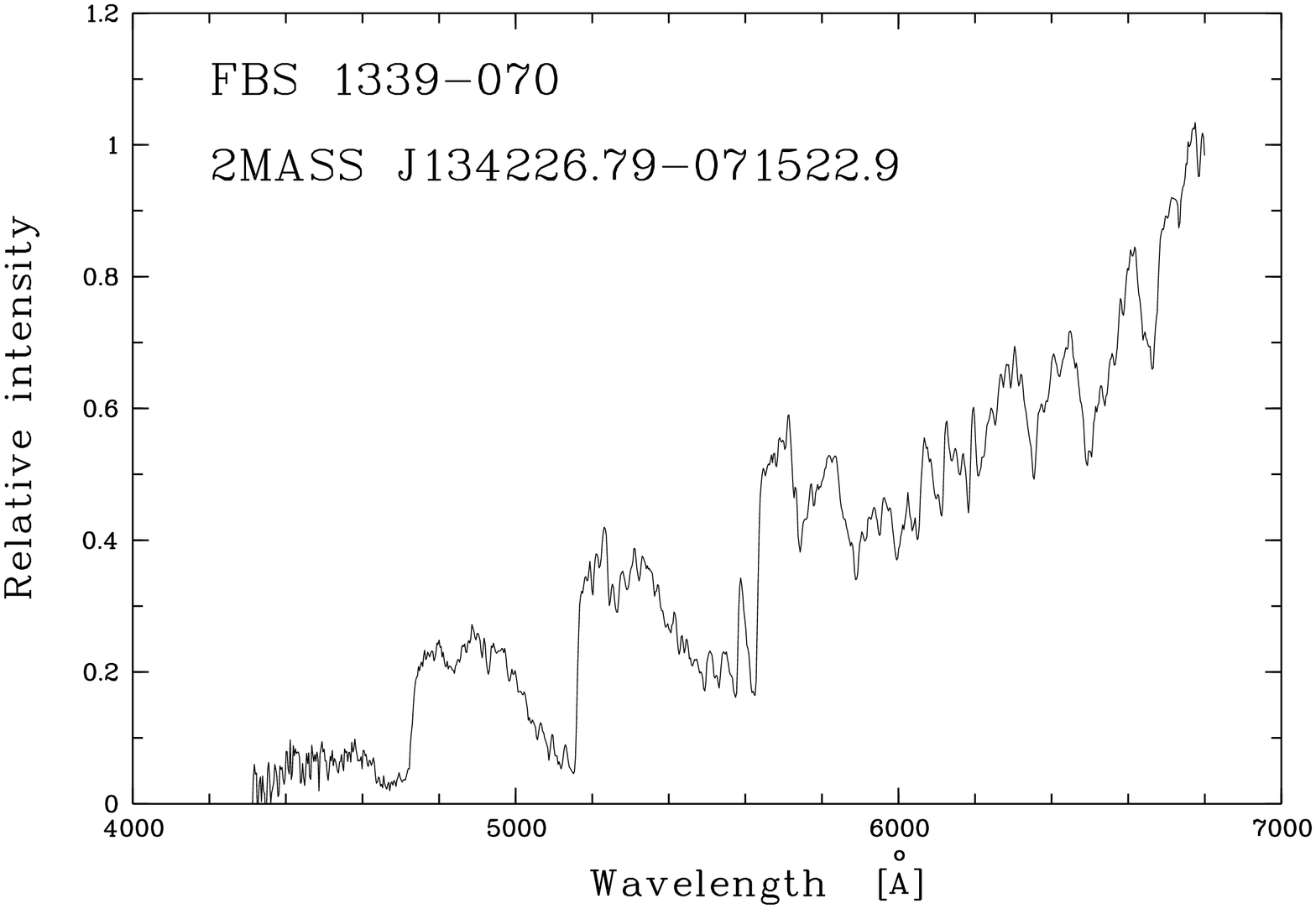}}}
\resizebox{8.5cm}{!}{\rotatebox{-90}{\includegraphics{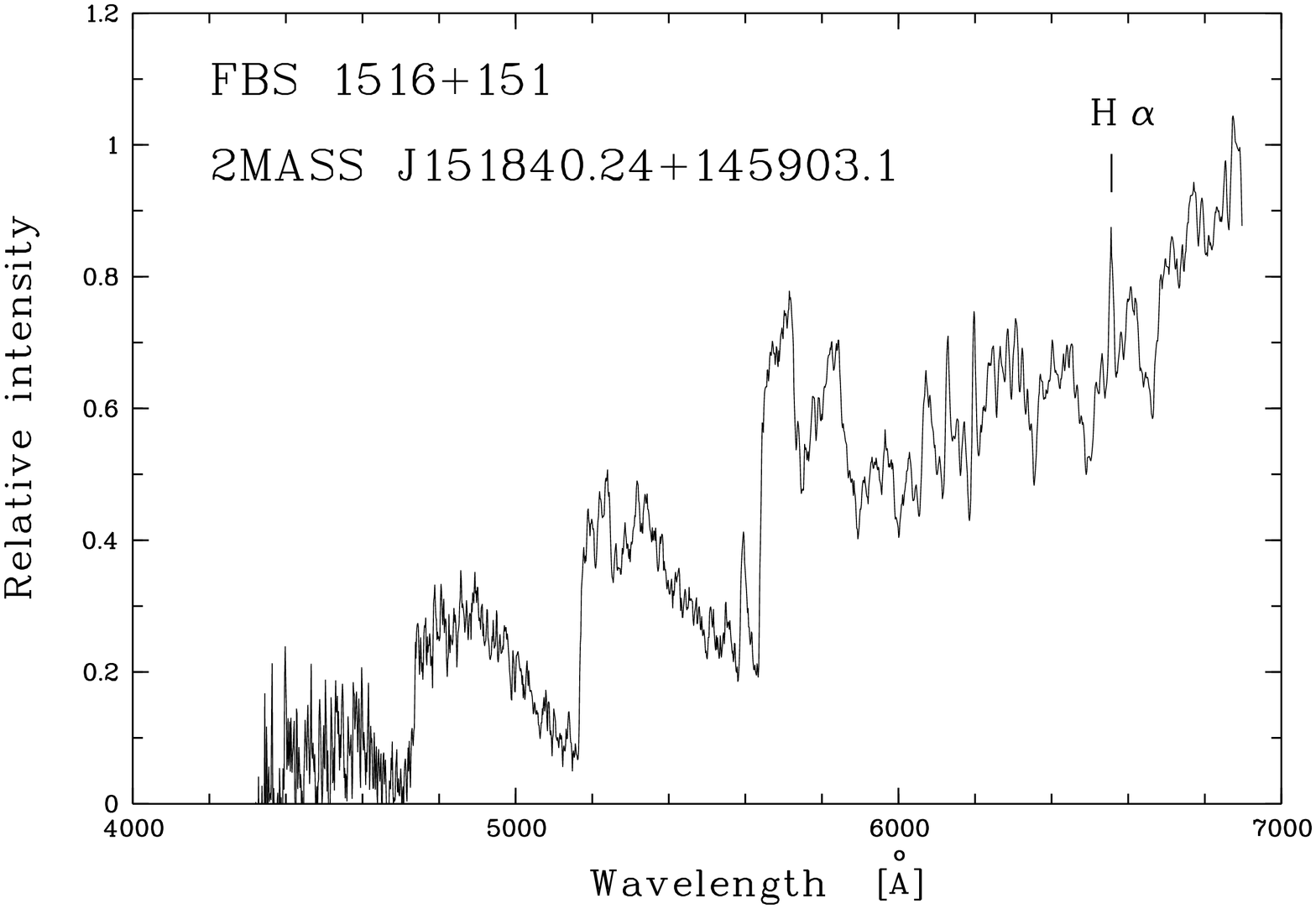}}}
\resizebox{8.5cm}{!}{\rotatebox{-90}{\includegraphics{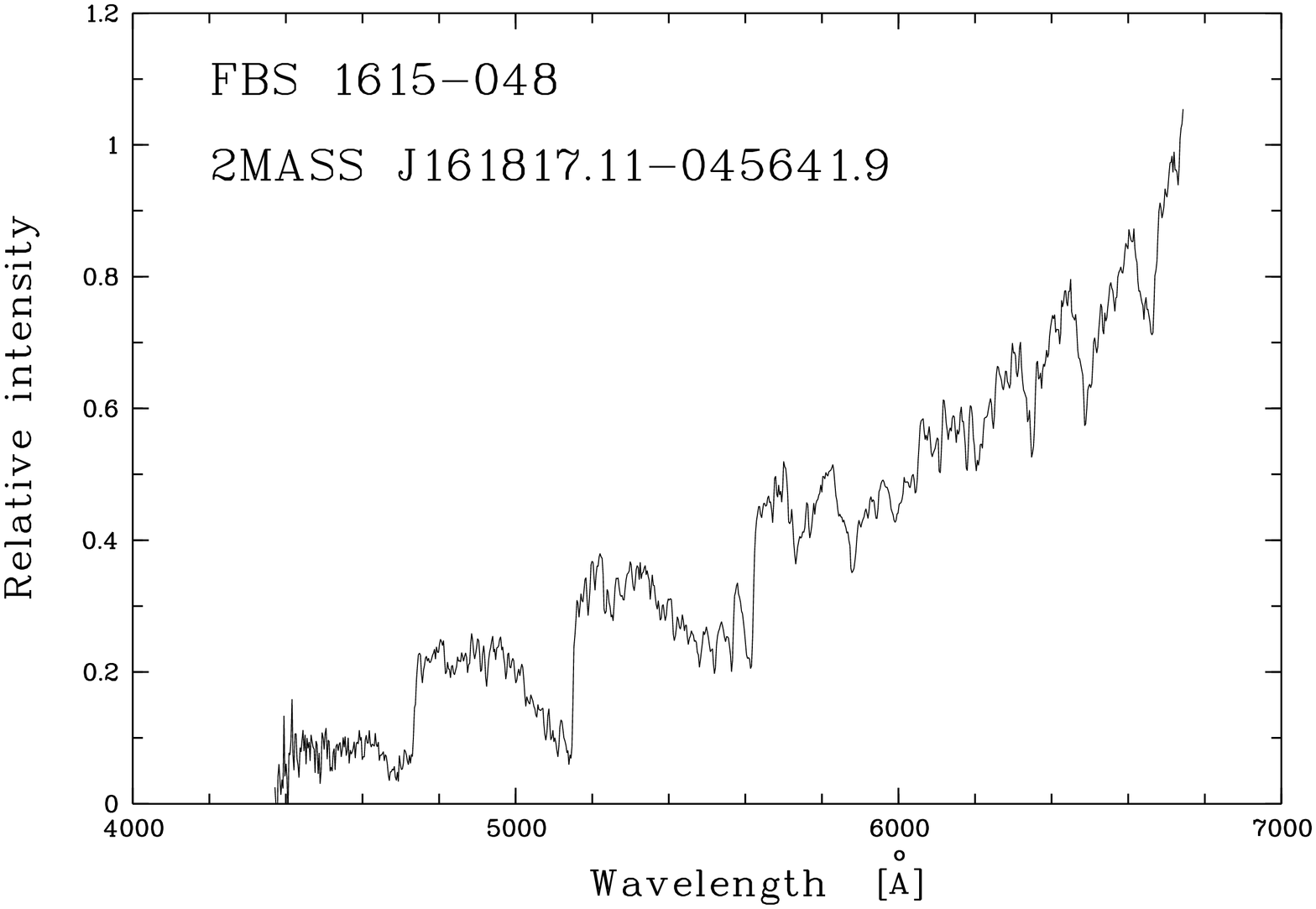}}}
\hspace{1.2cm}
\resizebox{8.5cm}{!}{\rotatebox{-90}{\includegraphics{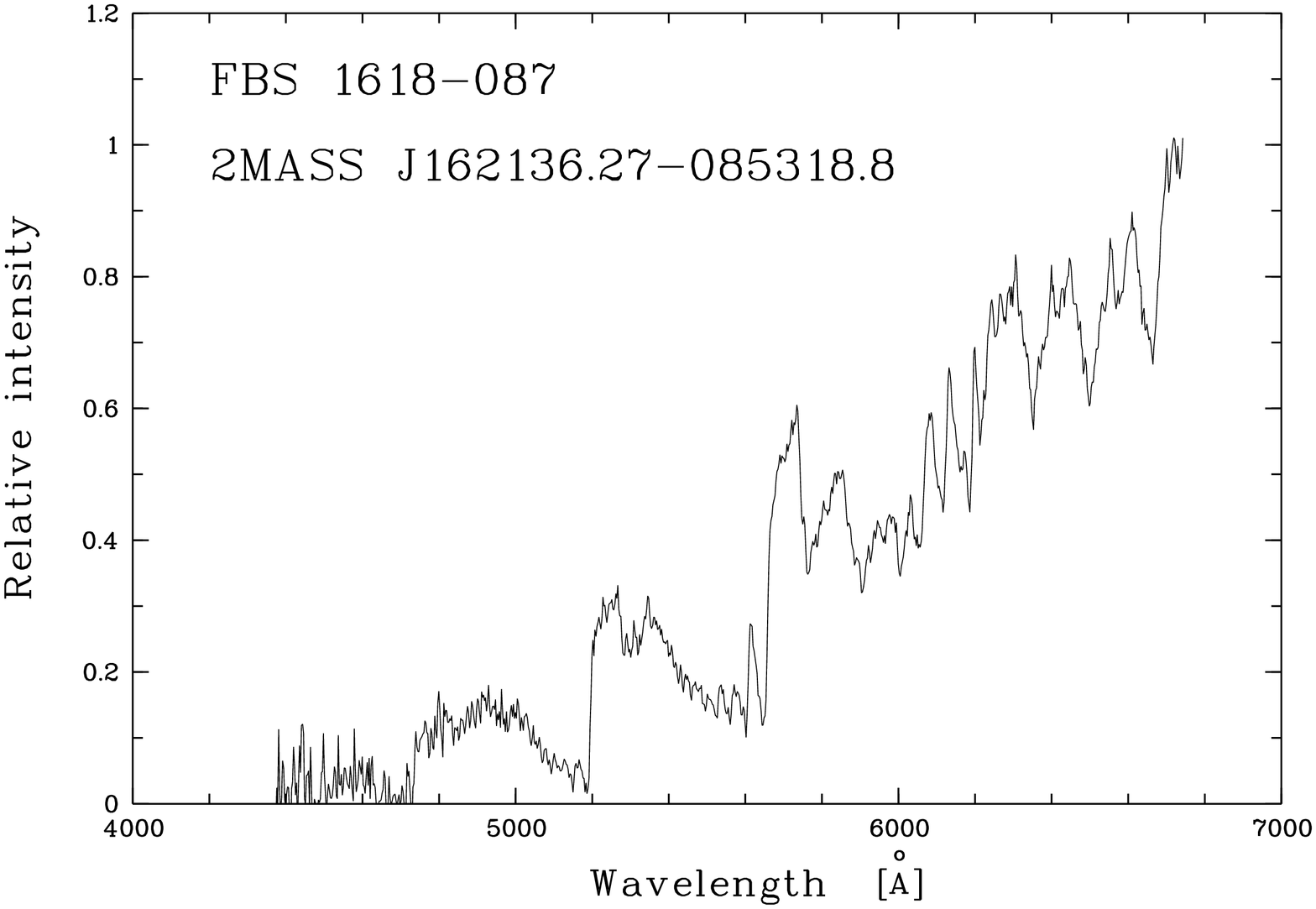}}}
\end{figure*}

\begin{figure*}
\caption{Spectra of Carina carbon stars}
\resizebox{8.5cm}{!}{\rotatebox{-90}{\includegraphics{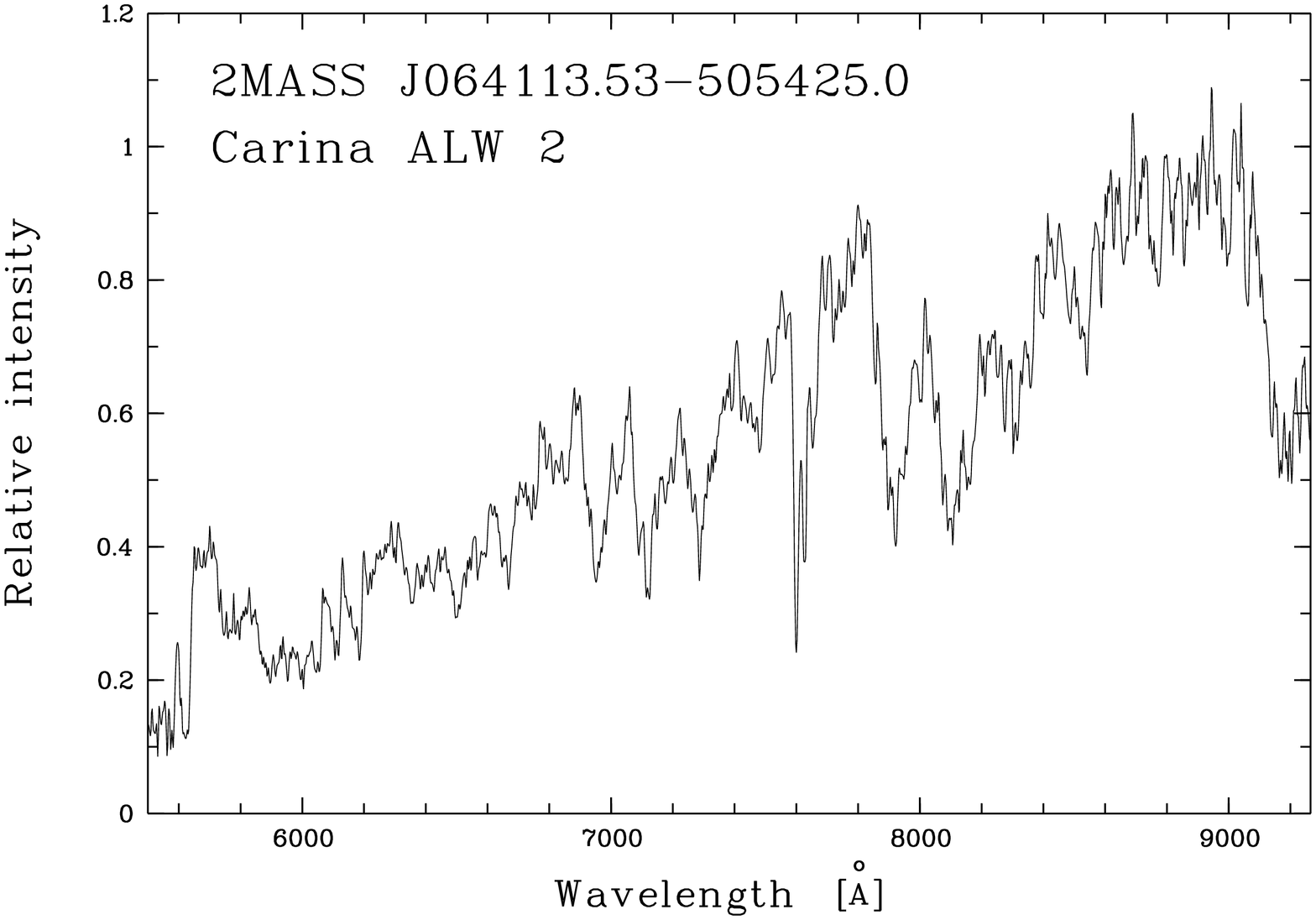}}}
\resizebox{8.5cm}{!}{\rotatebox{-90}{\includegraphics{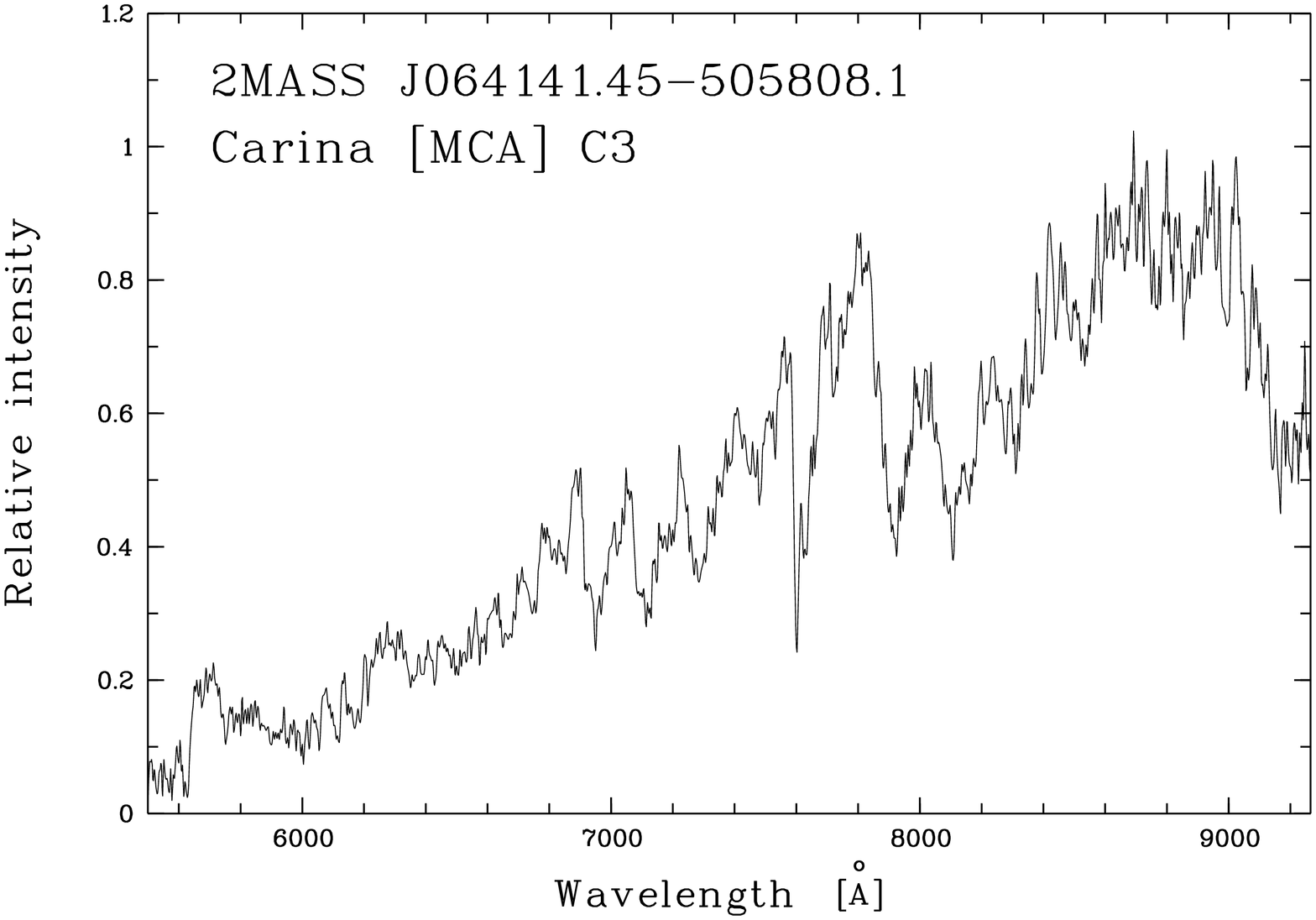}}}
\resizebox{8.5cm}{!}{\rotatebox{-90}{\includegraphics{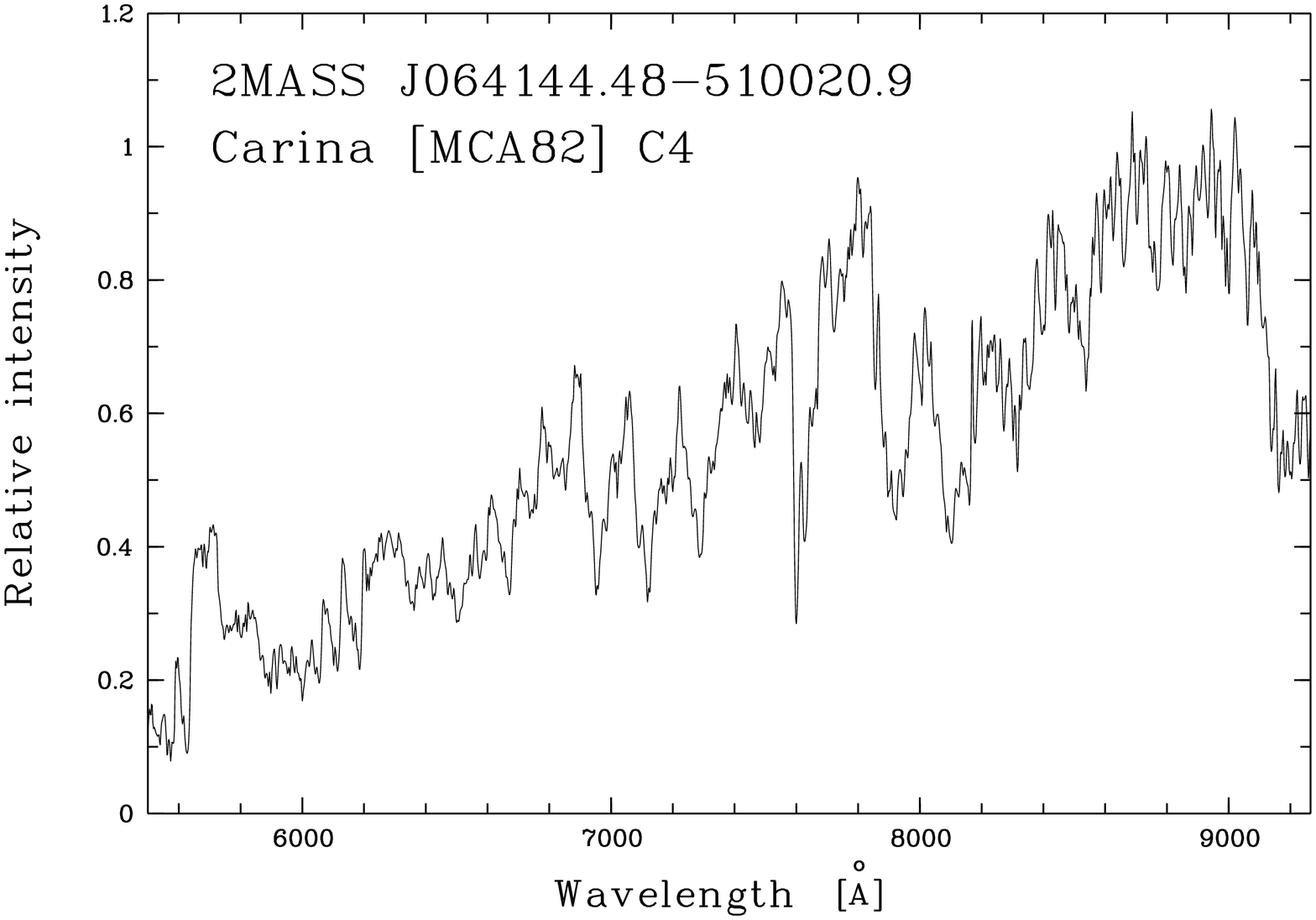}}}
\end{figure*}

\begin{figure*}
\caption{Spectra of Fornax carbon stars}
\resizebox{8.5cm}{!}{\rotatebox{-90}{\includegraphics{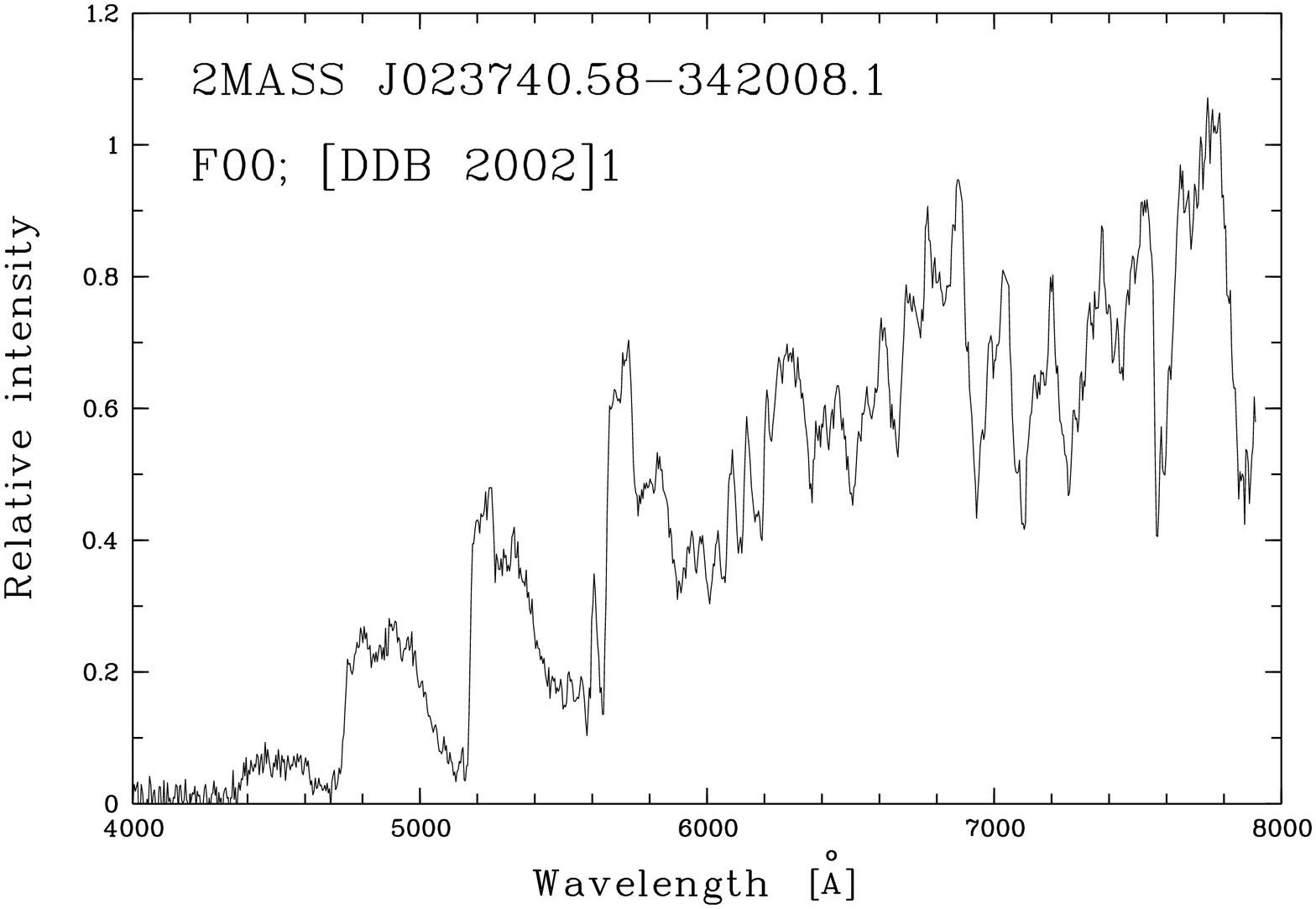}}}
\resizebox{8.5cm}{!}{\rotatebox{-90}{\includegraphics{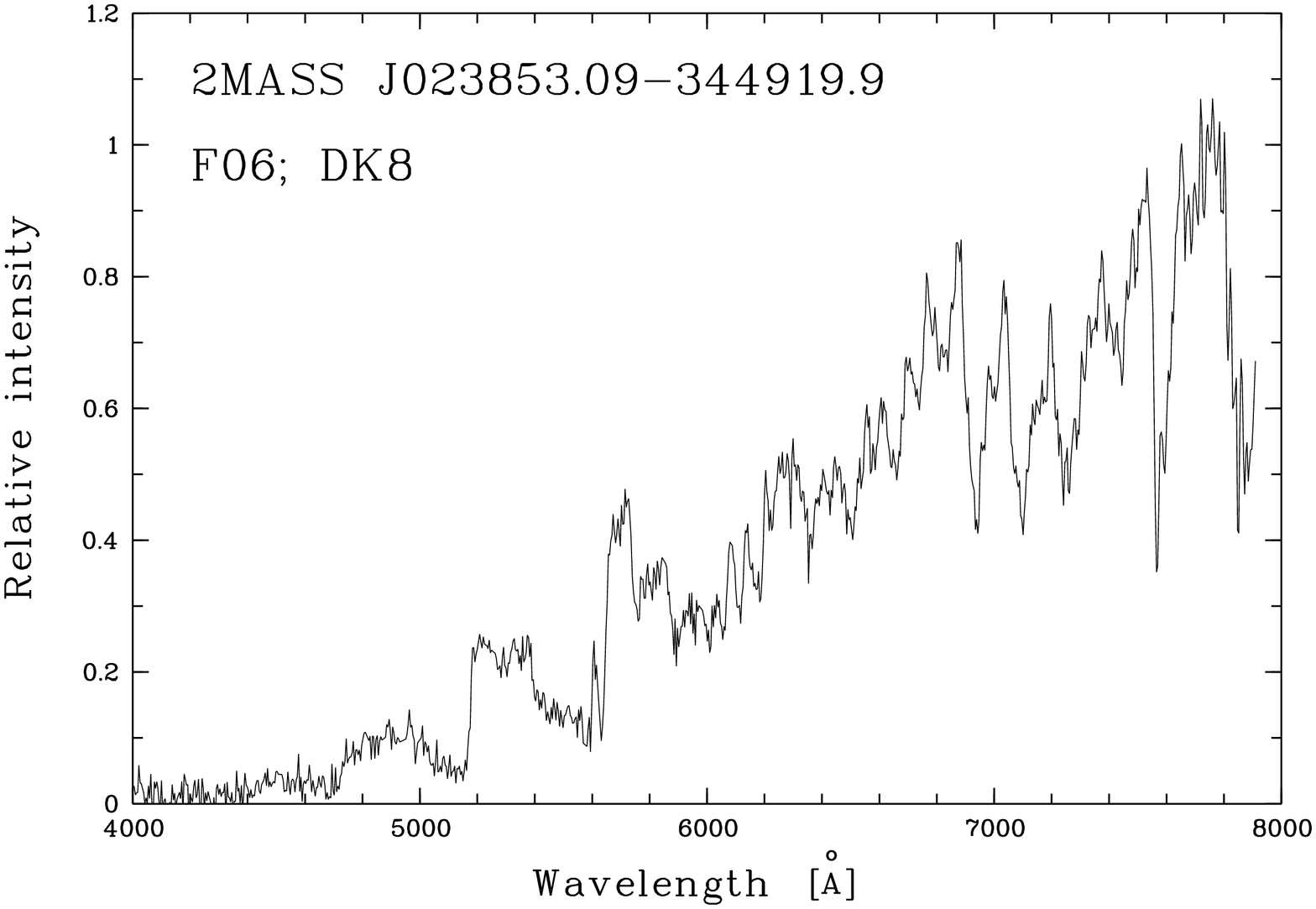}}}
\resizebox{8.5cm}{!}{\rotatebox{-90}{\includegraphics{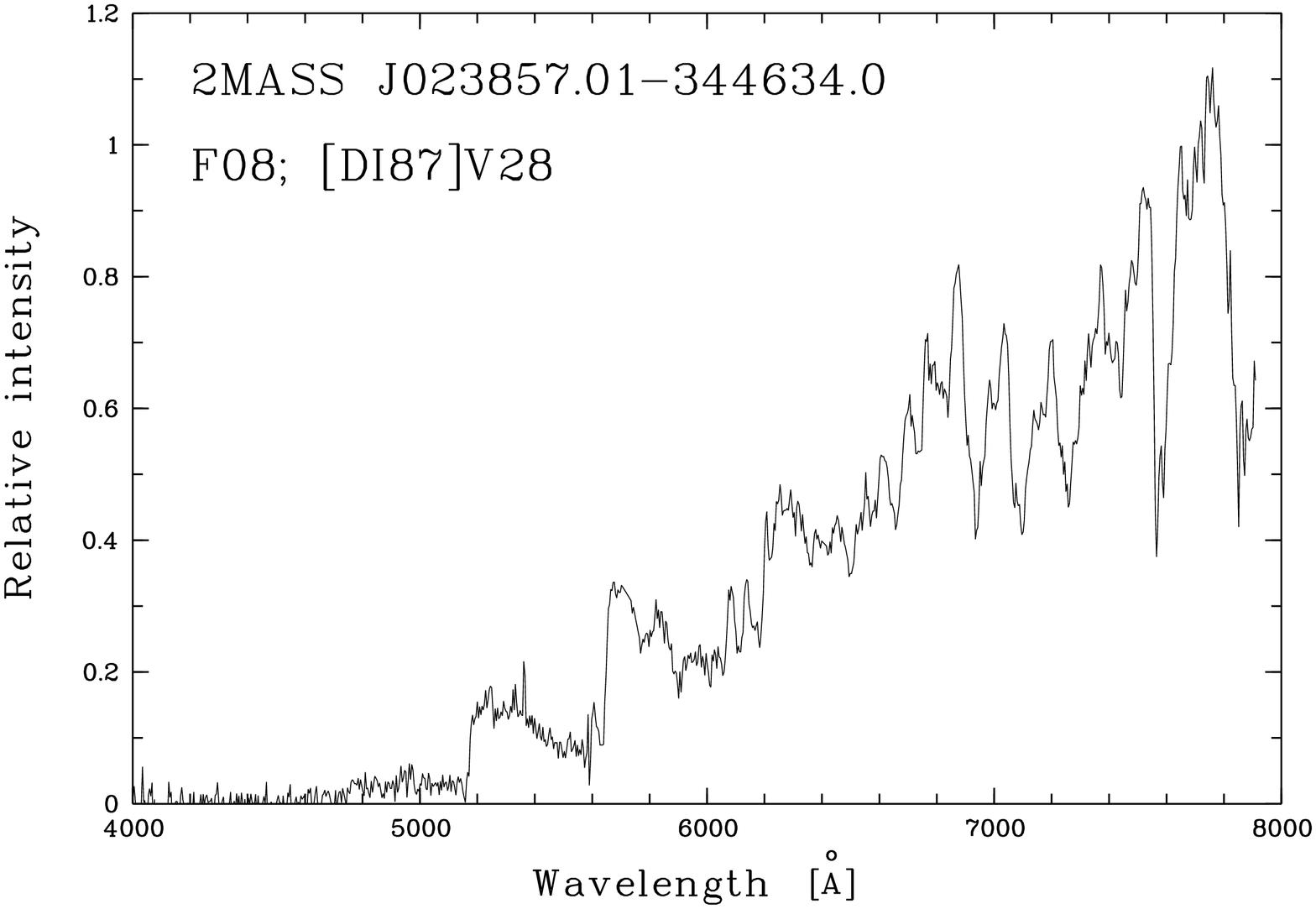}}}
\resizebox{8.5cm}{!}{\rotatebox{-90}{\includegraphics{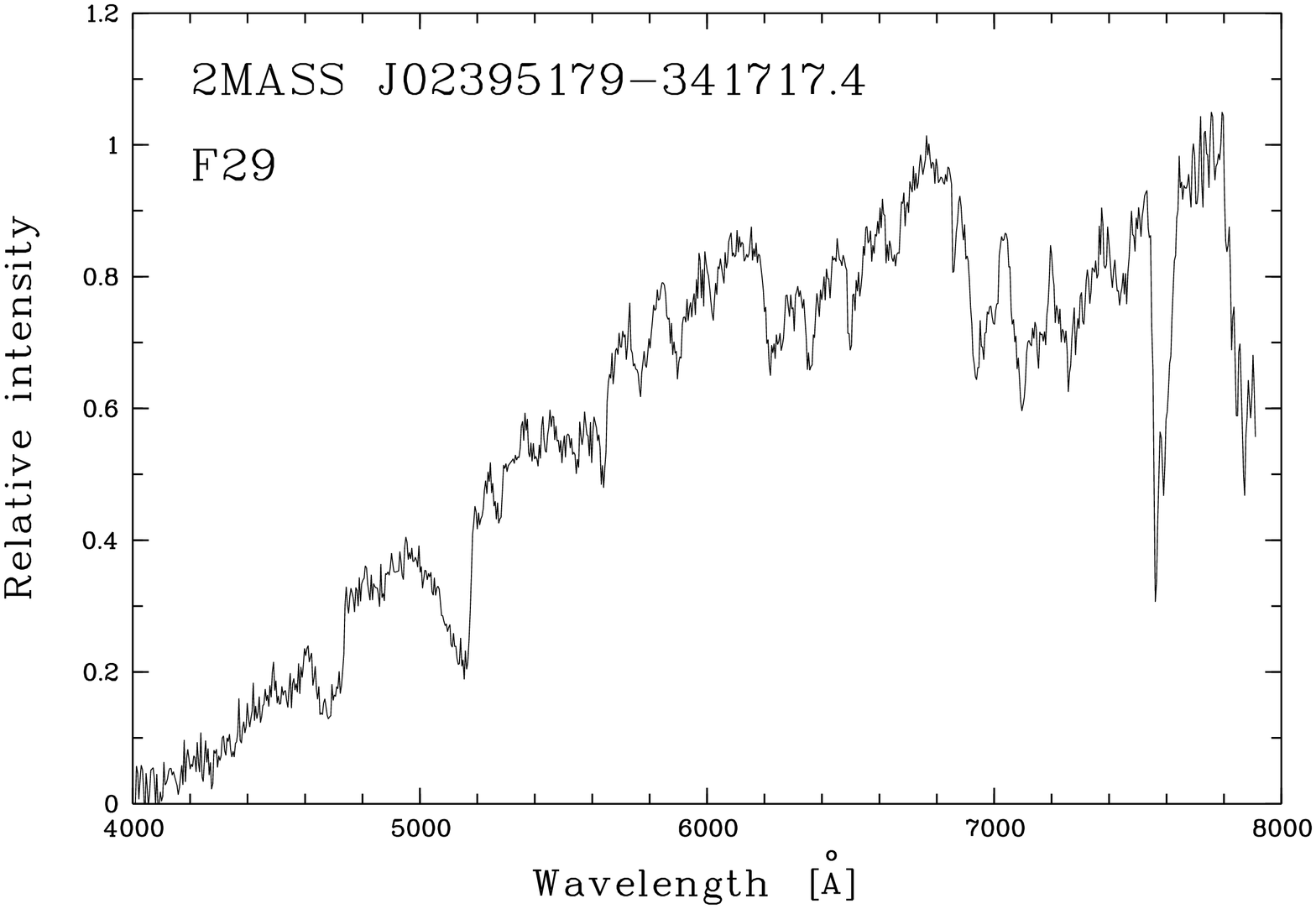}}}
\resizebox{8.5cm}{!}{\rotatebox{-90}{\includegraphics{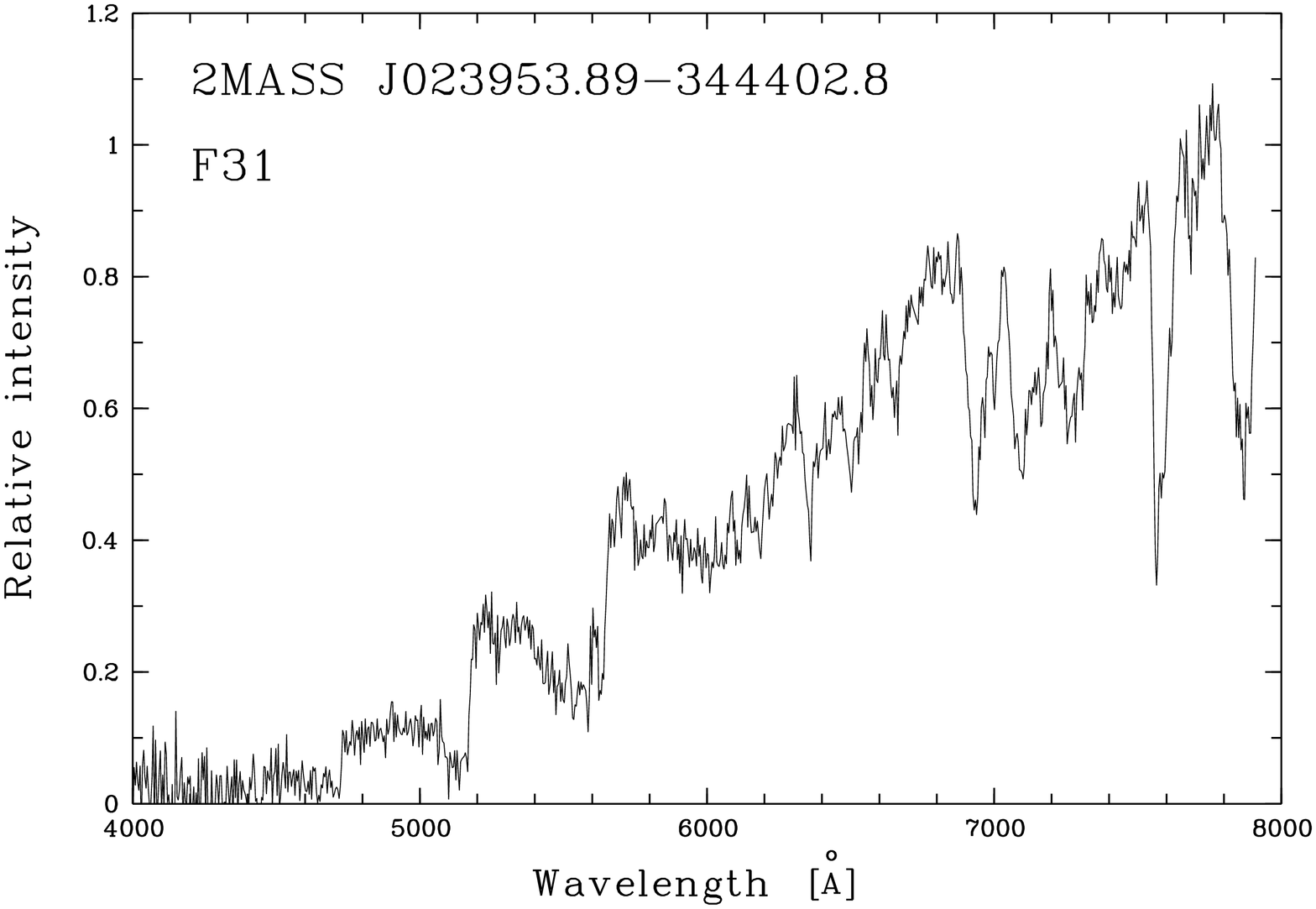}}}
\resizebox{8.5cm}{!}{\rotatebox{-90}{\includegraphics{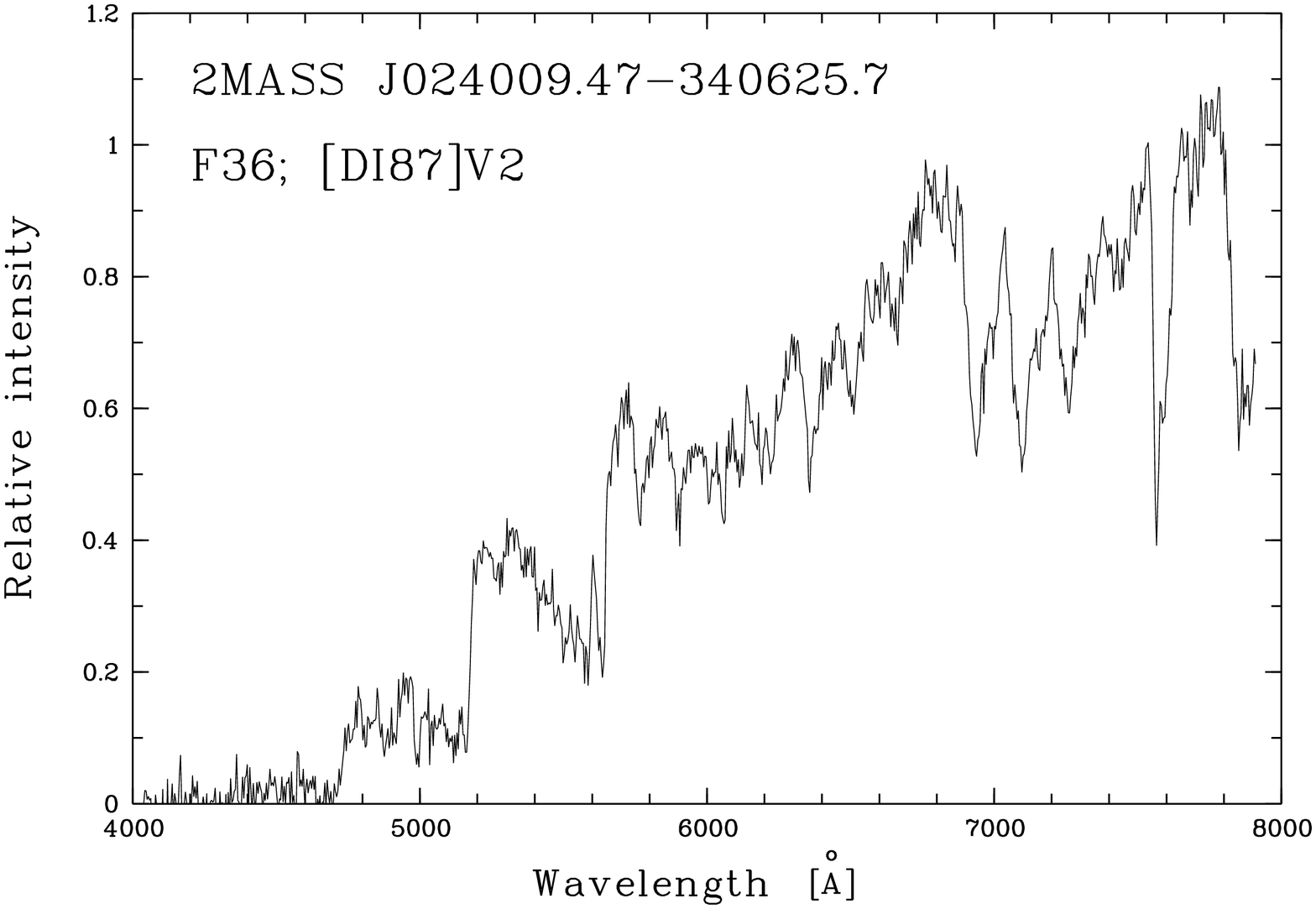}}}
\resizebox{8.5cm}{!}{\rotatebox{-90}{\includegraphics{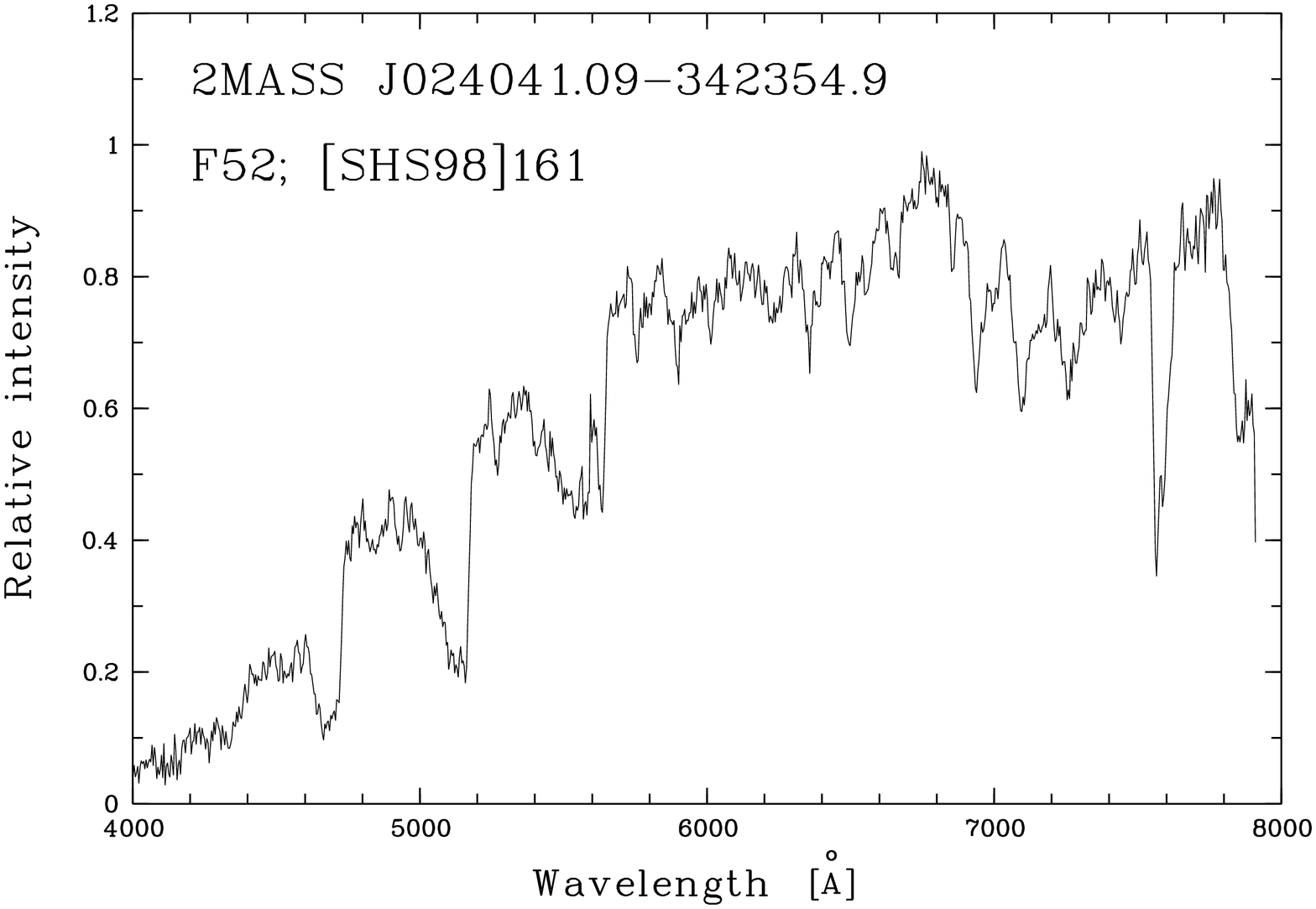}}}
\hspace{1.2cm}
\resizebox{8.5cm}{!}{\rotatebox{-90}{\includegraphics{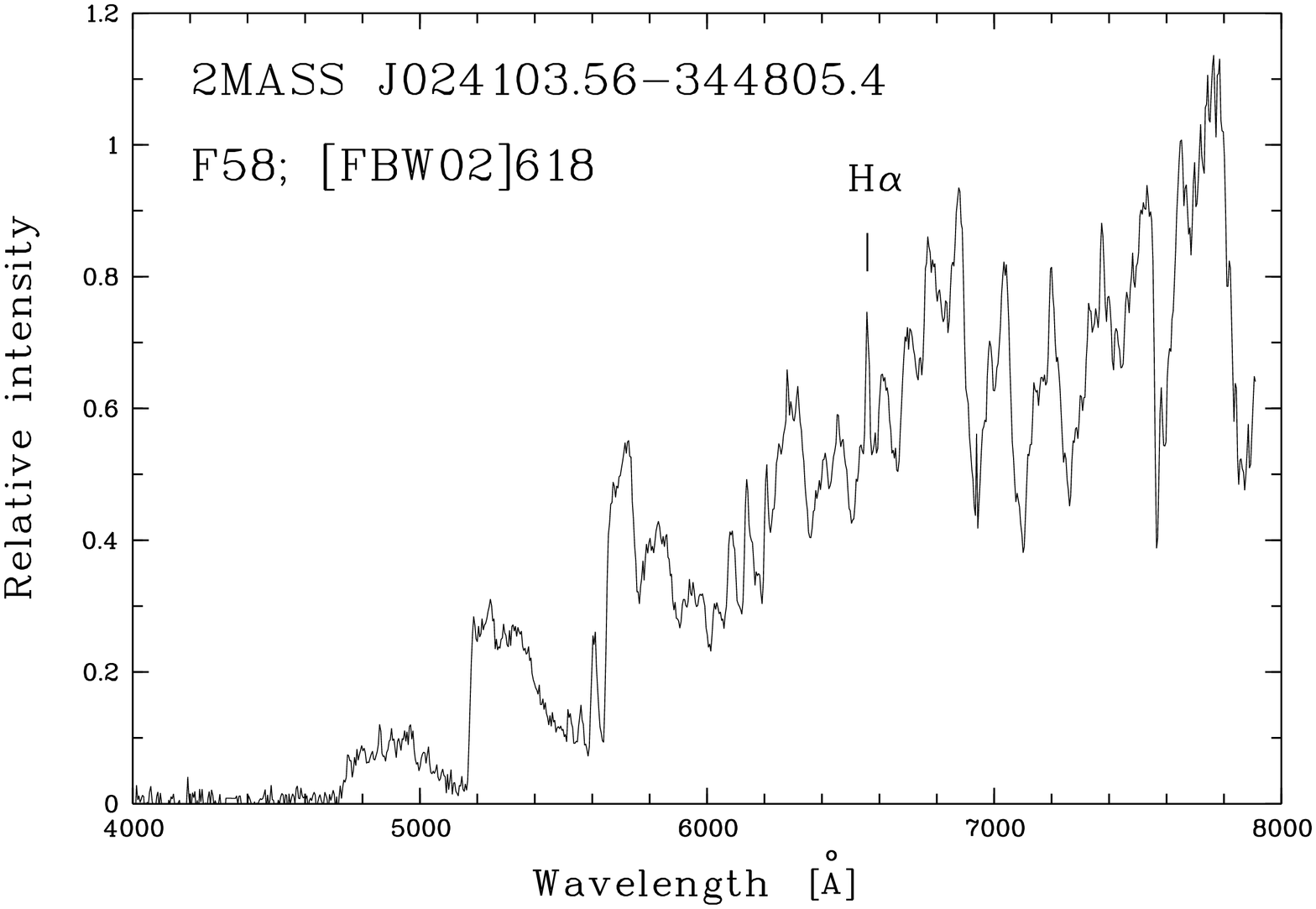}}}
\end{figure*}

% ========================== figs TAROT light curves ============= hou hou hou
        \begin{figure*}
%        \begin{center}
        \caption{TAROT light curves of halo carbon stars}
       \resizebox{8cm}{!}{\rotatebox{-90}{\includegraphics{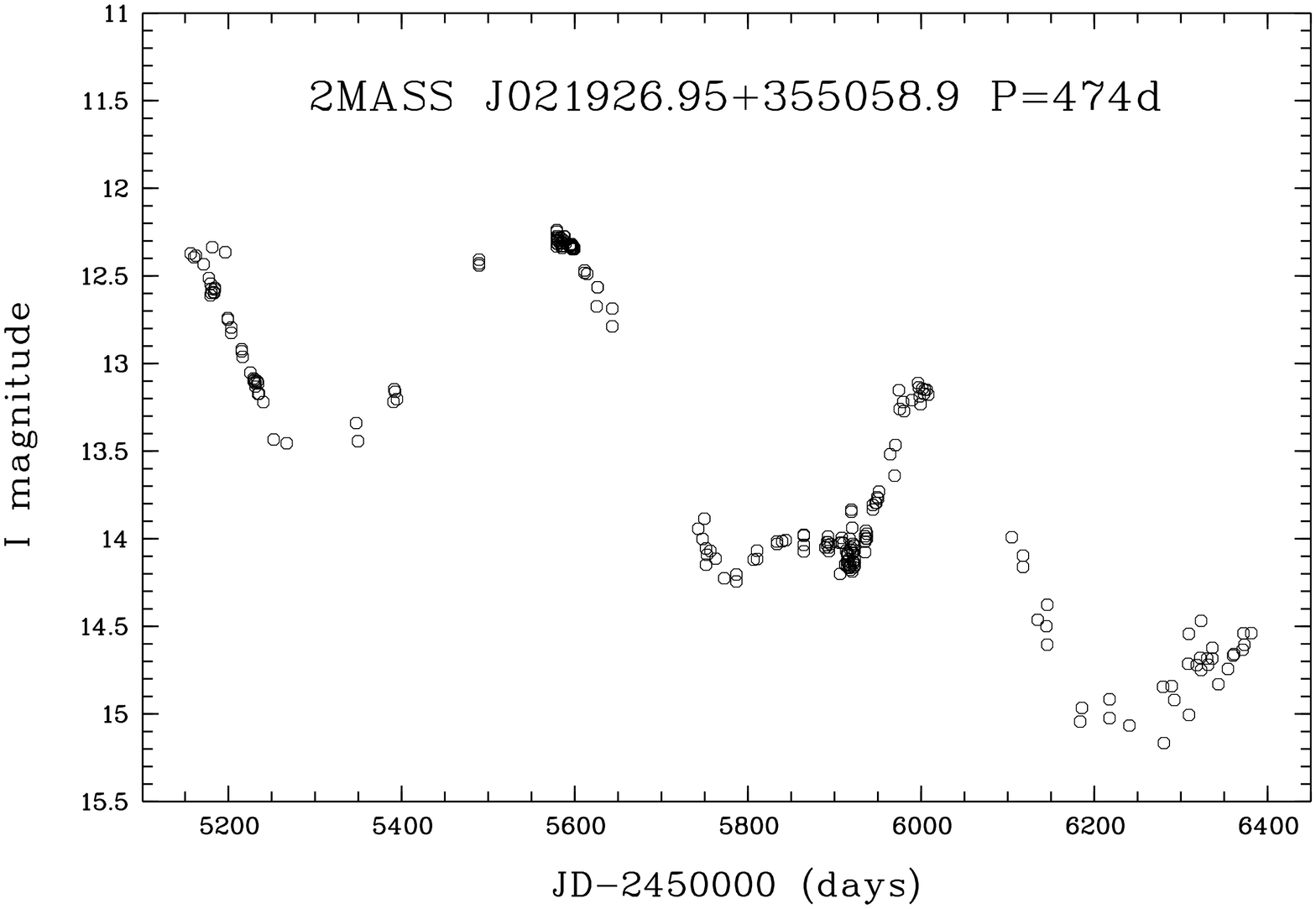}}}
       \resizebox{8cm}{!}{\rotatebox{-90}{\includegraphics{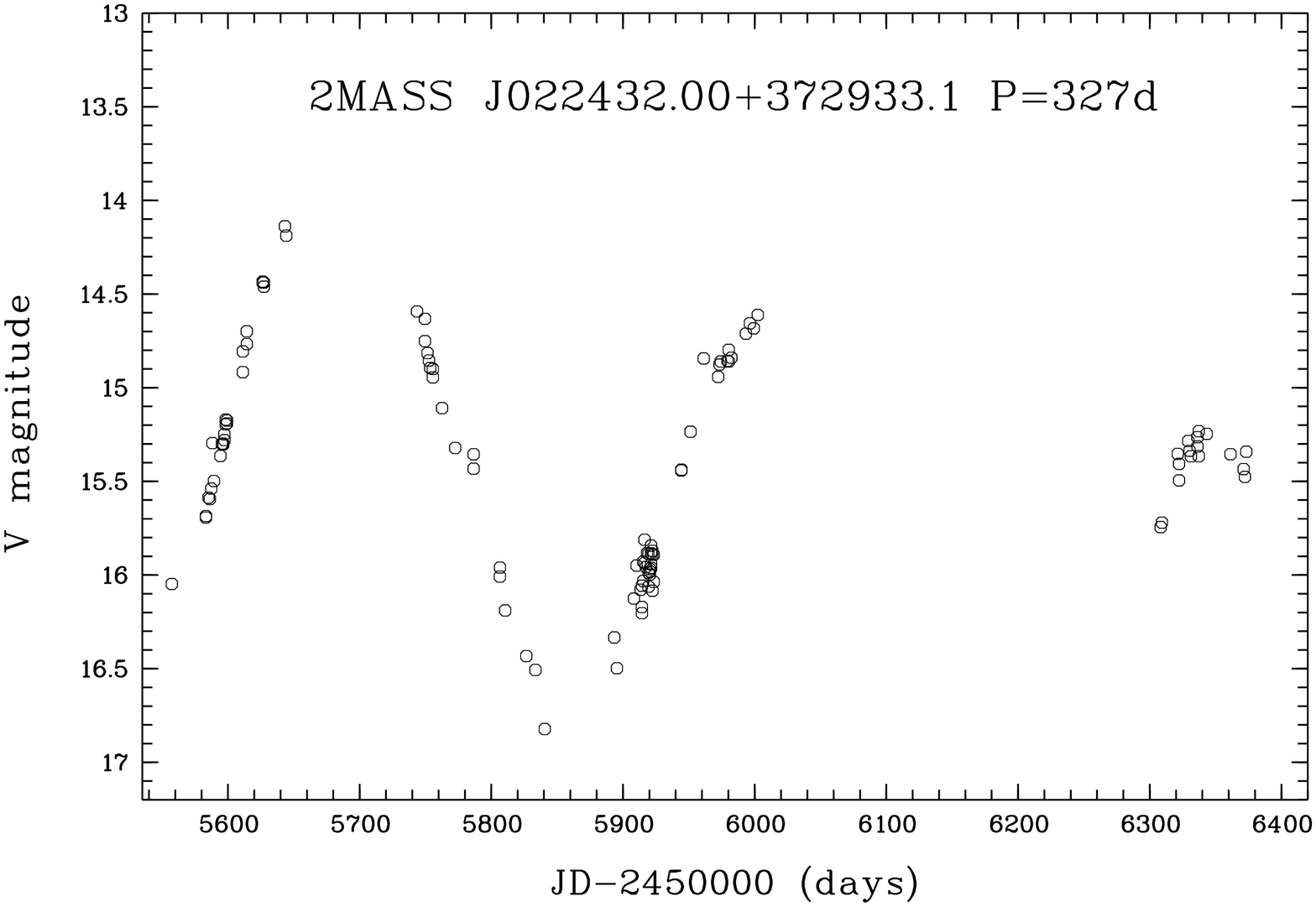}}}
     \vspace{2mm}

     \resizebox{8cm}{!}{\rotatebox{-90}{\includegraphics{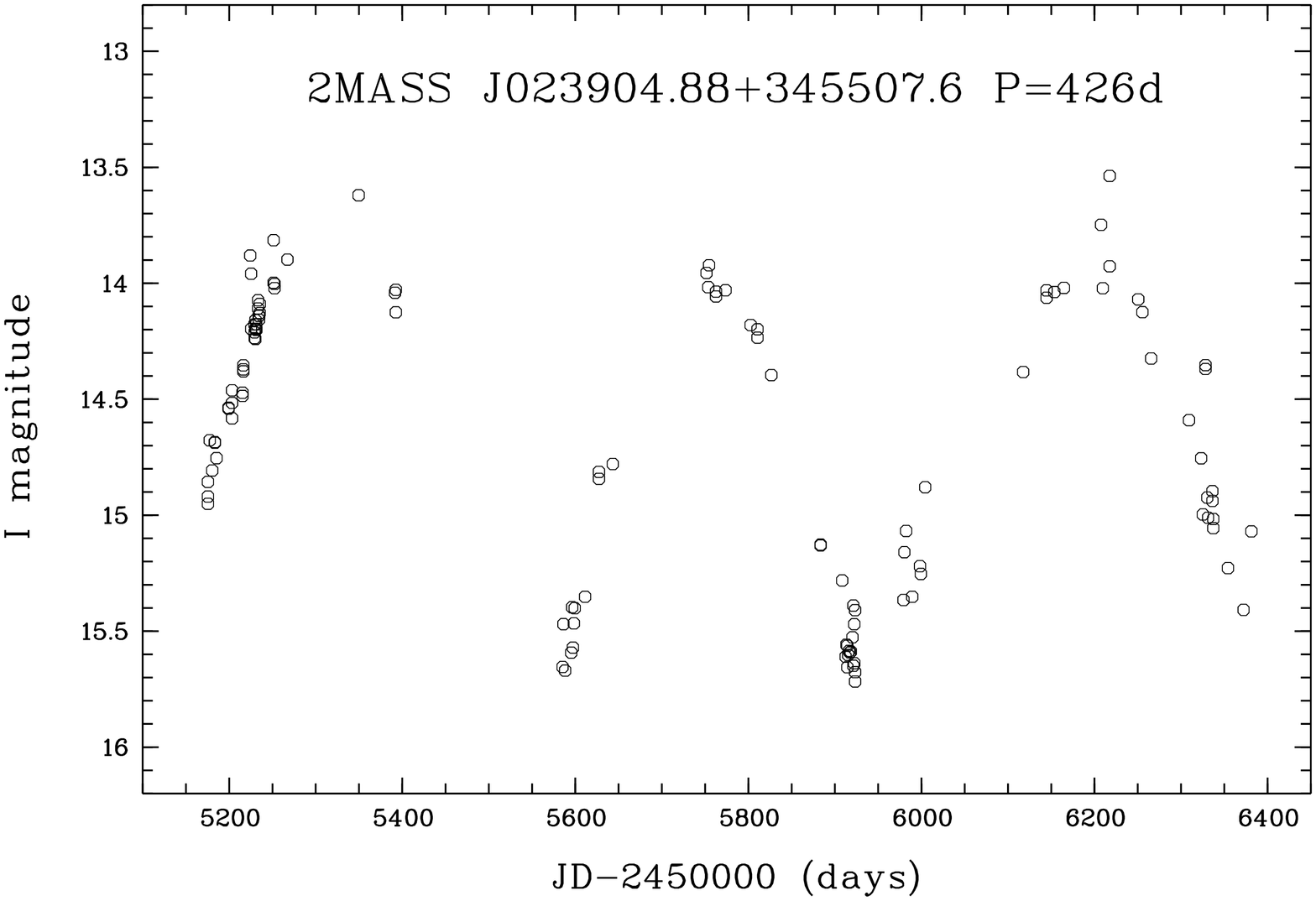}}}
     \resizebox{8cm}{!}{\rotatebox{-90}{\includegraphics{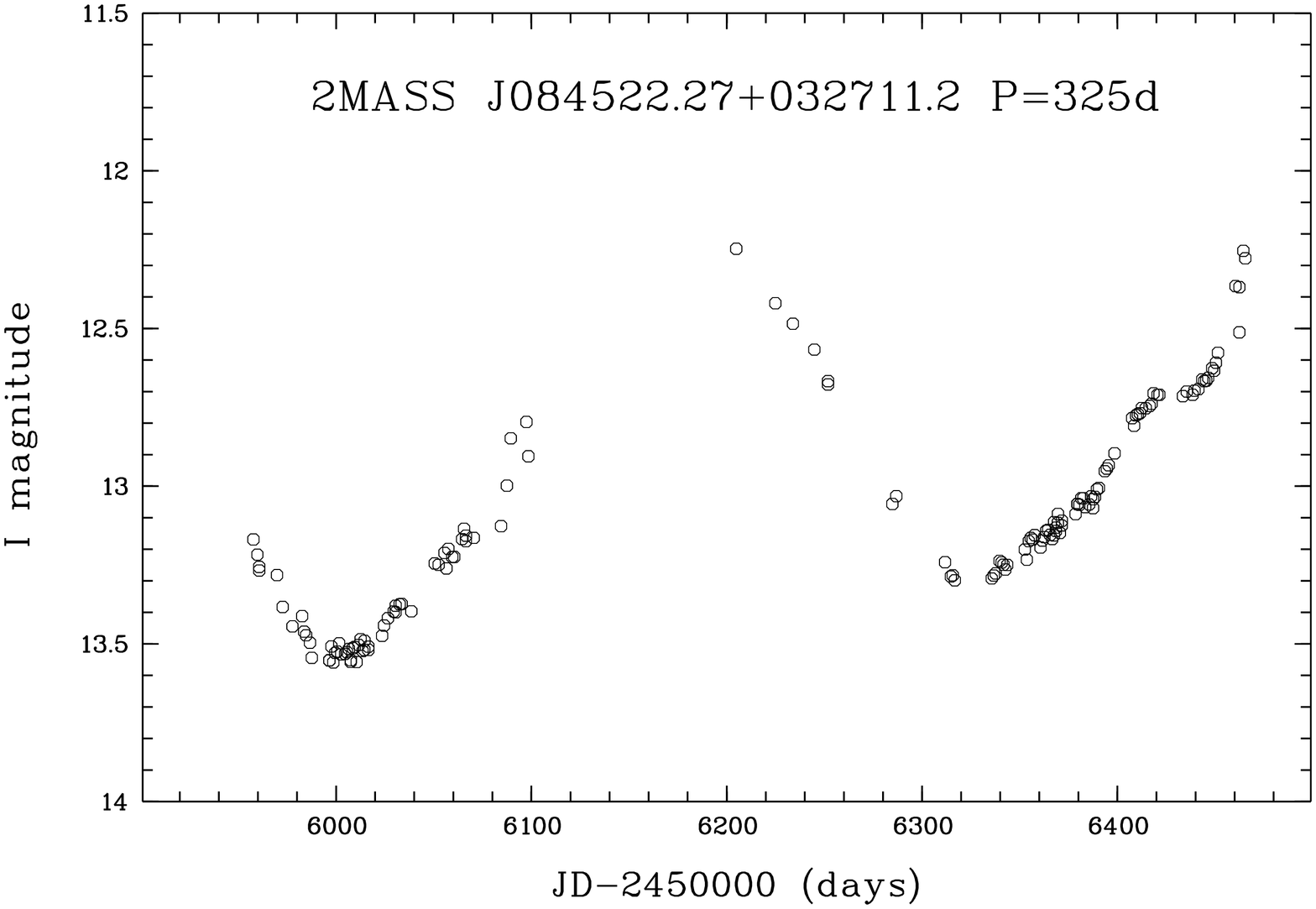}}}
     \vspace{2mm}

    \resizebox{8cm}{!}{\rotatebox{-90}{\includegraphics{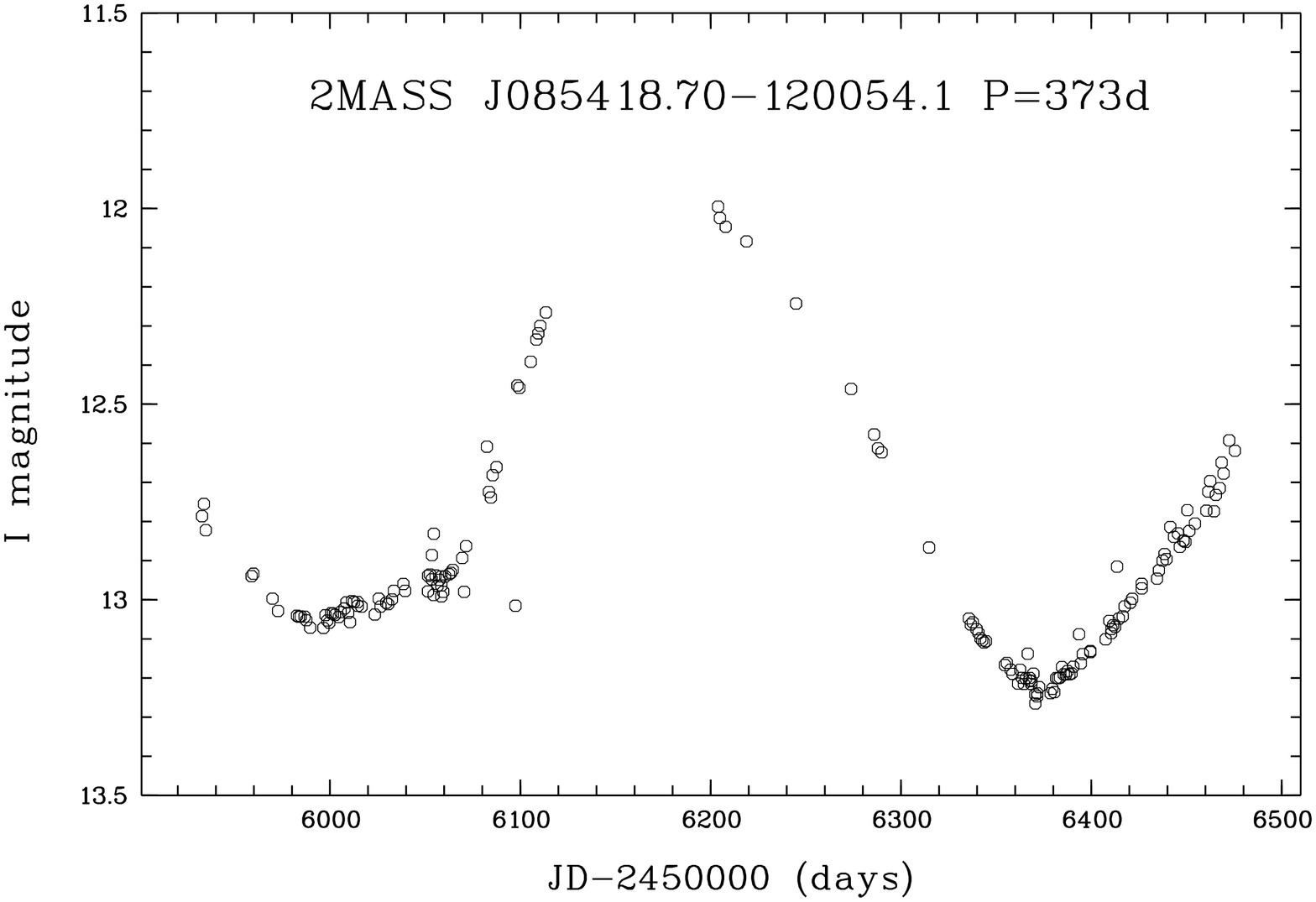}}}
    \resizebox{8cm}{!}{\rotatebox{-90}{\includegraphics{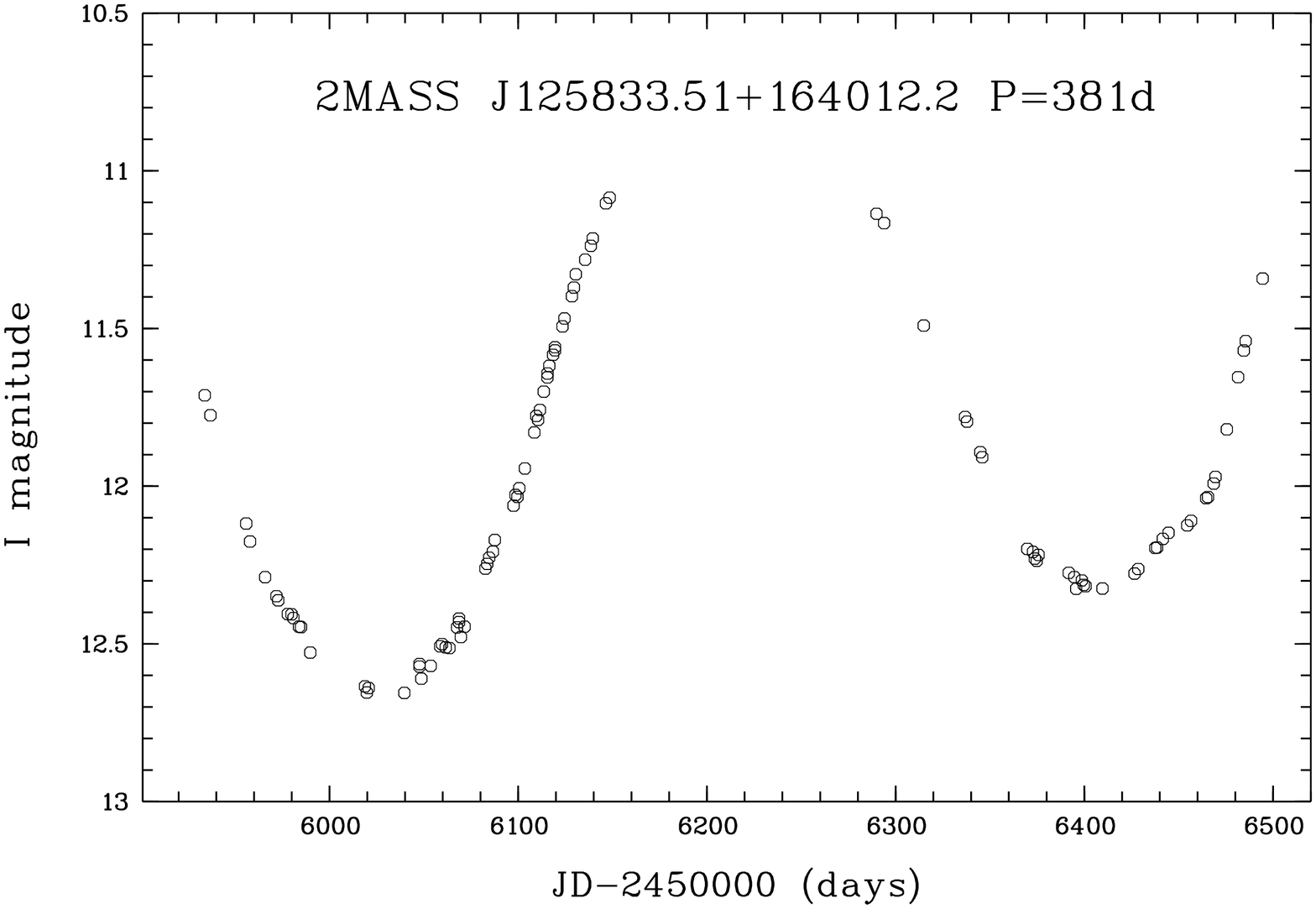}}}
     \vspace{2mm}

     \resizebox{8cm}{!}{\rotatebox{-90}{\includegraphics{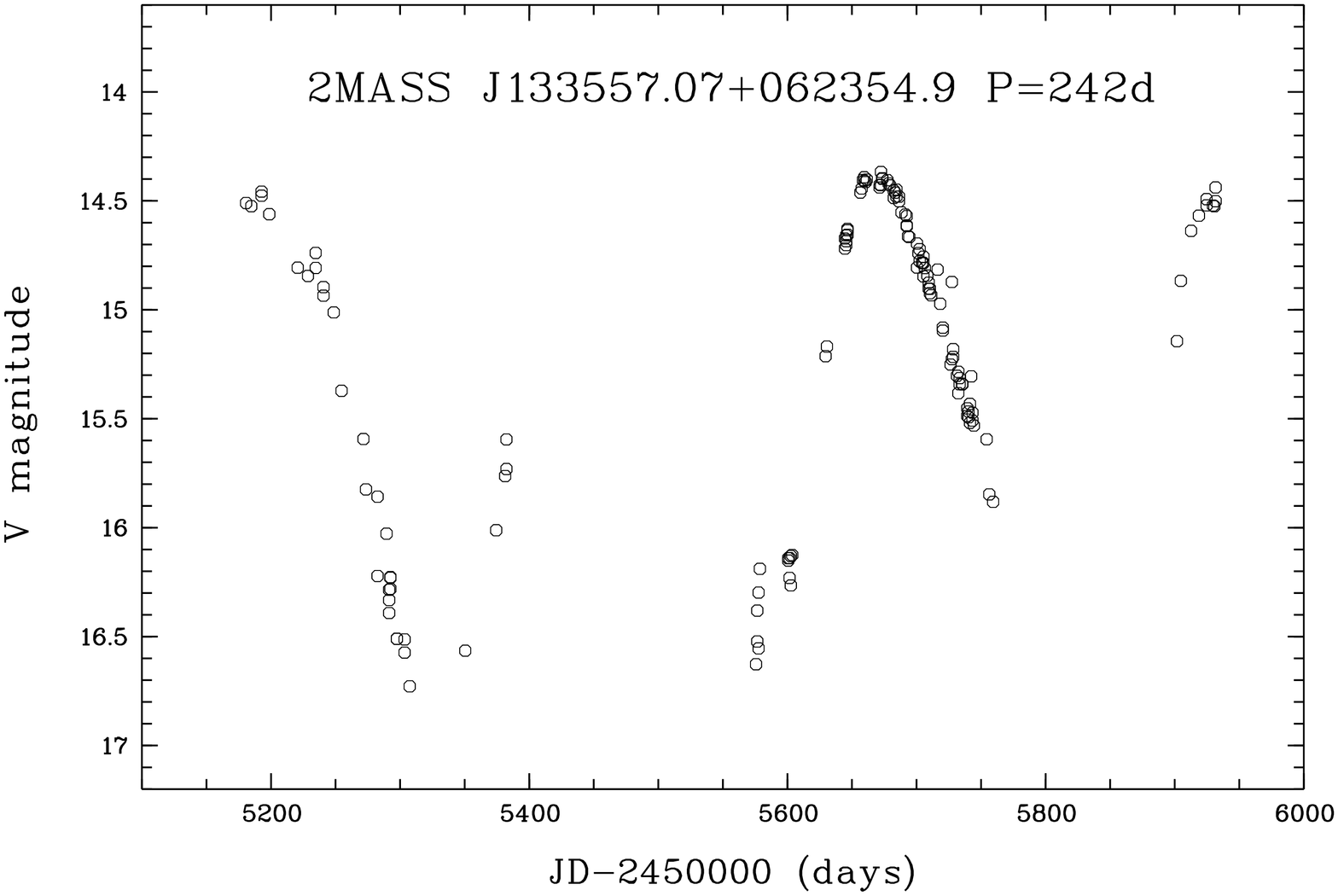}}}
     \resizebox{8cm}{!}{\rotatebox{-90}{\includegraphics{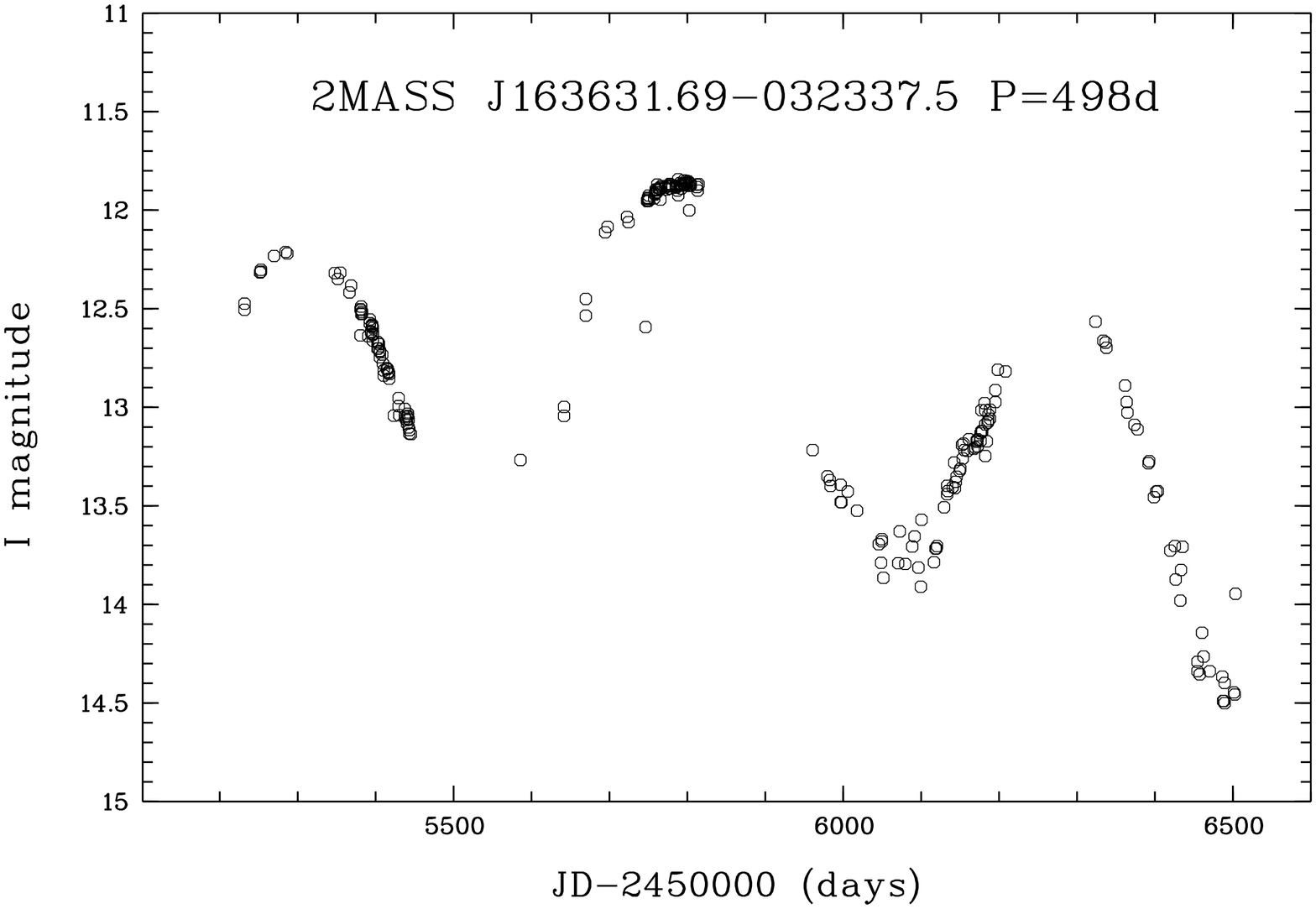}}}
 %   \end{center}
    \end{figure*}

     \begin{figure*}
 % \begin{center}
   \resizebox{8cm}{!}{\rotatebox{-90}{\includegraphics{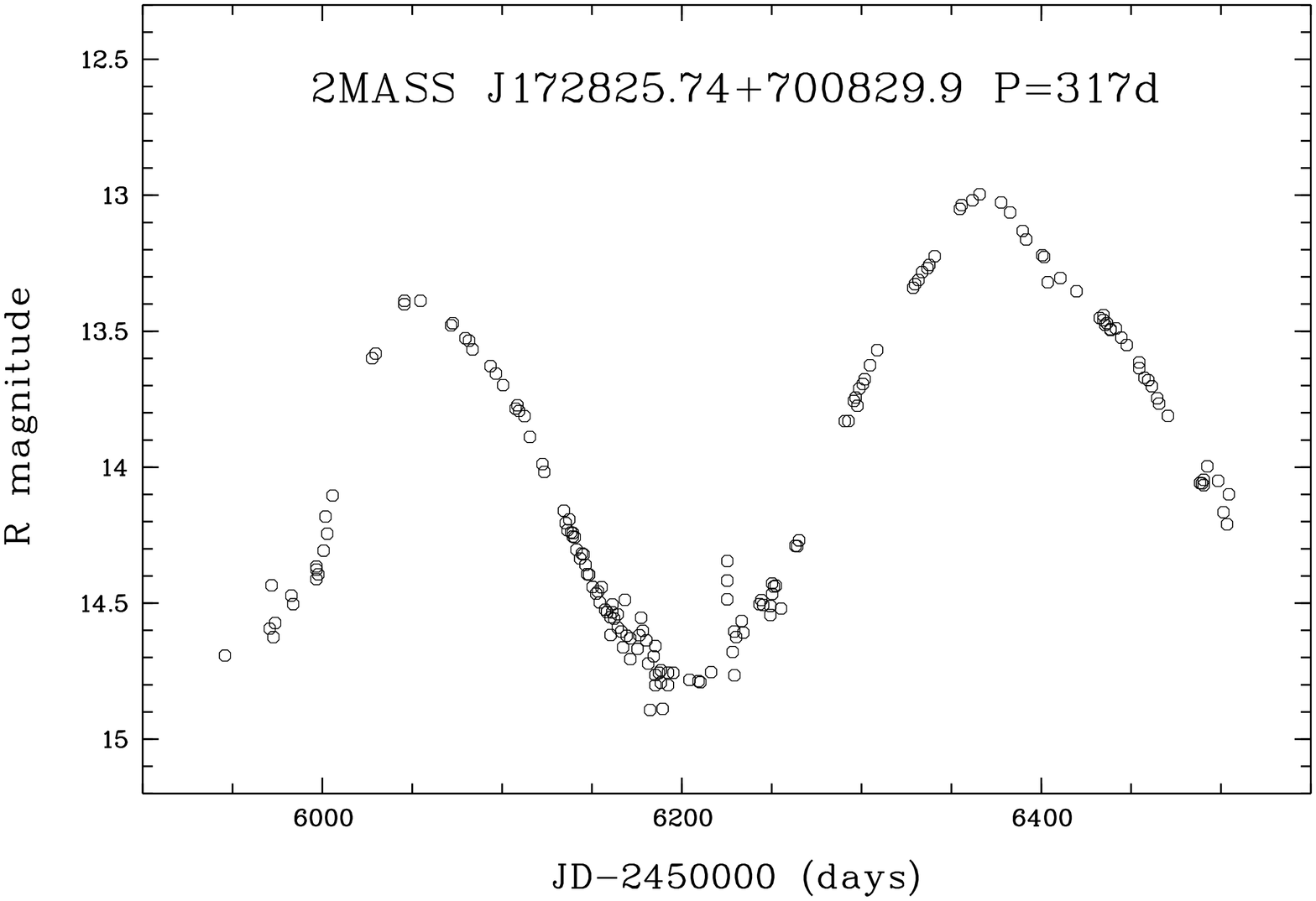}}}
   \resizebox{8cm}{!}{\rotatebox{-90}{\includegraphics{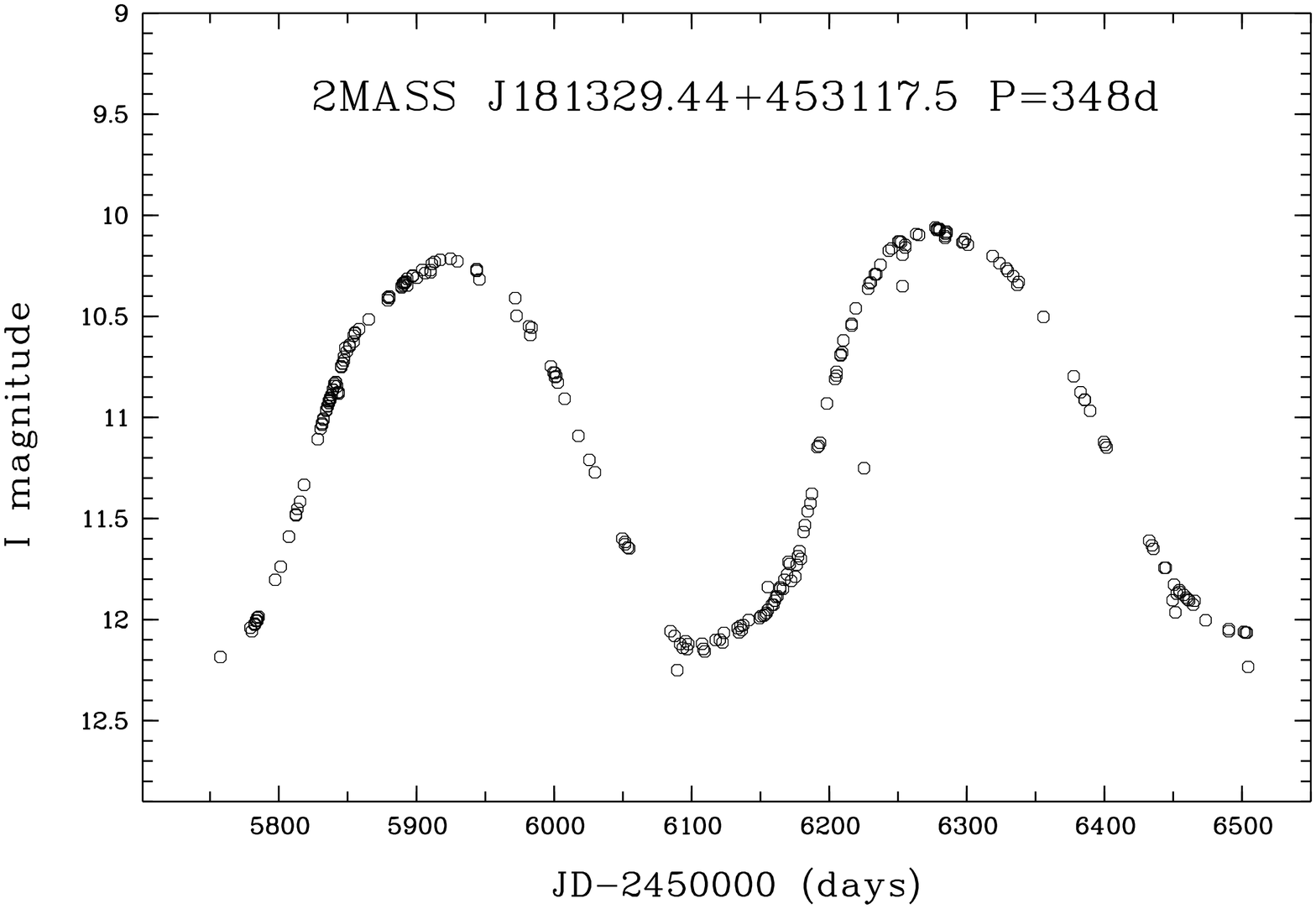}}}
   \vspace{2mm}

   \resizebox{8cm}{!}{\rotatebox{-90}{\includegraphics{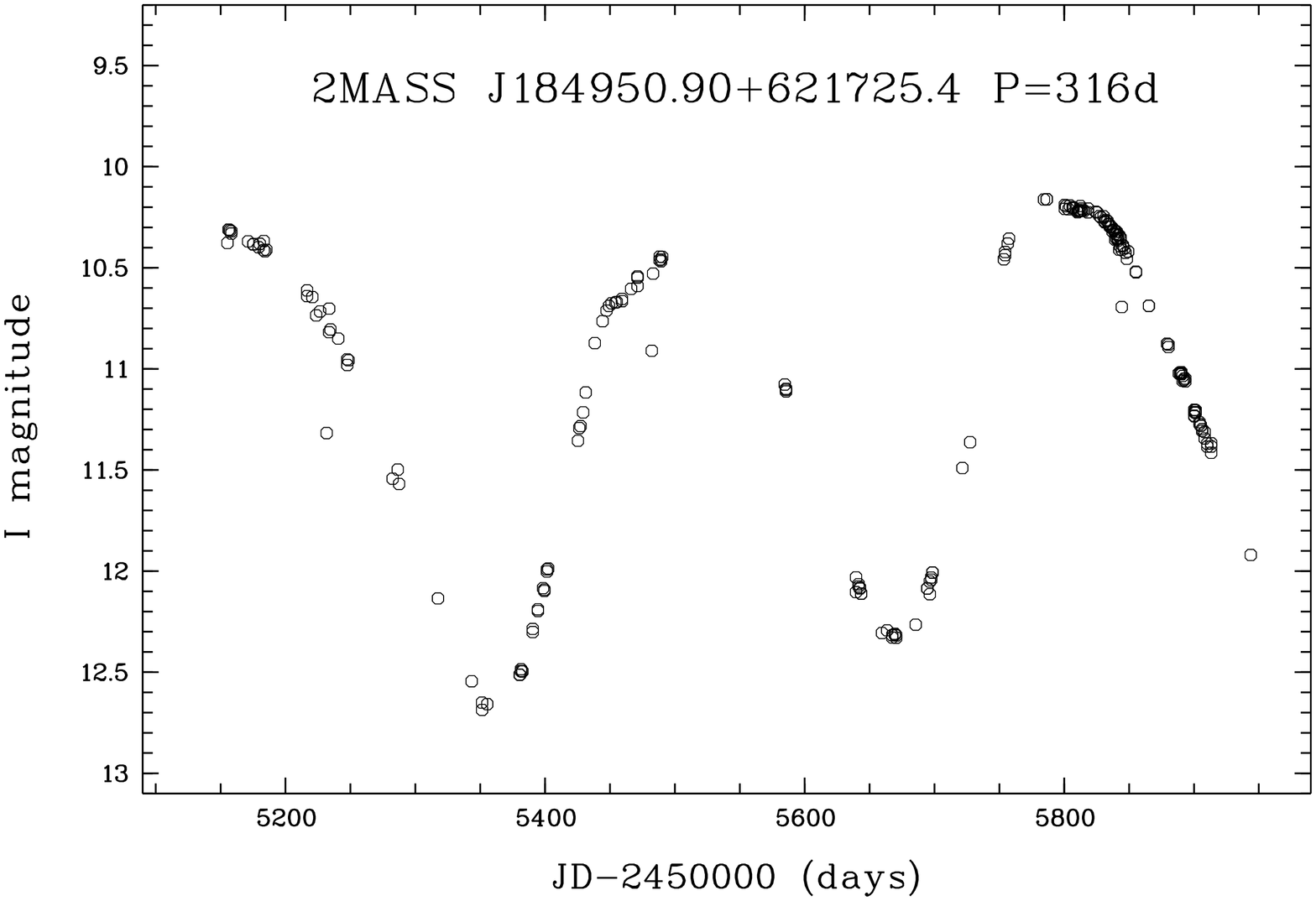}}}
   \resizebox{8cm}{!}{\rotatebox{-90}{\includegraphics{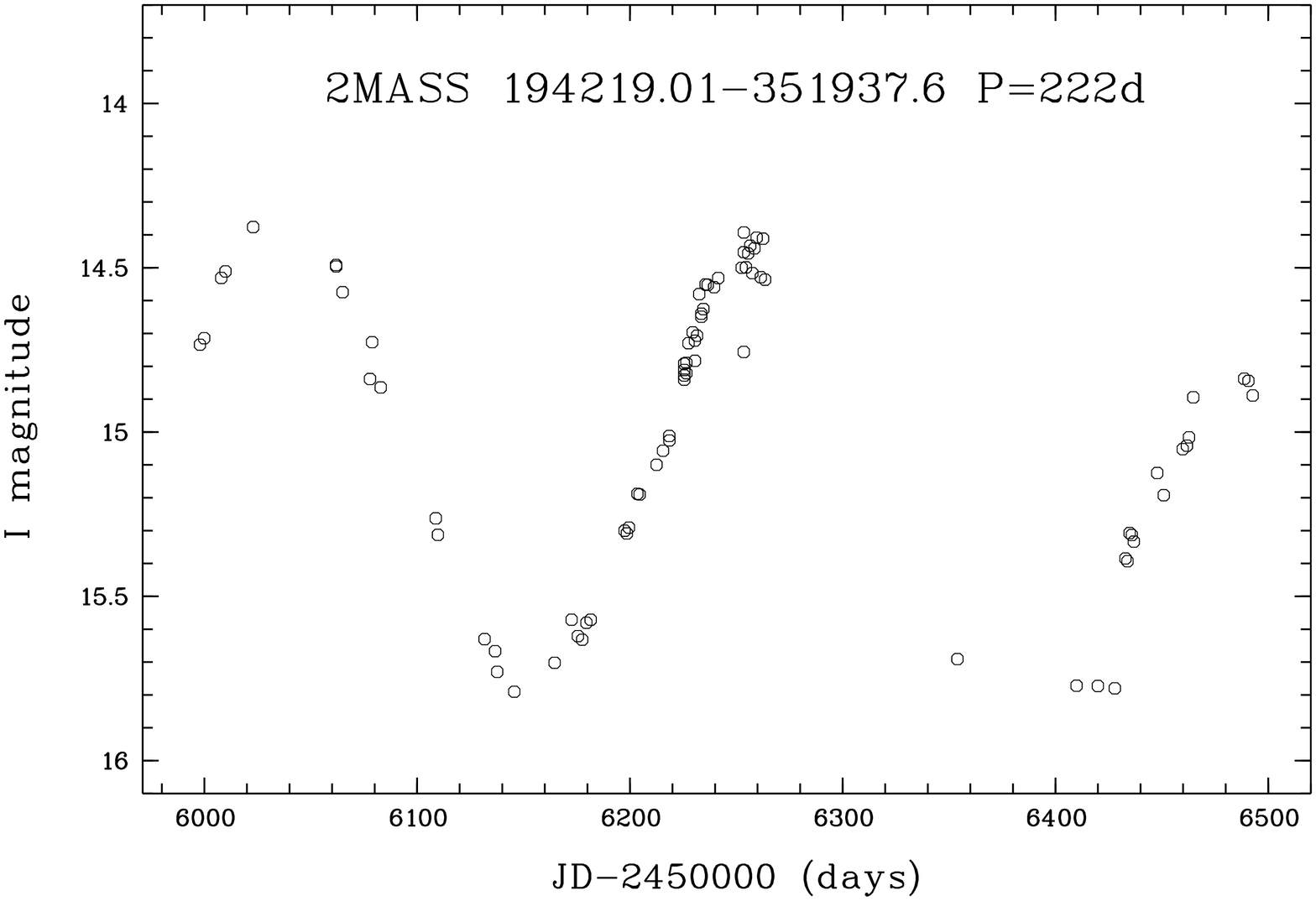}}}
   \vspace{2mm}

   \resizebox{8cm}{!}{\rotatebox{-90}{\includegraphics{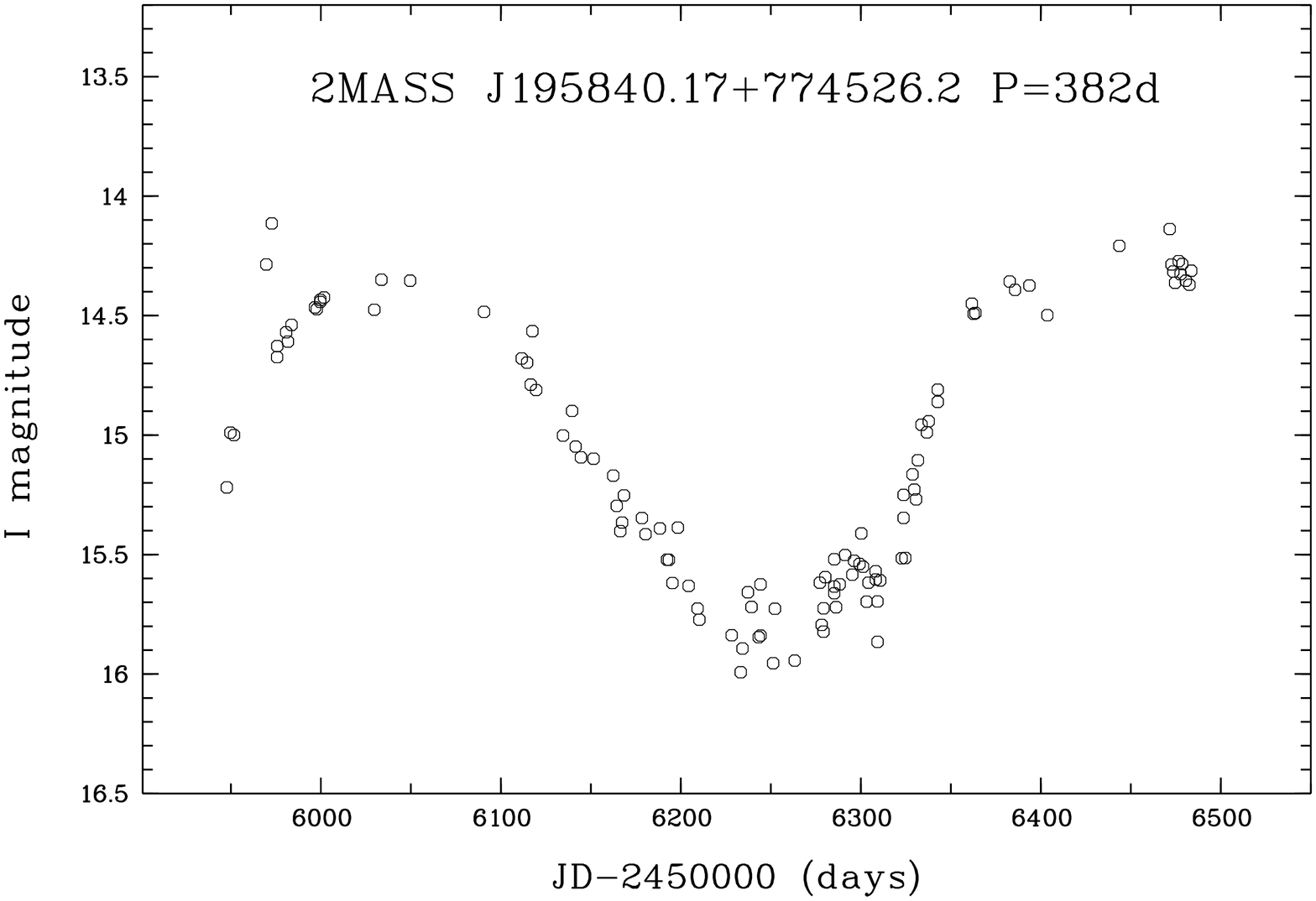}}}
   \resizebox{8cm}{!}{\rotatebox{-90}{\includegraphics{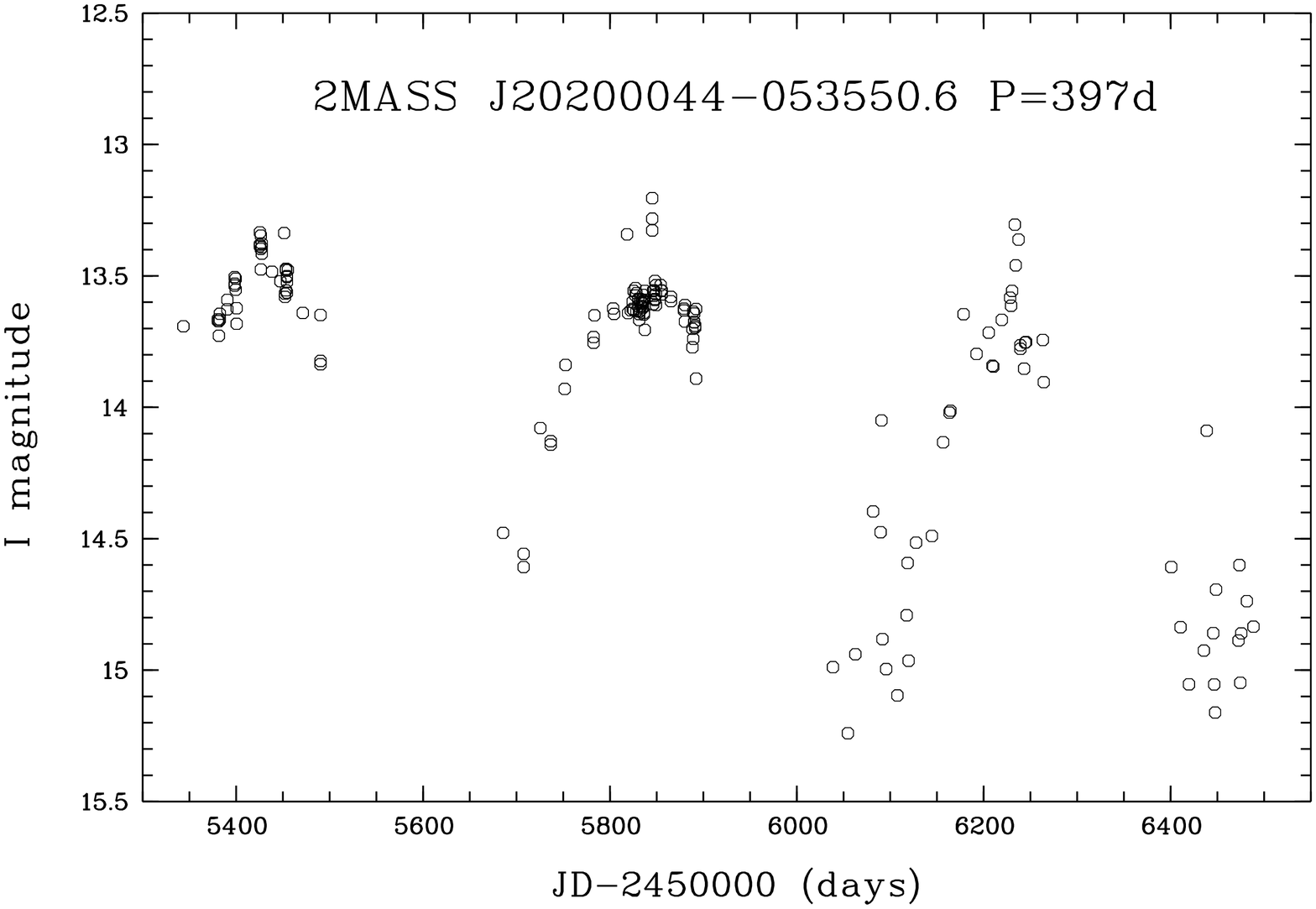}}}
   \vspace{2mm}

  \resizebox{8cm}{!}{\rotatebox{-90}{\includegraphics{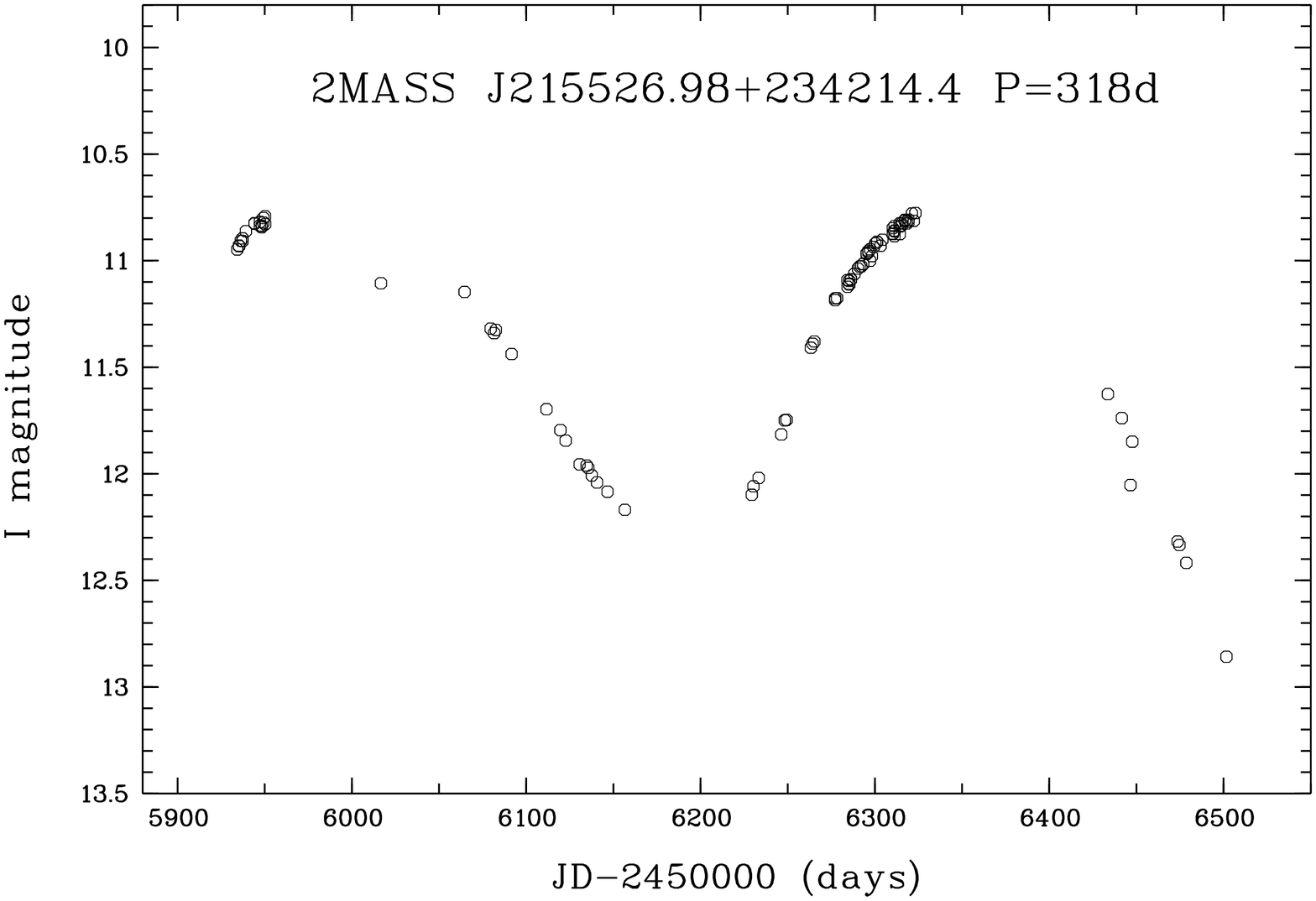}}}
  \resizebox{8cm}{!}{\rotatebox{-90}{\includegraphics{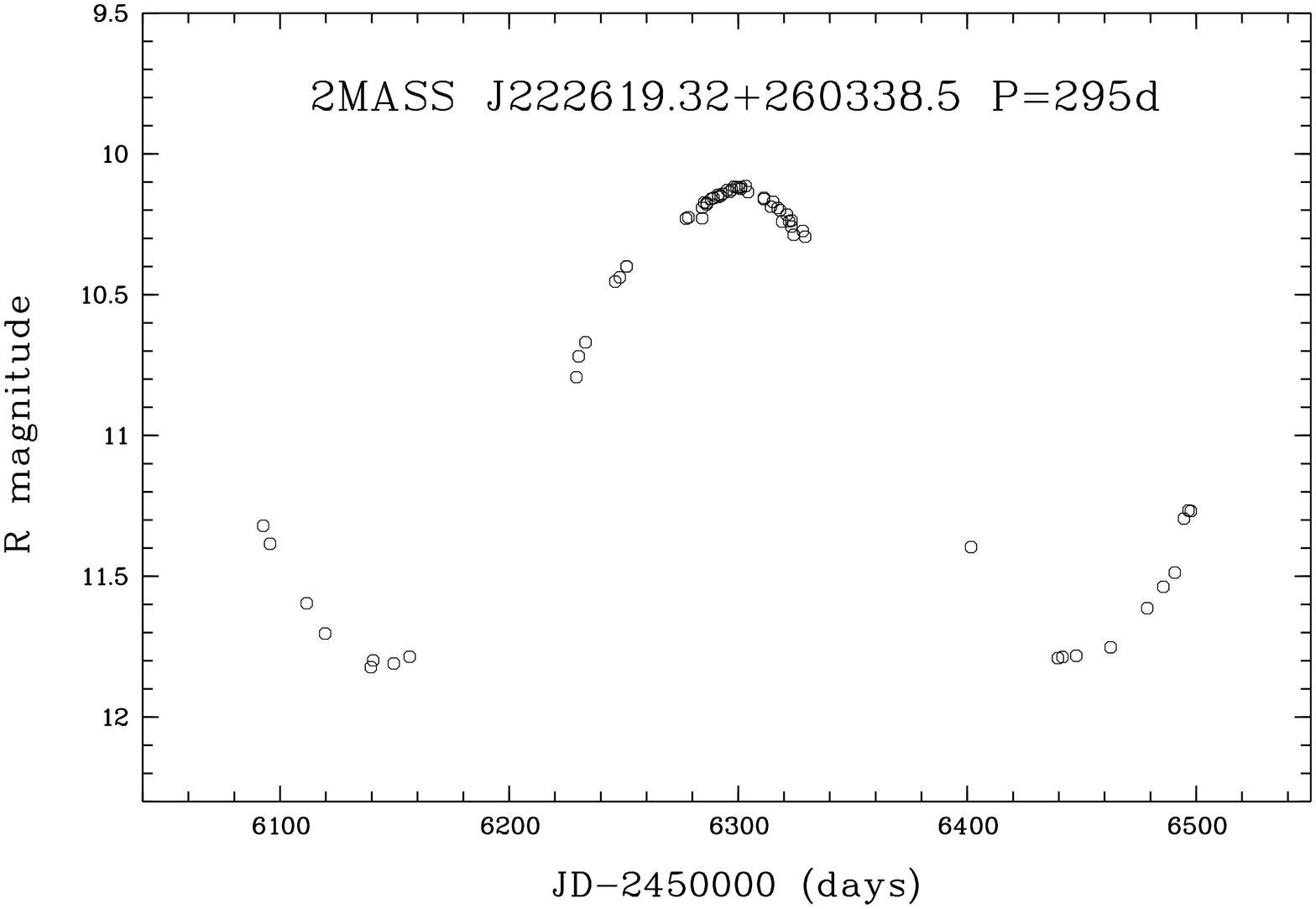}}}

 %  \end{center}
   \end{figure*}
%===============================================================================================

%\section{ Halo C stars with periodicity}

\begin{table*}[!ht]
\caption{Halo C stars with periods. $l$ and $b$ are given in degrees and  periods $P$ in days. 
The Catalina amplitude $A_C$ (in mag)
 is that of the fitted sinusoid. Mira is indicated in the M/SR column when $A_C > 1.5$\,mag. 
 We consider all others as SRa type variables. In the Note column, L means that  LINEAR data were  
used in addition to those of Catalina.}
\begin{flushleft}
\begin{tabular}{llrrrrrccc}
\noalign{\smallskip}
\hline
\hline
\noalign{\smallskip}
 Name &  ~~~~2MASS Ident. &  $l$~~~~   & $b$~~~~  & $K$ & $J-K$   &  $P$ & $A_C$~~~ & M/SR & Note  \\
\noalign{\smallskip}
\hline
\noalign{\smallskip}

APM 0000+3021      & \object{2MASS J000300.86$+$303823.3} & 110.769  & $-$31.086 & 8.90 & 2.00 &    305.3 & 0.8&      &\\
m31                & \object{2MASS J001655.77$-$440040.6} & 323.074  & $-$71.742 & 7.07 & 2.75 &    441.9 & 2.1& Mira &\\
m59                & \object{2MASS J003504.78$+$010845.8} & 114.351  & $-$61.453 &13.28 & 1.34 &    168.9 & 0.6&      & L\\
FBS 0158+095       & \object{2MASS J020056.14$+$094535.6} & 149.845  & $-$49.446 & 7.17 & 3.10 &    393.2 & 1.6& Mira &\\
APM 0207$-$0211    & \object{2MASS J021012.06$-$015739.0} & 163.123  & $-$58.546 &10.01 & 1.50 &    160.8 & 0.9&      &\\
m86                & \object{2MASS J022431.99$+$372933.1} & 142.664  & $-$21.786 & 8.75 & 2.74 &    337.0 & 1.6& Mira &\\
m87                & \object{2MASS J023904.88$+$345507.6} & 146.697  & $-$22.937 & 8.39 & 3.88 &    424.0 & 1.9& Mira &\\ 
m64                & \object{2MASS J030011.34$+$164940.2} & 162.049  & $-$36.082 &10.29 & 1.34 &    220.0 & 0.5&      &\\  
IRAS 03242$+$1429  & \object{2MASS J032659.92$+$143956.9} & 169.816  & $-$33.690 & 8.21 & 3.41 &    373.6 & 1.6& Mira &\\
APM 0340$+$0701    & \object{2MASS J034328.79$+$071058.8} & 179.778  & $-$36.062 & 8.32 & 1.46 &    296.8 & 1.0&      &\\
m65                & \object{2MASS J034828.13$+$165703.2} & 172.347  & $-$28.475 &11.10 & 2.00 &    223.9 & 1.1&      &\\
m88                & \object{2MASS J035955.96$+$091904.4} & 181.024  & $-$31.584 &10.11 & 2.64 &    255.2 & 1.5& Mira &\\
m69                & \object{2MASS J042638.78$+$142516.2} & 181.245  & $-$23.296 &11.20 & 2.01 &    318.8 & 1.3&      &\\
m103               & \object{2MASS J062806.04$-$531105.3} & 261.889  & $-$24.788 &11.27 & 1.48 &    222.2 & 0.6&      &\\
APM 0713$+$5016    & \object{2MASS J071736.25$+$501041.6} & 167.306  & $+$24.529 & 7.73 & 1.68 &    267.5 & 1.3&      &\\
FBS 0729$+$269     & \object{2MASS J073232.74$+$264715.6} & 192.427  & $+$20.319 & 8.15 & 1.73 &    148.8 & 1.2&      &\\
SDSS Green $\#$134 & \object{2MASS J075855.89$+$335838.1} & 186.898  & $+$27.973 &13.21 & 1.39 &    209.7 & 1.2&      &\\
m03                & \object{2MASS J082915.12$+$182307.4} & 206.225  & $+$29.566 & 7.07 & 1.68 &    323.3 & 1.4&      &\\
HE 1008$-$0636     & \object{2MASS J082929.03$+$104624.2} & 214.181  & $+$26.613 & 8.14 & 2.12 &    294.0 & 1.0&      &\\
m04                & \object{2MASS J085418.70$-$120054.1} & 239.096  & $+$20.472 & 8.02 & 2.73 &    388.7 & 1.2&      &\\

IRAS 08546$+$1732  & \object{2MASS J085725.83$+$172051.9} & 210.261  & $+$35.437 &10.71 & 4.41 &    394.8 & 1.2&      &L\\
SDSS Green $\#$232 & \object{2MASS J090546.36$+$202438.1} & 207.519  & $+$38.347 &12.29 & 2.82 &    257.6 & 1.9& Mira &L\\
m03                & \object{2MASS J091505.21$+$191737.8} & 209.815  & $+$40.043 & 8.50 & 1.70 &    257.1 & 1.1&      &L\\
HE 1008$-$0636     & \object{2MASS J101036.99$-$065113.6} & 248.123  & $+$38.348 & 8.53 & 1.42 &    334.9 & 1.3&      &\\
m04                & \object{2MASS J101525.93$-$020431.8} & 244.495  & $+$42.428 &11.95 & 2.10 &    340.8 & 1.4&      &L \\

SDSS Green $\#$429 & \object{2MASS J111320.64$+$221116.0} & 220.056  & $+$67.212 &14.50 & 1.35 &    184.3 & 0.9&      &L \\
m07                & \object{2MASS J111719.00$-$172915.4} & 273.180  & $+$39.885 &11.22 & 1.43 &    314.8 & 1.2&      &\\

m08                & \object{2MASS J120925.03$+$151618.4} & 261.330  & $+$74.639 & 9.83 & 1.35 &    256.0 & 0.7&      &L \\
APM 1211$-$0844    & \object{2MASS J121416.95$-$090050.0} & 287.661  & $+$52.753 &11.88 & 1.35 &    156.8 & 1.1&      &\\
APM 1225$-$0011    & \object{2MASS J122740.05$-$002750.7} & 290.267  & $+$61.822 &10.55 & 2.21 &    284.4 & 1.8& Mira &L\\

2MASS Gizis $\#$32 & \object{2MASS J123829.27$-$405609.0} & 300.297  & $+$21.869 &12.73 & 2.64 &    265.9 & 2.1& Mira &\\
APM 1241$+$0237    & \object{2MASS J124337.31$+$022130.2} & 298.280  & $+$65.159 &11.45 & 1.54 &    515.6 & 0.3&      &L \\
m09                & \object{2MASS J124904.78$+$132035.4} & 300.526  & $+$76.204 &11.17 & 1.43 &    238.5 & 1.0&      &L \\
APM 1249+0146      & \object{2MASS J125149.87$+$013001.8} & 303.159  & $+$64.372 &10.20 & 1.87 &    278.9 & 1.4&      &L \\
SDSS Green $\#$638 & \object{2MASS J130155.85$+$083631.8} & 311.052  & $+$71.315 &12.07 & 2.64 &    263.5 & 1.1&      &L \\

m91                & \object{2MASS J133557.08$+$062355.0} & 331.973  & $+$66.727 &10.57 & 1.70 &    243.1 & 1.3&      &L \\
m41                & \object{2MASS J134723.04$-$344723.4} & 315.767  & $+$26.684 &10.10 & 1.98 &    307.3 & 0.9&      &\\
APM 1350$+$0101    & \object{2MASS J135301.31$+$004714.1} & 334.780  & $+$59.795 &11.08 & 1.57 &    239.3 & 0.6&      &L \\
m10                & \object{2MASS J135602.37$-$013626.2} & 333.930  & $+$57.324 &11.31 & 1.59 &    215.0 & 0.4&      &L \\
APM 1429$-$0518    & \object{2MASS J143228.75$-$053117.8} & 343.501  & $+$49.225 &11.24 & 2.75 &    289.9 & 1.5& Mira &  \\

m42                & \object{2MASS J143758.36$-$041336.1} & 346.372  & $+$49.445 &10.90 & 1.36 &    260.7 & 0.5&      & \\
SDSS Green $\#$853 & \object{2MASS J144631.07$-$005500.3} & 352.235  & $+$50.600 &11.05 & 1.36 &    229.4 & 1.6& Mira &L \\
m92                & \object{2MASS J144644.19$+$051238.1} & 359.497  & $+$54.876 &11.35 & 2.29 &    238.1 & 1.2&      &L \\
m44                & \object{2MASS J145726.98$+$051603.5} &   2.520  & $+$52.890 &11.44 & 2.80 &    217.7 & 1.3&      &L \\
m12                & \object{2MASS J150106.93$-$053138.8} & 351.628  & $+$44.738 &11.50 & 2.09 &    253.9 & 0.8&      &\\

FBS 1502$+$359     & \object{2MASS J150455.30$+$354757.6} &  58.321  & $+$60.407 & 9.69 & 2.32 &    309.9 & 1.2&      &L \\
APM 1511$-$0342    & \object{2MASS J151341.56$-$035348.0} & 356.357  & $+$43.734 &11.31 & 1.43 &    224.1 & 0.7&      &\\
m13                & \object{2MASS J151511.07$-$133227.9} & 348.096  & $+$36.427 &10.80 & 1.80 &    241.9 & 1.7& Mira &\\
APM 1519$-$0614    & \object{2MASS J152236.57$-$062534.2} & 356.004  & $+$40.379 &11.42 & 1.37 &    195.2 & 0.8&      &\\
m45                & \object{2MASS J152244.43$-$123749.5} & 350.551  & $+$35.893 &11.43 & 1.65 &    233.7 & 0.4&      &\\

m15                & \object{2MASS J172825.74$+$700829.9} & 100.829  & $+$32.411 & 9.02 & 2.52 &    301.3 & 0.9&      &L \\
m52                & \object{2MASS J193734.13$-$353237.7} &   3.862  & $-$24.196 & 9.12 & 2.14 &    367.6 & 0.7&      &\\
m16                & \object{2MASS J194219.01$-$351937.7} &   4.398  & $-$25.059 &10.04 & 2.62 &    230.4 & 1.5& Mira &\\
m17                & \object{2MASS J194221.31$-$321104.1} &   7.701  & $-$24.132 & 9.98 & 1.98 &    232.6 & 1.0&      &\\
m18                & \object{2MASS J194850.65$-$305831.7} &   9.433  & $-$25.076 &10.21 & 3.39 &    337.0 & 1.8& Mira &\\

m19                & \object{2MASS J195330.18$-$383559.3} &   1.518  & $-$28.070 & 9.26 & 2.03 &    256.0 & 1.6& Mira &\\
m106               & \object{2MASS J200144.00$-$302446.5} &  10.954  & $-$27.553 &10.36 & 1.30 &    166.1 & 0.9&      &\\
m99                & \object{2MASS J200303.83$-$194903.9} &  22.189  & $-$24.332 & 9.11 & 1.42 &    186.5 & 1.5& Mira &\\
m100               & \object{2MASS J202000.43$-$053550.7} &  38.153  & $-$22.230 & 8.61 & 3.43 &    424.5 & 1.8& Mira &\\
m109               & \object{2MASS J203347.68$-$463620.6} & 353.398  & $-$36.543 & 8.92 & 1.59 &    240.3 & 0.9&      &\\

m23                & \object{2MASS J220514.58$+$000846.0} &  60.313  & $-$41.674 & 9.30 & 2.24 &    219.5 & 1.0&      &L \\
m24                & \object{2MASS J220653.67$-$250628.2} &  26.546  & $-$53.173 & 8.95 & 1.98 &    330.9 & 1.6& Mira &\\
m25                & \object{2MASS J221709.92$-$260703.3} &  25.636  & $-$55.642 & 8.88 & 2.16 &    311.7 & 1.6& Mira &\\
m83                & \object{2MASS J222301.20$+$221656.5} &  83.298  & $-$28.940 & 9.52 & 2.09 &    322.6 & 1.4&      &\\
APM 2225$-$1401    & \object{2MASS J222810.68$-$134622.4} &  47.457  & $-$54.048 & 9.90 & 1.97 &    286.1 & 1.4&      &\\
 
m27                & \object{2MASS J231935.54$-$185623.9} &  49.282  & $-$67.385 & 9.97 & 1.53 &    147.6 & 0.9&      &\\

\noalign{\smallskip}
\hline
\end{tabular}
\end{flushleft}
\end{table*}

% -------------------
   \clearpage

  \begin{figure*}
  \caption{Atlas of light curves with LINEAR (colored in cyan) and Catalina (in magenta), and fitted sinusoids.}

  \vspace{10mm}
  \resizebox{17cm}{!}{\rotatebox{00}{\includegraphics{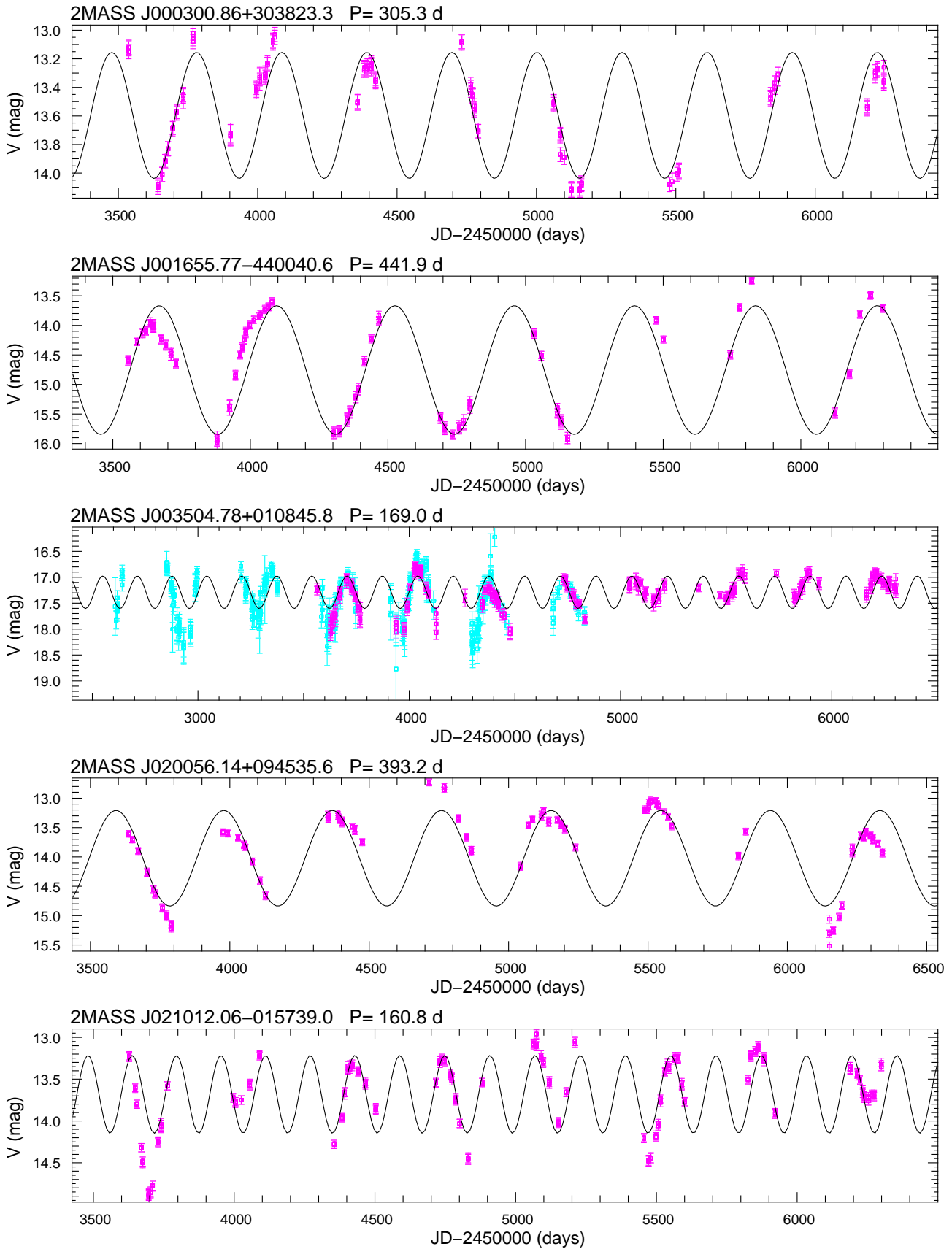}}}
  \end{figure*}

  \begin{figure*}
  \resizebox{17cm}{!}{\rotatebox{00}{\includegraphics{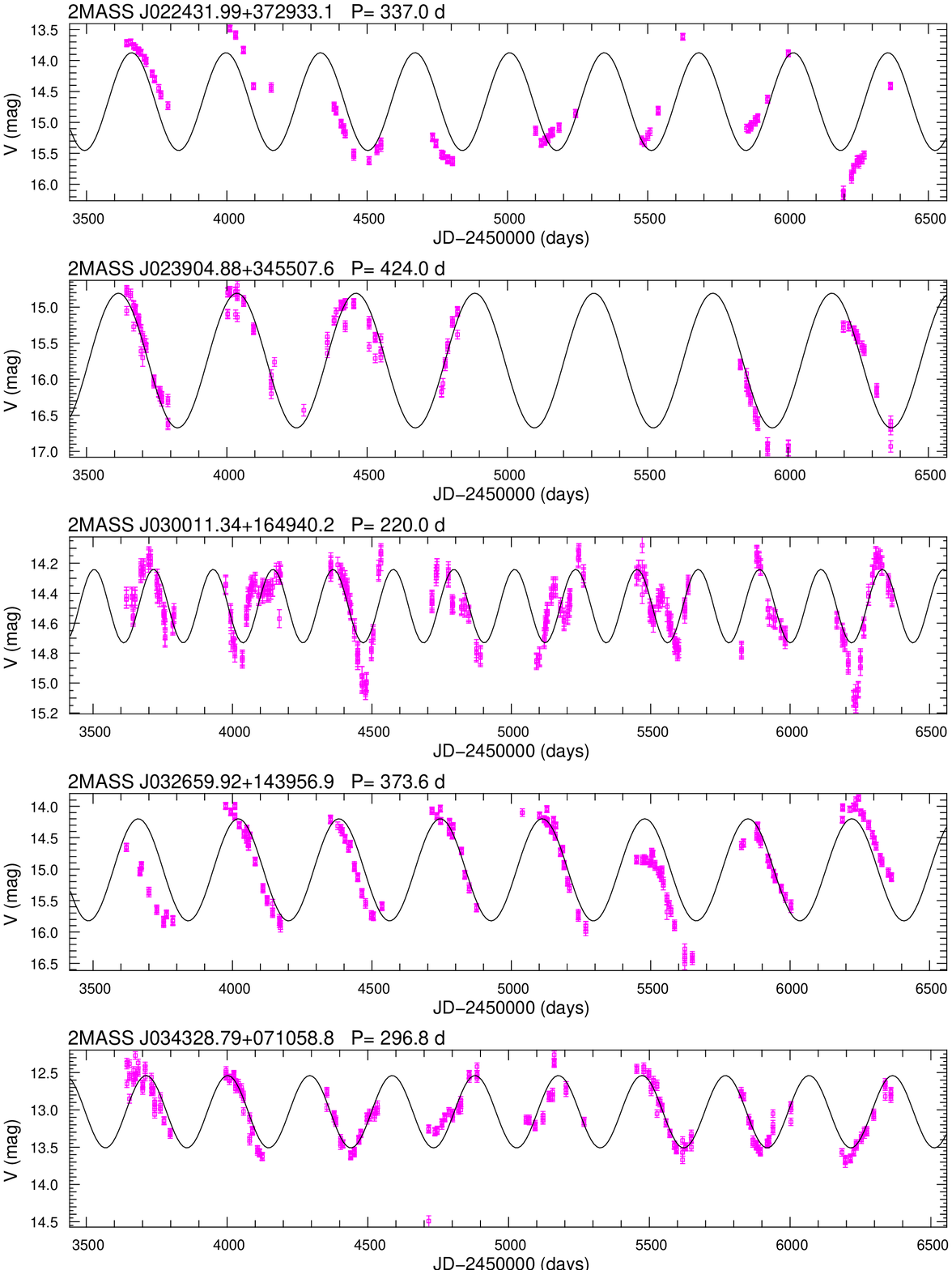}}}
  \end{figure*}

  \begin{figure*}
  \resizebox{17cm}{!}{\rotatebox{00}{\includegraphics{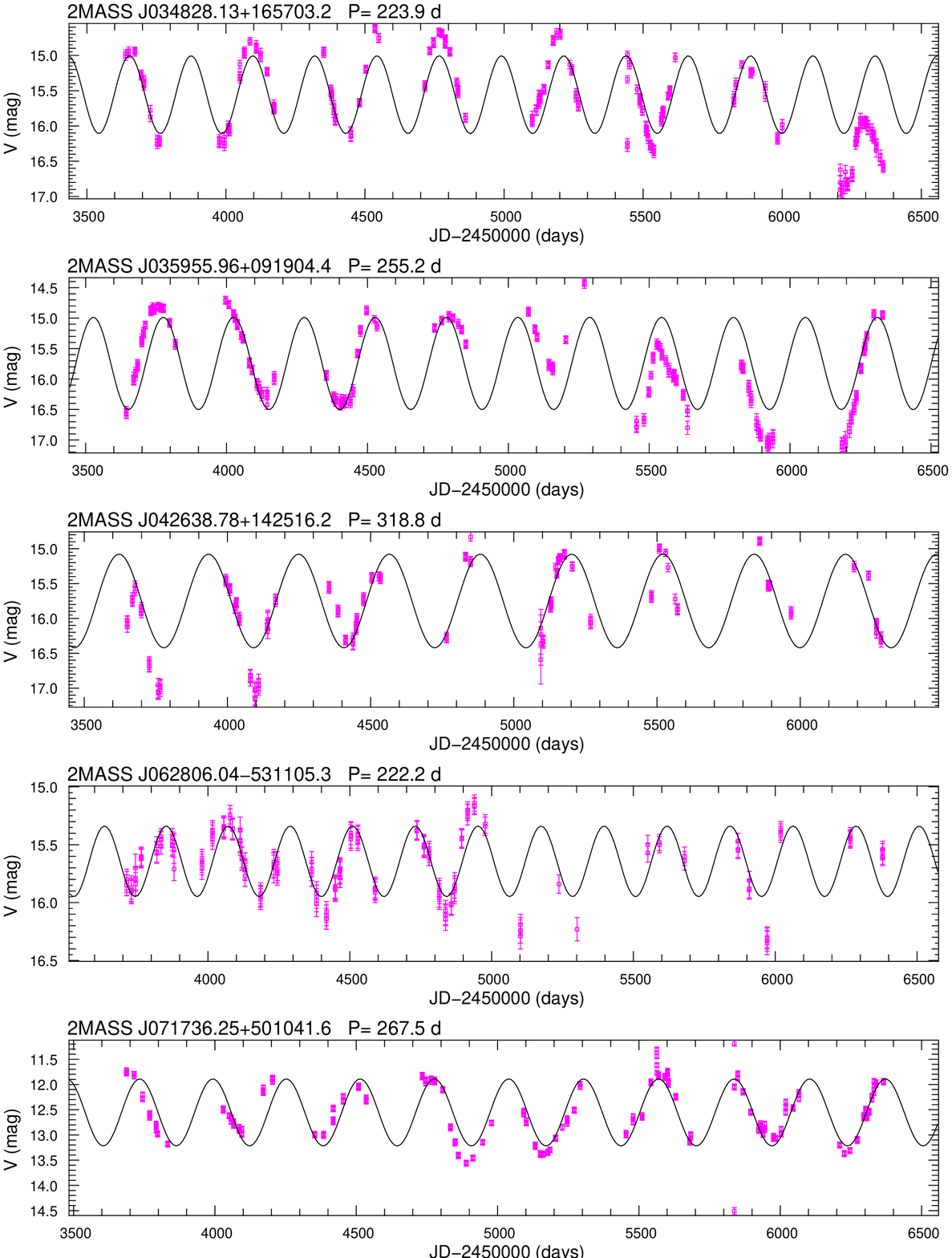}}}
  \end{figure*}

  \begin{figure*}
  \resizebox{17cm}{!}{\rotatebox{00}{\includegraphics{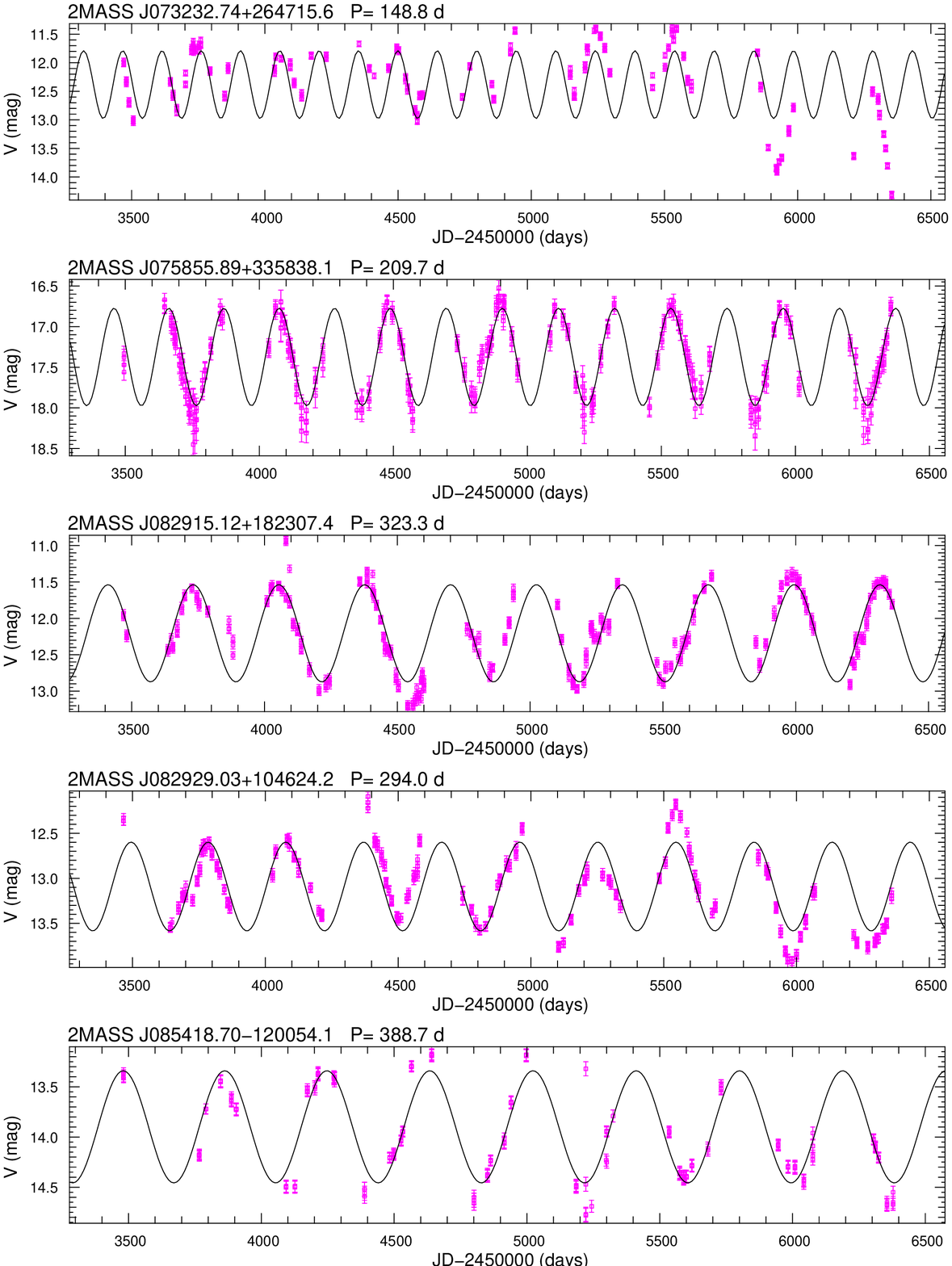}}}
  \end{figure*}

  \begin{figure*}
  \resizebox{17cm}{!}{\rotatebox{00}{\includegraphics{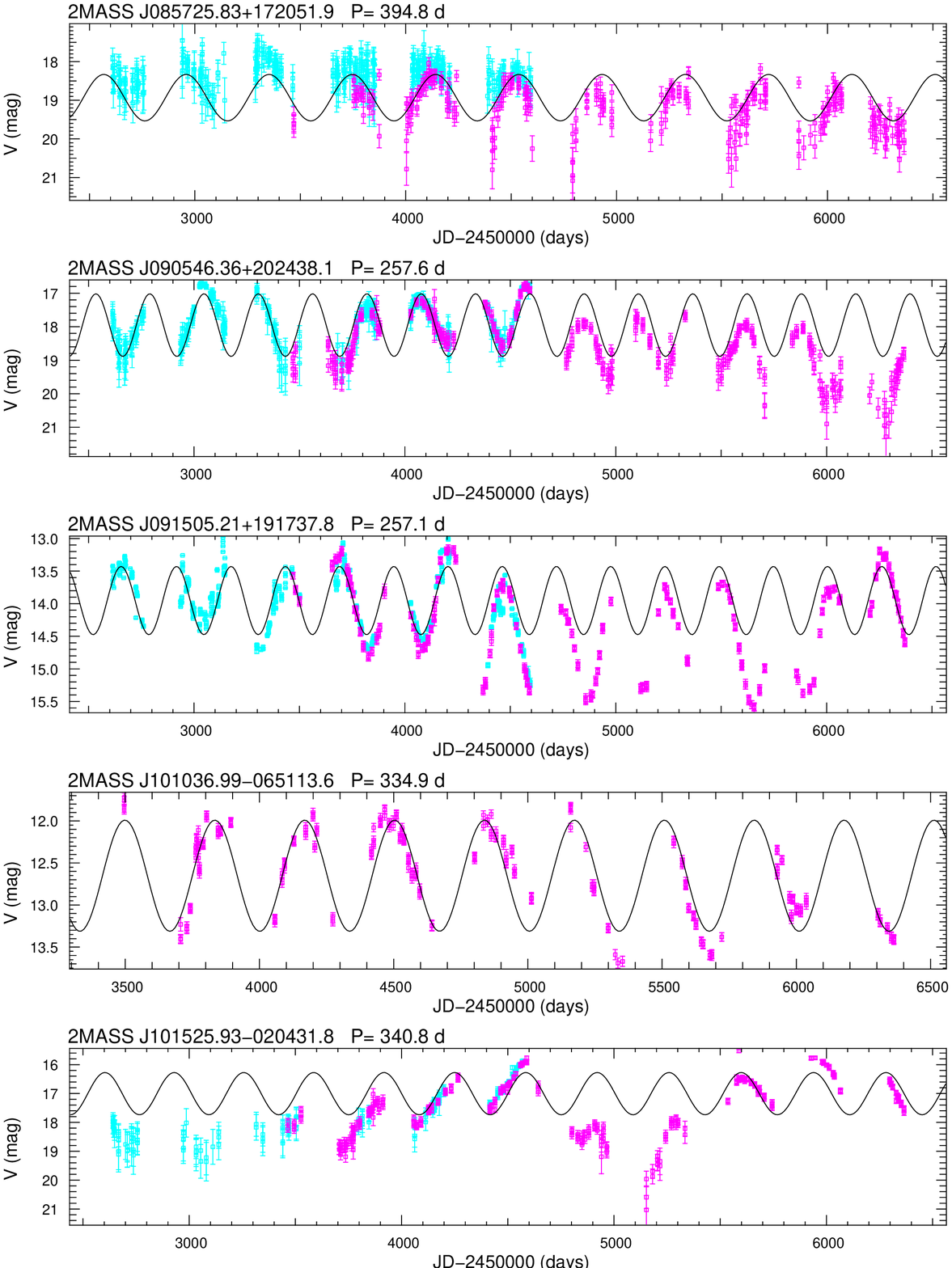}}}
  \end{figure*}

  \begin{figure*}
  \resizebox{17cm}{!}{\rotatebox{00}{\includegraphics{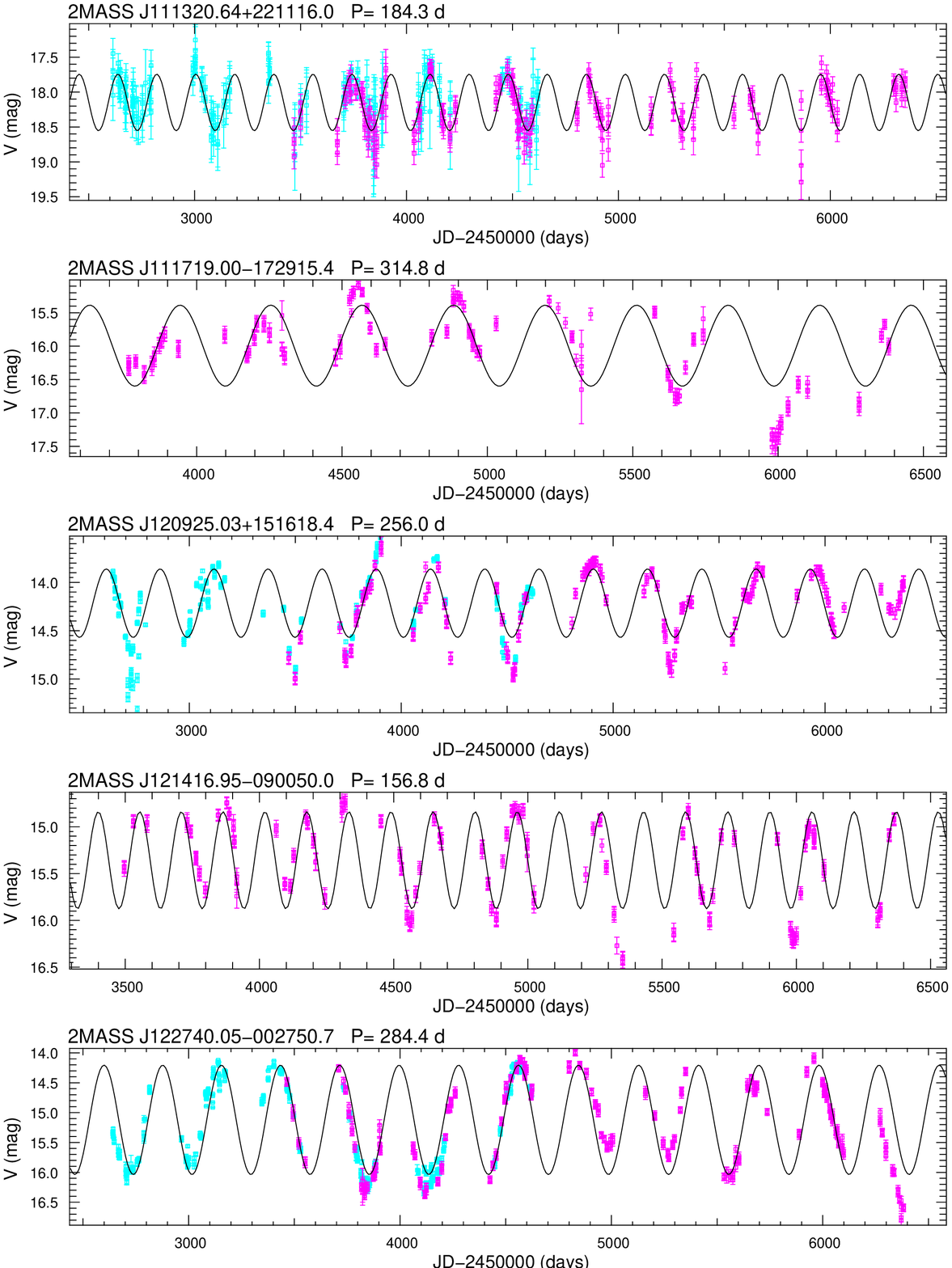}}}
  \end{figure*}

  \begin{figure*}
  \resizebox{17cm}{!}{\rotatebox{00}{\includegraphics{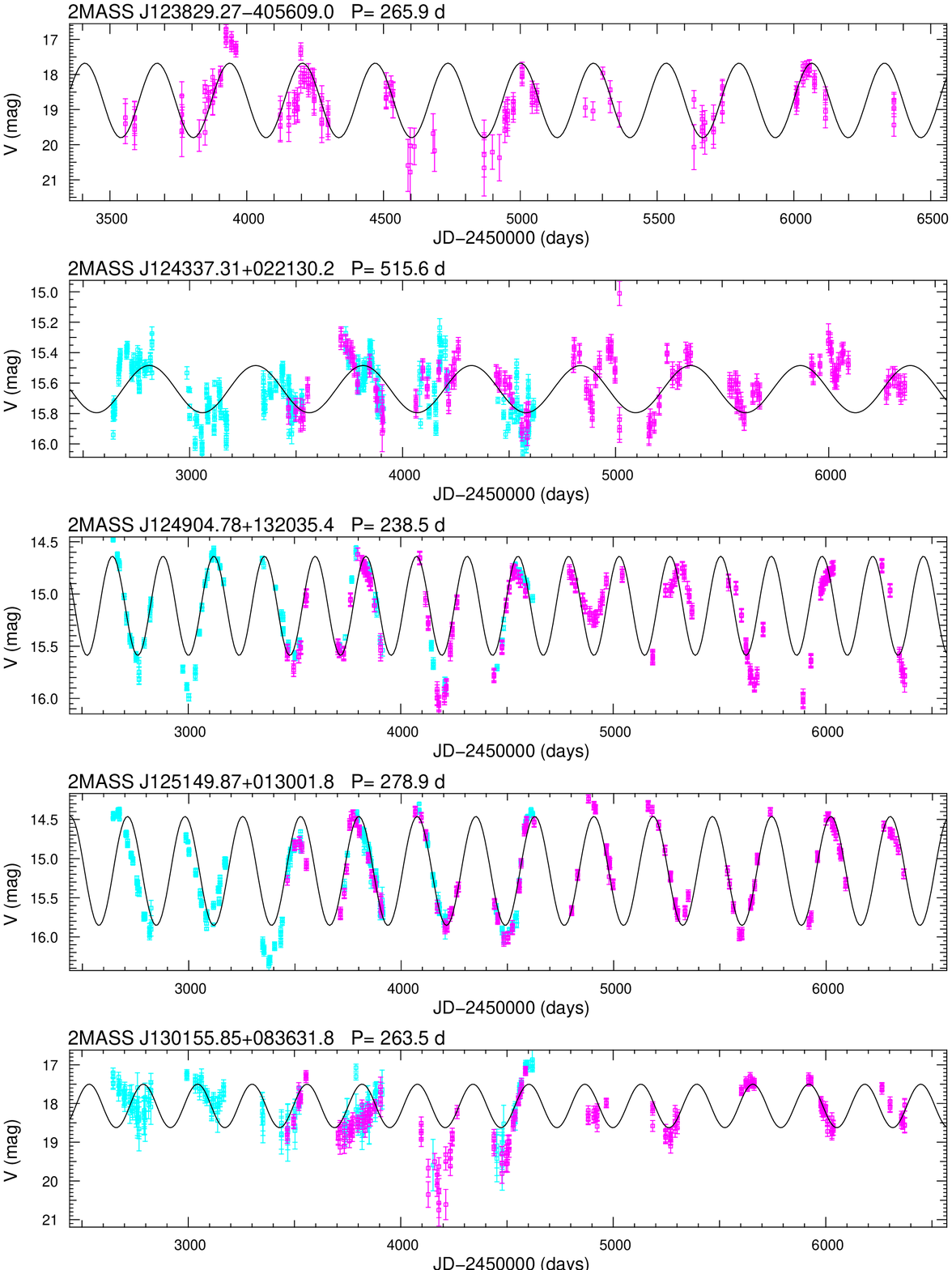}}}
  \end{figure*}

    \begin{figure*}
   \resizebox{17cm}{!}{\rotatebox{00}{\includegraphics{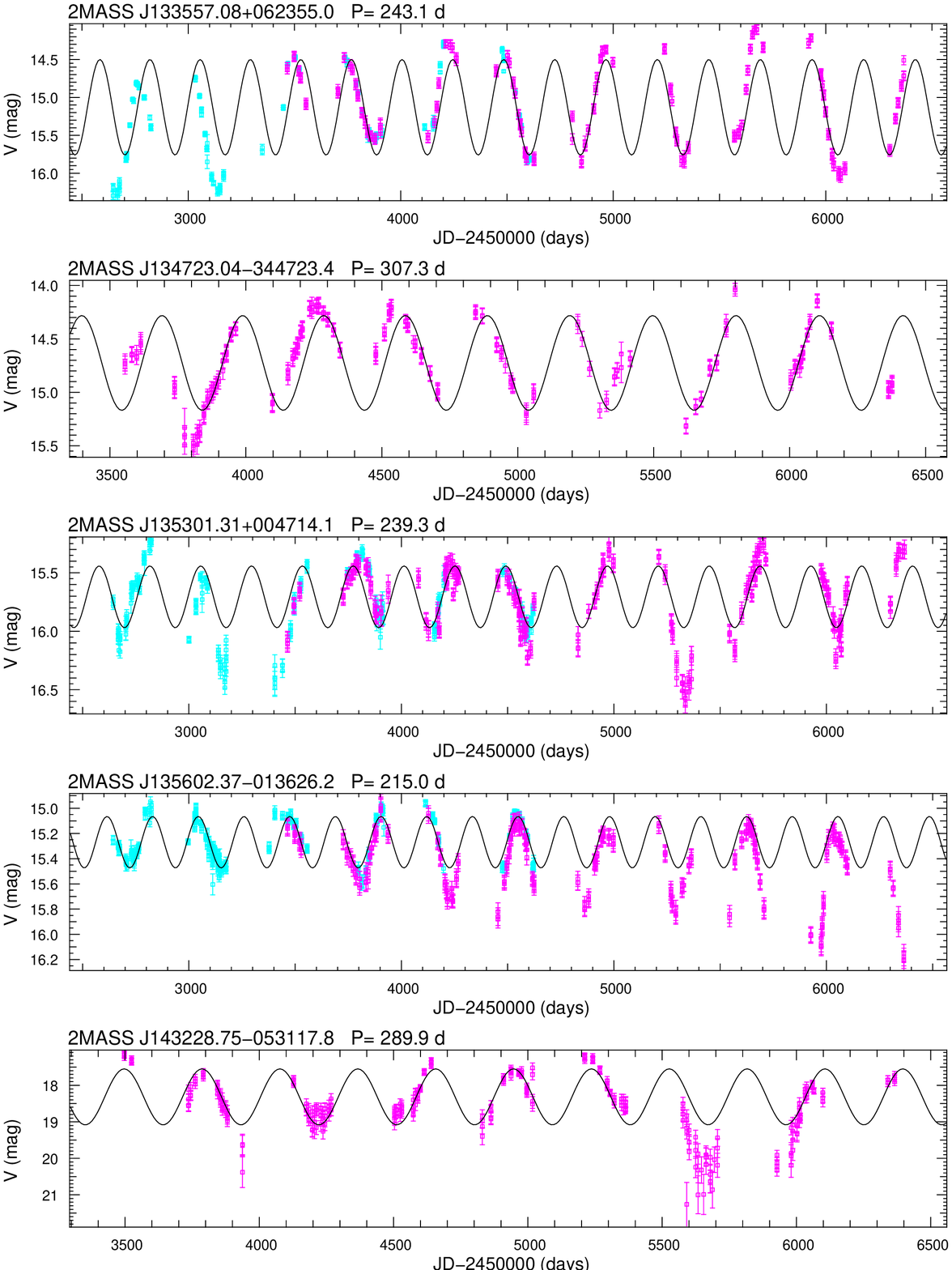}}}
   \end{figure*}

   \begin{figure*}
   \resizebox{17cm}{!}{\rotatebox{00}{\includegraphics{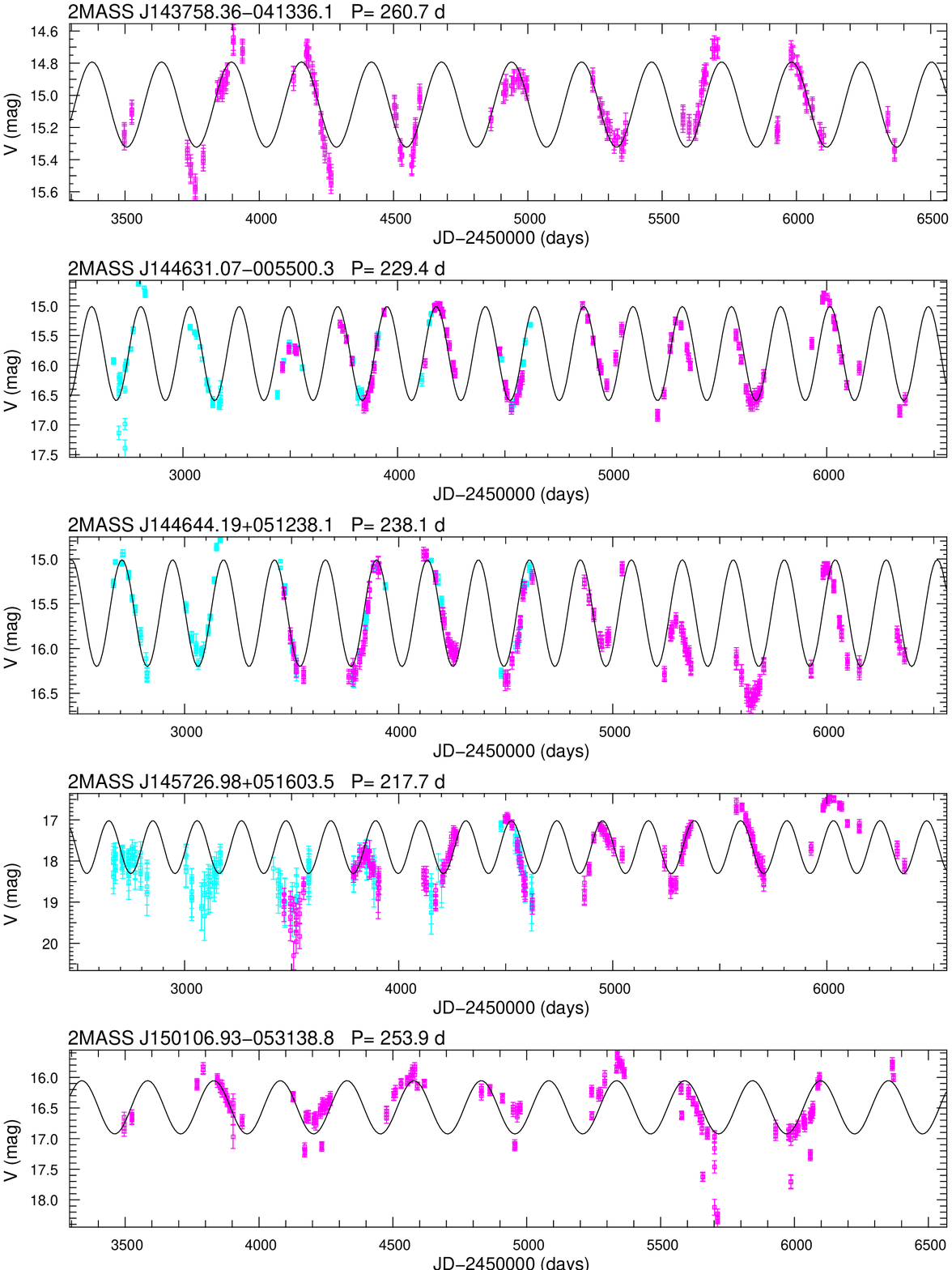}}}
   \end{figure*}

   \begin{figure*}
   \resizebox{17cm}{!}{\rotatebox{00}{\includegraphics{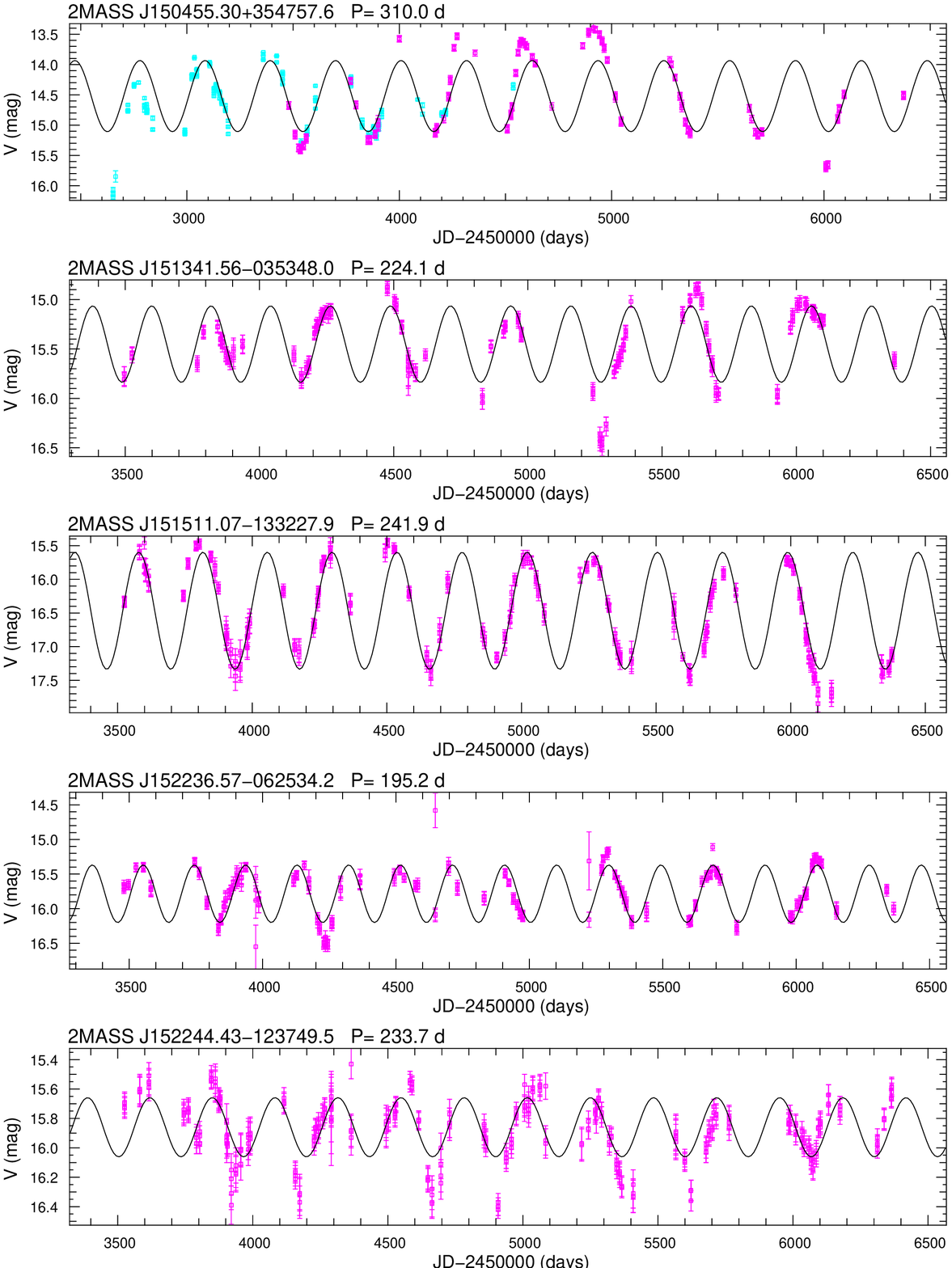}}}
   \end{figure*}

    \begin{figure*}
    \resizebox{17cm}{!}{\rotatebox{00}{\includegraphics{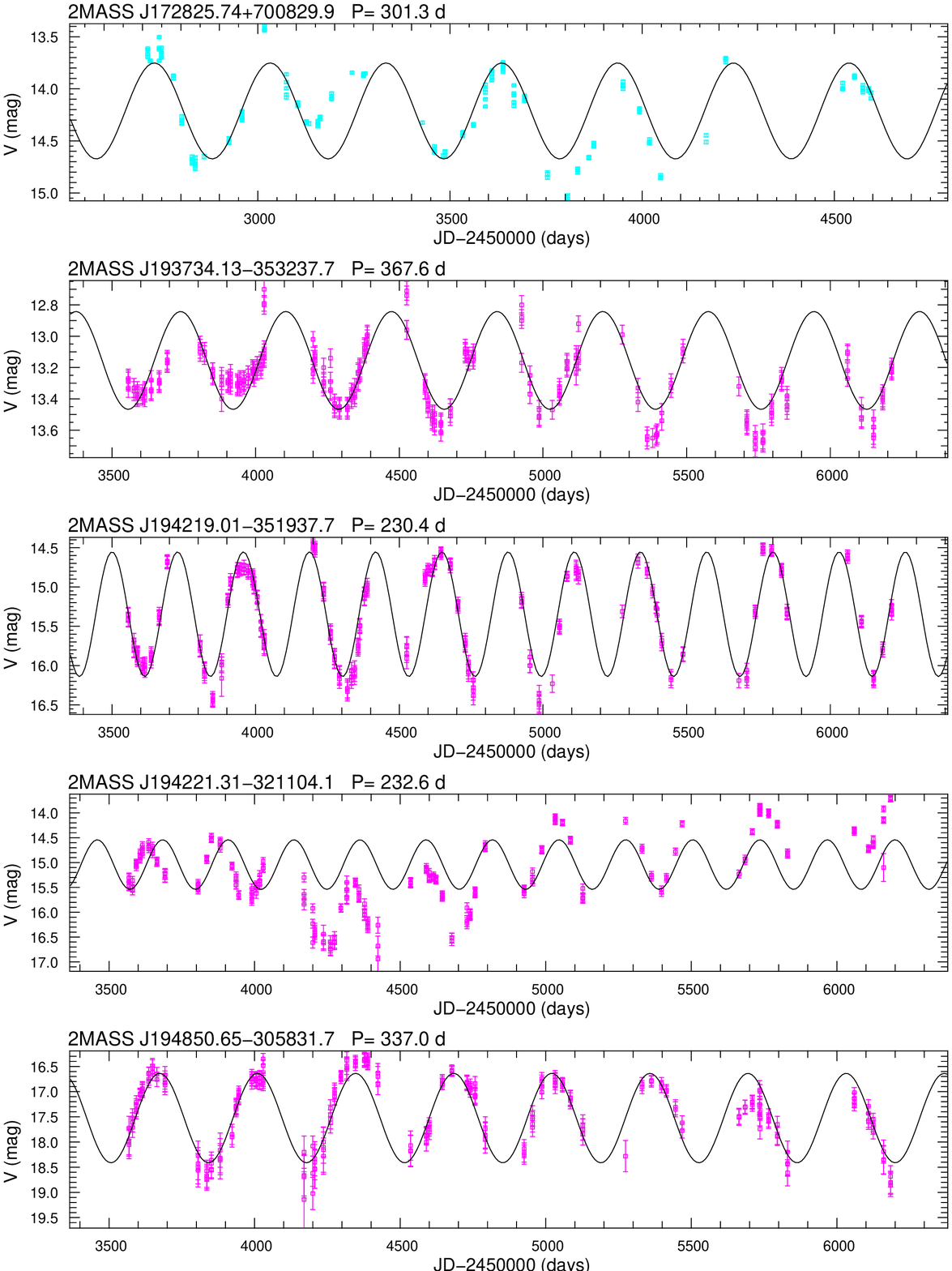}}}
    \end{figure*}

  \newpage

   \begin{figure*}
   \resizebox{17cm}{!}{\rotatebox{00}{\includegraphics{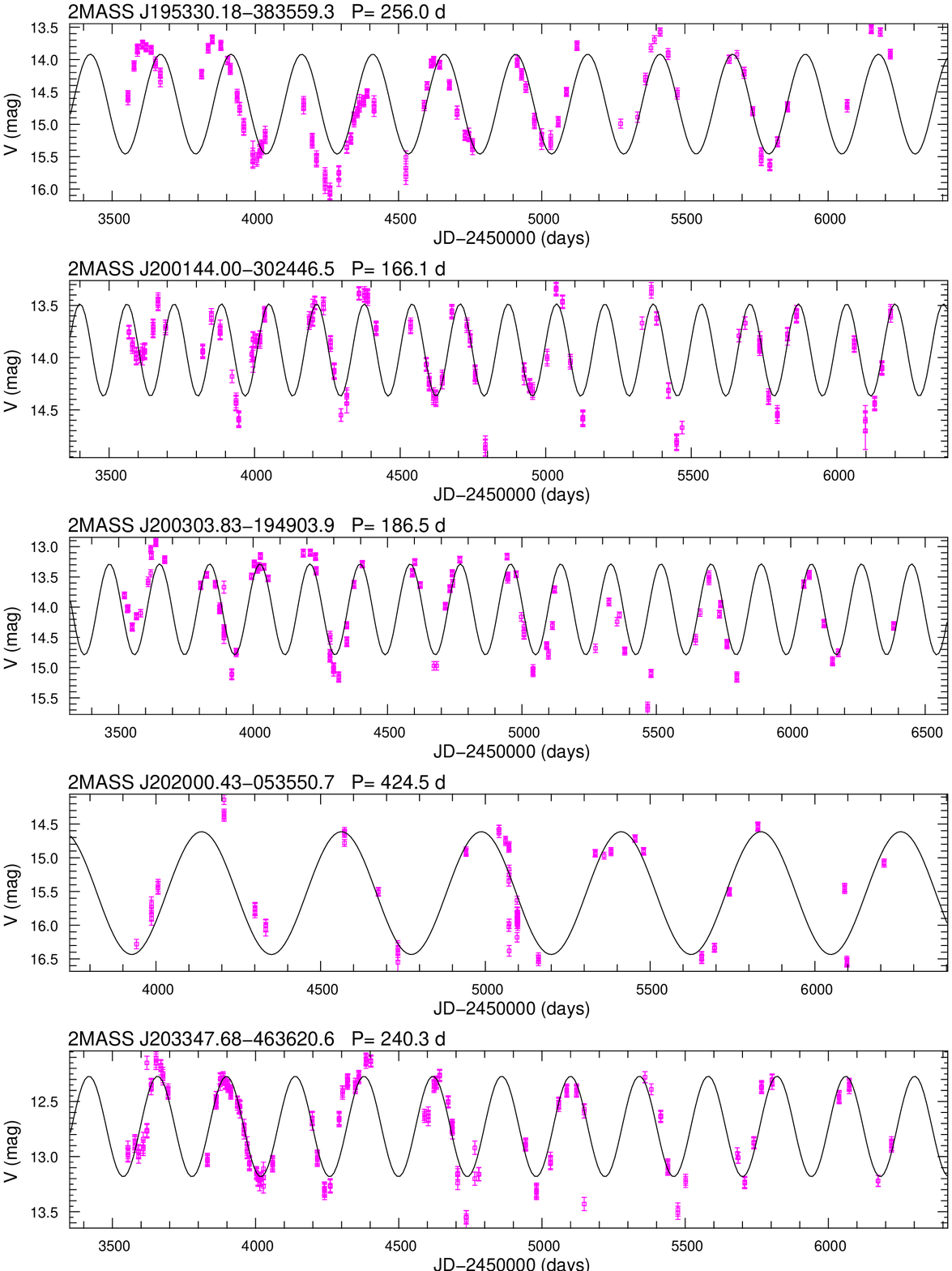}}}
    \end{figure*}

   \begin{figure*}
   \resizebox{17cm}{!}{\rotatebox{00}{\includegraphics{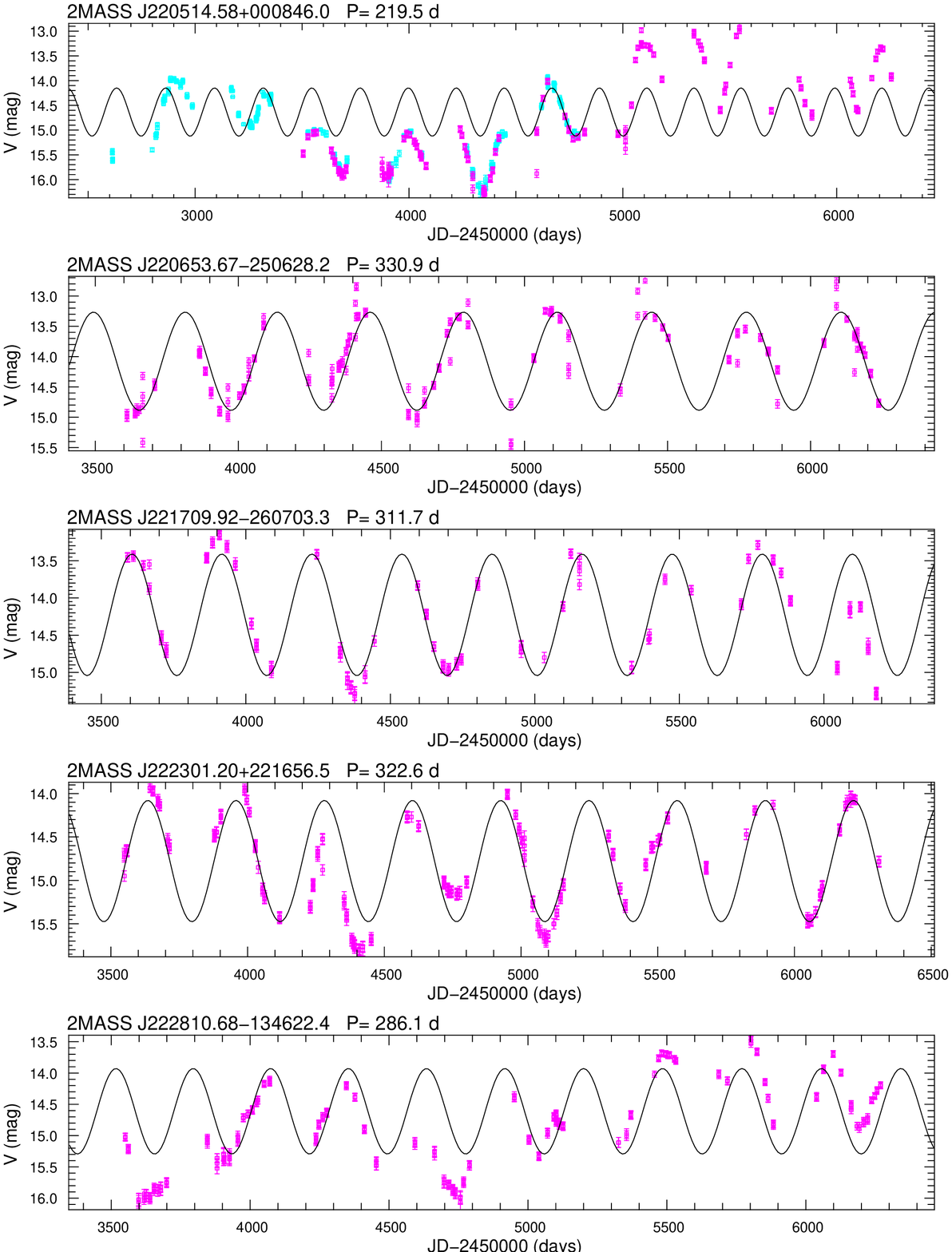}}}
   \end{figure*}

   \begin{figure*}
   \resizebox{17cm}{!}{\rotatebox{00}{\includegraphics{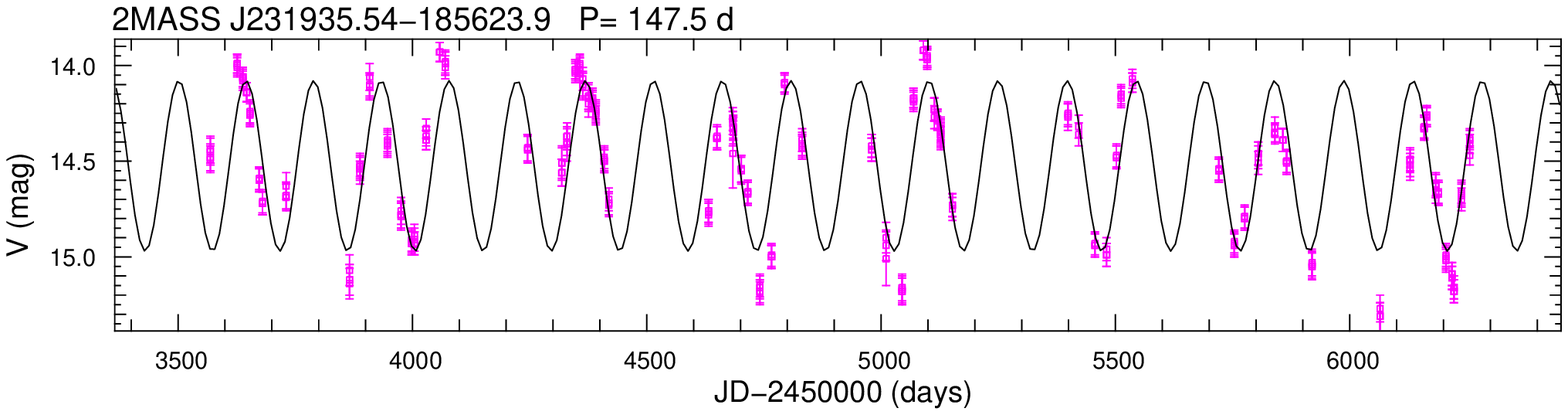}}}
   \end{figure*}

     \clearpage

%--------------------------------   figure pour objets de Fornax

  \begin{figure*}
  \caption{Light curves of Fornax C stars  and fitted sinusoids}
  \vspace{10mm}
  \resizebox{17cm}{!}{\rotatebox{00}{\includegraphics{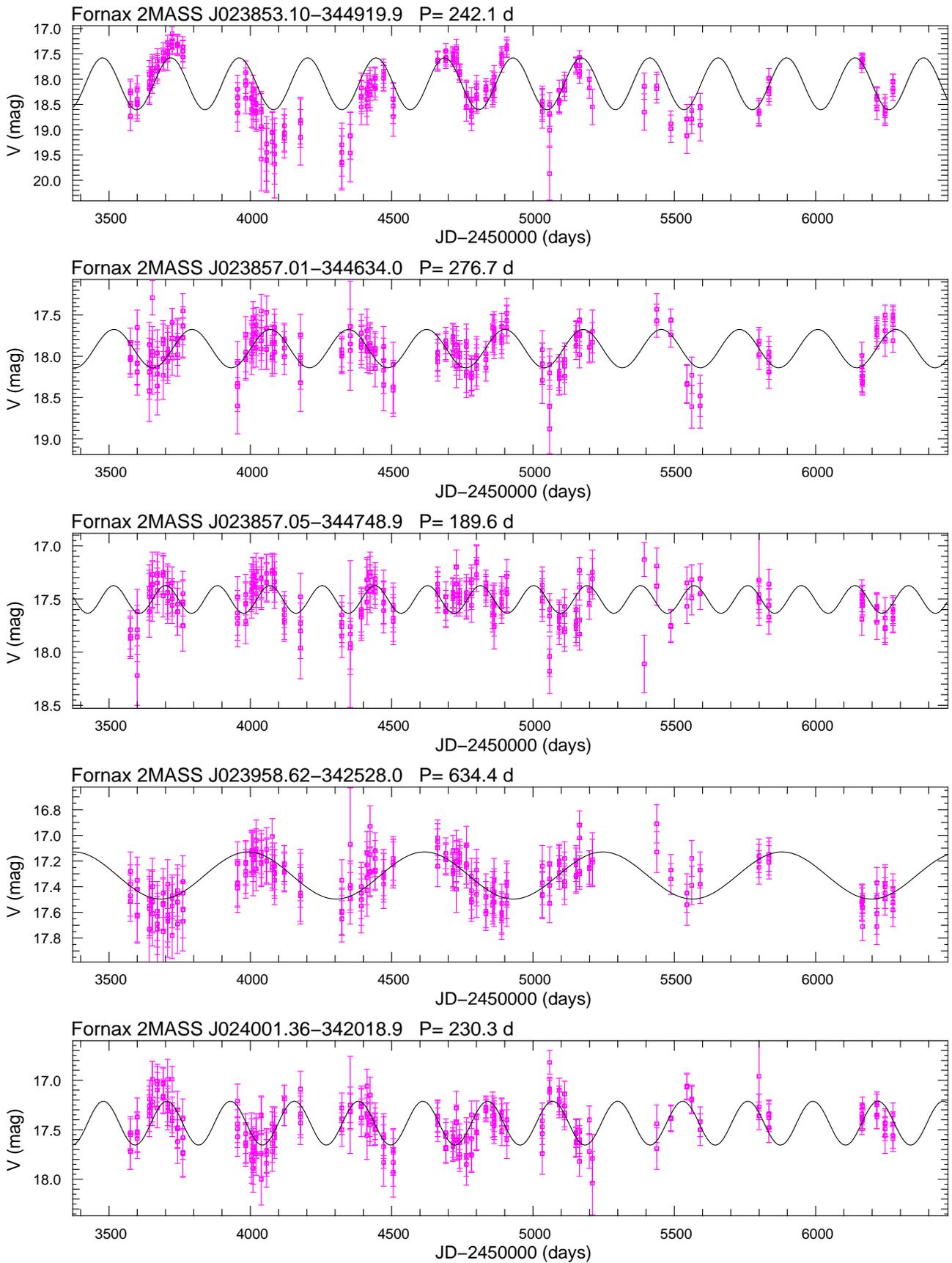}}}   
  \end{figure*}

  \begin{figure*}
  \resizebox{17cm}{!}{\rotatebox{00}{\includegraphics{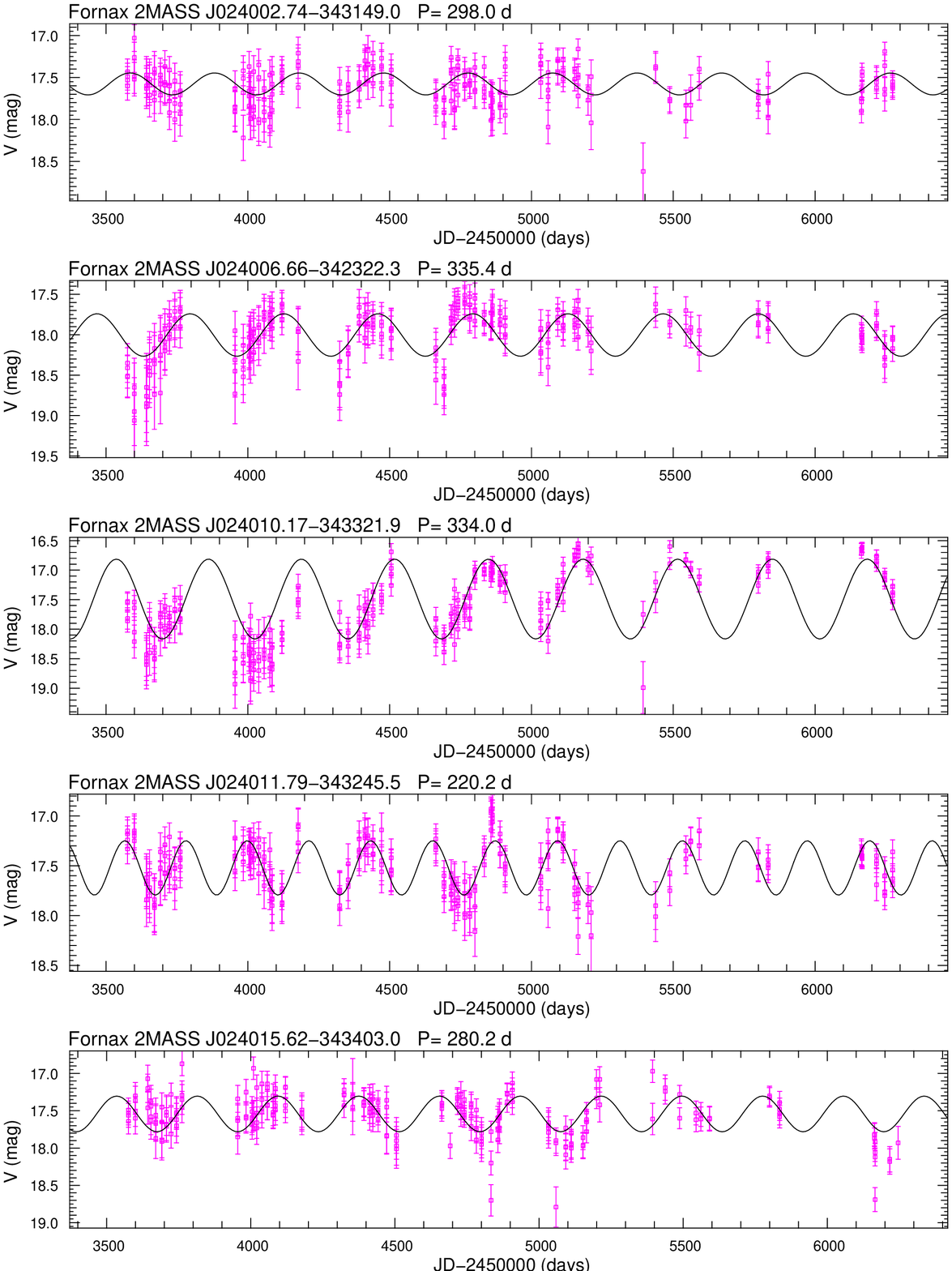}}}
  \end{figure*}

   \begin{figure*}
   \resizebox{17cm}{!}{\rotatebox{00}{\includegraphics{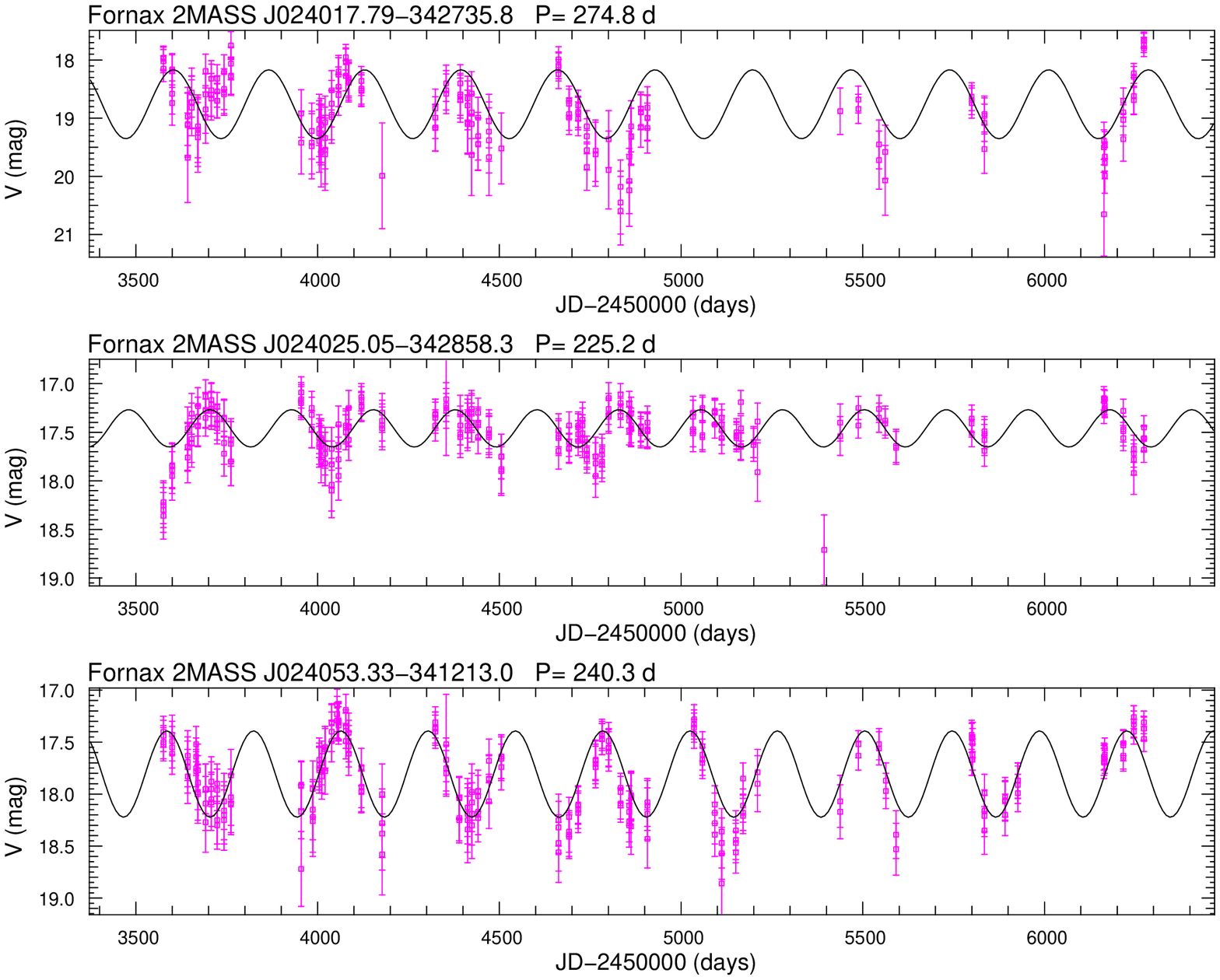}}}
   \end{figure*}

\end{document}